\documentclass[pra,twocolumn,showpacs,amsmath,amssymb,floatfix]{revtex4}

\usepackage{amsmath}
\usepackage{graphicx}
\usepackage{bm}

\begin{document}

\title{Casimir-Polder forces: a nonperturbative approach}

\author{Stefan Yoshi Buhmann}
\email{s.buhmann@tpi.uni-jena.de}

\author{Ludwig Kn\"{o}ll}

\author{Dirk-Gunnar Welsch}

\affiliation{Theoretisch-Physikalisches Institut,
Friedrich-Schiller-Universit\"{a}t Jena,
Max-Wien-Platz 1, 07743 Jena, Germany}

\author{\surname{Ho} Trung Dung}

\affiliation{Institute of Physics, National Center
for Sciences and Technology, 1 Mac Dinh Chi Street,
District 1, Ho Chi Minh city, Vietnam}

\date{\today}

\begin{abstract}
Within the frame of macroscopic QED in linear, causal media, we study
the radiation force of Casimir-Polder type acting on an atom which is
positioned near dispersing and absorbing magnetodielectric bodies
and initially prepared in an arbitrary electronic state. It is shown
that minimal and multipolar coupling lead to essentially the same
lowest-order perturbative result for the force acting on an atom in an
energy eigenstate. To go beyond perturbation theory, the calculations
are based on the exact center-of-mass equation of motion. For a
nondriven atom in the weak-coupling regime, the force as a function
of time is a superposition of force components that are related to the
electronic density-matrix elements at a chosen time. Even the force
component associated with the ground state is not derivable from a
potential in the ususal way, because of the position dependence of the
atomic polarizability. Further, when the atom is initially prepared
in a coherent superposition of energy eigenstates, then temporally
oscillating force components are observed, which are due to the
interaction of the atom with both electric and magnetic fields.
\end{abstract}


\maketitle

%%%%%%%%%%%%%%%%%%%%%%%%%%%%%%%%%%%%%%%%%%%%%%%%%%%%%%%%%%%%%%%%%%%%%%

\section{Introduction}
\label{sec1}

It is well known that in the presence of macroscopic bodies an atom in
the ground state (or in an excited energy eigenstate) is subject to a
nonvanishing force --- the Casimir-Polder (CP) force --- that results
from the vacuum fluctuations of the electromagnetic field. CP forces
play an important role in a variety of processes in physical
chemistry, atom optics, and cavity QED. Moreover, they hold the key to
a number of potential applications in nanotechnology such as the
construction of atomic-force microscopes \cite{Binnig86} or reflective
atom-optical elements \cite{Shimizu02}. Over the years, substantial
efforts have been made to improve the understanding of CP forces  (for
reviews, see Ref.~\cite{Dzyaloshinskii61}). Measuring CP forces acting
on individual particles is a challenging task. Since the early
observation of the deflection of thermal atomic beams by conducting
surfaces \cite{Raskin69}, measurement techniques and precision have
been improving continuously. More recent experiments have been
performed with atomic beams traversing between parallel plates
\cite{Anderson88}. Other methods include transmission grating
diffraction of molecular beams \cite{Grisenti99}, atomic quantum
reflection \cite{Shimizu01,Friedrich02}, evanescent-wave atomic mirror
techniques \cite{Landragin96}, and indirect measurements via
spectroscopic means \cite{Oria91}. Proposals have been made on
improvements of monitoring the CP interaction by using atomic
interferometry \cite{Gorlicki99}.

The theoretical approaches to the problem of determining the CP force
can be roughly divided into two categories. In the first,
first-principle approach explicit field quantization is performed
and perturbation theory is applied to calculate the body-induced
atomic energy shift, which is regarded as the potential of the force
in lowest-order perturbation theory \cite{Casimir48,Bullough70,
Renne,Milloni,Tikochinski,Bostrom00,Marvin82,Wu}. The calculations
have typically been based on macroscopic QED, by beginning with
a normal-mode decomposition and including the bodies via the
well-known conditions of continuity at the surfaces of discontinuity.
Since in such a (noncausal) approach the frequency dependence of the
bodies' response to the field cannot be properly taken into account,
material dispersion and absorption are commonly ignored. As has been
shown recently \cite{Buhmann03}, the problem does not occur within the
frame of a generalized quantization scheme that properly takes into
account a Kramers-Kronig consistent response of the bodies to the
field. Clearly, the problem can also be circumvented in microscopic
QED, where the bodies are treated on a microscopic level by adopting,
e.g., harmonic-oscillator models \cite{Renne}. In the second,
semiphenomenological approach, the problem is circumvented by basing
the calculations on linear response theory (LRT), without explicitly
quantizing the electromagnetic field \cite{McLachlan,Agarwal75,
WylieSipe,Girard,Kryszewski92,Fichet,Boustimi,Henkel02}. In the ansatz
for the force, either the field quantities or both the field and the
atomic entities are expressed in terms of correlation functions, which
in turn are related, via the fluctuation-dissipation theorem, to
response functions.

At first glance one would expect the result obtained from exploiting
LRT to be more generally valid than the QED result obtained in
lowest-order perturbation theory. In fact, this is not the case. In
both approaches, it is not the exact atomic polarizability that enters
the expression for the (ground-state) CP force but the approximate
expression which is obtained in lowest-order perturbation theory and
which effectively corresponds to the atomic polarizability in free
space. Since the structure of the electromagnetic field is changed in
the presence of macroscopic bodies, the atomic polarizability is
expected to change as well. It is well known that the atomic level
shifts and broadenings sensitively depend on the material
surroundings. In particular, when an atom is situated very close to a
body, the effect can be quite significant (see, e.g.,
Refs.~\cite{Ho01,Bondarev02}), thereby changing the atomic
polarizability. As a result, a position-dependent polarizability is
expected to occur, which prevents the CP force from being derivable
from a potential in the usual way.

A way to derive a more rigorous expression for the CP force is to base
the calculations on the exact quantum-mechanical center-of-mass
equation of motion of the atom as we shall do in this paper. The
calculations are performed for both minimal and multipolar coupling,
and contact is made with earlier studies of the center-of-mass motion
of an atom in free space, with special emphasis on the so-called
R\"{o}ntgen interaction term that appears in the multipolar
Hamiltonian \cite{Healy77,Baxter93,Lembessis93,Wilkens93,Guillot02}.
After taking the expectation value with respect to the internal
(electronic) quantum state of the atom and the quantum state of the
medium-assisted electromagnetic field, the resulting force formula
can be used to calculate the time-dependent force acting on a
nondriven or driven atom that is initially prepared in an arbitrary
(internal) quantum state. In this paper, the force formula is further
evaluated for the case of a nondriven, initially arbitrarily prepared
atom, by assuming weak atom-field coupling treated in Markovian
approximation. It is worth noting that the theory, being based on the
quantized version of the macroscopic Maxwell field, with the bodies
being described in terms of spatially varying, Kramers-Kronig
consistent complex permittivities and permeabilities
\cite{Knoll01,Ho03}, also applies to left-handed materials
\cite{Veselago68} where standard quantization runs into difficulties.

The paper is organized as follows. After a brief sketch of the
quantization scheme (Sec.~\ref{sec2}), in Sec.~\ref{sec3}
attention is focused on the perturbative treatment of the CP force
acting on an atom in an energy eigenstate, and previous results
\cite{Buhmann03} obtained for dielectric surroundings of the atom are
extended to magnetodielectric surroundings, including left-handed
materials. In Sec.~\ref{sec4} the exact center-of-mass Heisenberg
equation of motion of an atom and the Lorentz force therein are
studied, and Sec.~\ref{sec5} is devoted to the calculation of the
average force, with special emphasis on a nondriven atom in the
weak-coupling regime. Finally, a summary and some concluding remarks
are given in Sec.~\ref{sec6}.

%%%%%%%%%%%%%%%%%%%%%%%%%%%%%%%%%%%%%%%%%%%%%%%%%%%%%%%%%%%%%%%%%%%%%%

\section{Sketch of the quantization scheme}
\label{sec2}

\subsection{Minimal coupling}
\label{sec2.1}

In Coulomb gauge, the minimal-coupling Hamiltonian of an atomic system
(e.g., an atom or a molecule) consisting of nonrelativistic charged
particles interacting with the electromagnetic field in the presence
of macroscopic dispersing and absorbing bodies reads \cite{Ho03}
\begin{eqnarray}
\label{eq1}
      \hat{H}&=&
      \sum_{\lambda=e,m}\int\mathrm{d}^3r
      \int_0^{\infty}\mathrm{d}\omega\,
      \hbar\omega\,\hat{\mathbf{f}}_\lambda^{\dagger}
      (\mathbf{r},\omega)
      \hat{\mathbf{f}}_\lambda(\mathbf{r},\omega)
\nonumber\\
      &&+\sum_{\alpha}\frac{1}{2 m_{\alpha}}
      \left[\hat{\mathbf{p}}_{\alpha}
      -q_{\alpha}\hat{\mathbf{A}}(\hat{\mathbf{r}}_{\alpha})\right]^2
\nonumber\\
      &&
      + {\textstyle\frac{1}{2}}\int\mathrm{d}^3r\,
      \hat{\rho}_A(\mathbf{r})
      \hat{\varphi}_\mathrm{A}(\mathbf{r})
      +\int\mathrm{d}^3r\,\hat{\rho}_A(\mathbf{r})
      \hat{\varphi}(\mathbf{r}),
      \quad
\end{eqnarray}
where
\begin{equation}
\label{eq2}
      \hat{\rho}_\mathrm{A}(\mathbf{r})
      =\sum_{\alpha}q_{\alpha}\delta(\mathbf{r} 
      -\hat{\mathbf{r}}_\alpha)
\end{equation}
and
\begin{equation}
\label{eq3}
      \hat{\varphi}_\mathrm{A}(\mathbf{r})
      =  \int \mathrm{d}^3{r}'
      \frac{\hat{\rho}_\mathrm{A}(\mathbf{r}')}
      {4\pi\varepsilon_0|\mathbf{r}-\mathbf{r}'|}
\end{equation}
are the charge density and scalar potential of the particles,
respectively. The particle labeled $\alpha$ has charge $q_\alpha$,
mass $m_\alpha$, position $\hat{\mathbf{r}}_{\alpha}$, and canonically
conjugated momentum $\hat{\mathbf{p}}_{\alpha}$. The fundamental
Bosonic fields $\hat{\mathbf{f}}_\lambda({\mathbf{r}},\omega)$ [and 
$\hat{\mathbf{f}}_\lambda^\dagger(\mathbf{r},\omega)$] which can be
related to noise polarization (for $\lambda$ $\!=$ $\!e$) and noise magnetization
(for $\lambda$ $\!=$ $\!m$), respectively, are the dynamical variables
for describing the system composed of the electromagnetic field and
the medium including the dissipative system responsible for
absorption,
\begin{align}
\label{eq4}
      &\hspace{-1ex}
      \left[\hat{f}_{\lambda i}(\mathbf{r},\omega),
      \hat{f}^\dagger_{\lambda' i'}(\mathbf{r}',\omega')\right]
      =\delta_{\lambda\lambda'}
      \delta_{ii'}\delta(\mathbf{r}-\mathbf{r}')
      \delta(\omega-\omega'),
\\
\label{eq5}
      &\hspace{-1ex}
      \left[\hat{f}_{\lambda i}(\mathbf{r},\omega),
      \hat{f}_{\lambda' i'}(\mathbf{r}',\omega')\right]
      =0.
\end{align}
Note that the first term on the right-hand side of Eq.~(\ref{eq1})
is the energy of that system. Further 
$\hat{\mathbf{A}}(\mathbf{r})$ and $\hat{\varphi}(\mathbf{r})$ are
the vector and scalar potentials of the medium-assisted
electromagnetic field, respectively, which in Coulomb gauge are
expressed in terms of the fundamental fields 
$\hat{\mathbf{f}}_\lambda(\mathbf{r},\omega)$ [and 
$\hat{\mathbf{f}}_\lambda^\dagger(\mathbf{r},\omega)$] as
\begin{gather}
\label{eq6}
      \hat{\mathbf{A}}(\mathbf{r}) =
      \int_0^\infty \mathrm{d} \omega \, (i\omega)^{-1}
      \underline{\hat{\mathbf{E}}}{^\perp}(\mathbf{r},\omega)
      + \mathrm{H.c.},
\\[1ex]
\label{eq7}
      -\bm{\nabla} \hat{\varphi}(\mathbf{r})
      = \int_0^\infty \mathrm{d} \omega\,
      \underline{\hat{\mathbf{E}}}{^\parallel}(\mathbf{r},\omega)
      + \mathrm{H.c.},
\end{gather}
where
\begin{eqnarray}
\label{eq8}
     \underline{\hat{\mathbf{E}}}(\mathbf{r},\omega)=
      \sum_{\lambda=e,m}\int \mathrm{d}^3r'\,
      \bm{G}_\lambda(\mathbf{r},\mathbf{r}',\omega)
      \hat{\mathbf{f}}_\lambda(\mathbf{r}',\omega),
\end{eqnarray}
\begin{equation}
\label{eq9}
     \bm{G}_e(\mathbf{r},\mathbf{r}',\omega) = i\,\frac{\omega^2}{c^2}
     \sqrt{\frac{\hbar}{\pi\varepsilon_0}\,
     \mathrm{Im}\,\varepsilon(\mathbf{r}',\omega)}\,
     \bm{G}(\mathbf{r},\mathbf{r}',\omega),
\end{equation}
\begin{eqnarray}
\label{eq10}
     \bm{G}_m(\mathbf{r},\mathbf{r}',\omega)
     &=& -i\,\frac{\omega}{c}
     \sqrt{-\frac{\hbar}{\pi\varepsilon_0}\,
     \mathrm{Im}\,\kappa(\mathbf{r}',\omega)}\nonumber\\
     &&\times\bigl[\bm{G}(\mathbf{r},\mathbf{r}',\omega)\!\times\!\!
     \overleftarrow{\bm{\nabla}}_{\!\!\mathbf{r}'}\bigr]\!,
\end{eqnarray}
with $\bigl[\bm{G}(\mathbf{r},\mathbf{r}',\omega)\!\times\!
\overleftarrow{\bm{\nabla}}_{\!\!\mathbf{r}'}\bigr]_{ij}$
$\!=$ $\!\epsilon_{jkl}\partial'_l 
G_{ik}(\mathbf{r},\mathbf{r}',\omega)
$ and $\kappa(\mathbf{r},\omega)$ $\!=$
$\!\mu^{-1}(\mathbf{r},\omega)$.
Here and in the following, transverse and longitudinal vector fields
are denoted by $\perp$ and $\parallel$, respectively, e.g.,
\begin{equation}
\label{eq11}
      \hat{\underline{\mathbf{E}}}{^{\perp(\parallel)}}
      (\mathbf{r},\omega)
      = \int \mathrm{d}^3 r' \, \bm{\delta}
      ^{\perp(\parallel)}(\mathbf{r}-\mathbf{r}')
      {} \hat{\underline{\mathbf{E}}}(\mathbf{r}',\omega),
\end{equation}
with
\begin{equation}
\label{eq12}
 \delta_{ij}^\parallel(\mathbf{r})
 =-\partial_i\partial_j\left(\frac{1}{4\pi r}\right)
\end{equation}
and
\begin{equation}
\label{eq13}
 \delta_{ij}^\perp(\mathbf{r})
 =\delta(\mathbf{r})\delta_{ij}-\delta_{ij}^\parallel(\mathbf{r})
\end{equation}
being the longitudinal and transverse dyadic $\delta$ functions,
respectively.

In Eqs.~(\ref{eq9}) and (\ref{eq10}), 
$\bm{G}(\mathbf{r},\mathbf{r}',\omega)$ is the (classical) Green
tensor, which in the case of magnetodielectric matter obeys the equation
\begin{equation}
\label{eq14}
      \left[
      \bm{\nabla}\times\kappa(\mathbf{r},\omega)
      \bm{\nabla}\times
      -\frac{\omega^2}{c^2}\,\varepsilon(\mathbf{r},\omega)
      \right]
      \bm{G}(\mathbf{r},\mathbf{r}',\omega)
      = \bm{\delta}(\mathbf{r}-\mathbf{r}')
\end{equation}
together with the boundary condition 
\begin{equation}
\label{eq14.1}
\bm{G}(\mathbf{r},\mathbf{r}',\omega)\to 0 \quad
\mbox{for }|\mathbf{r}-\mathbf{r}'|\to \infty.
\end{equation}
Note that the (relative) permittivity $\varepsilon(\mathbf{r},\omega)$
and permeability $\mu(\mathbf{r},\omega)$ of the (inhomogeneous)
medium are complex functions of frequency, whose real and imaginary
parts satisfy the Kramers-Kronig relations. Since
for absorbing media we have $\mathrm{Im}\,
\varepsilon(\mathbf{r},\omega)$
$\!>$ $\!0$ and $\mathrm{Im}\,\mu(\mathbf{r},\omega)$ $\!>$
$\!0$ $\!\Rightarrow$ $\!\mathrm{Im}\,\kappa(\mathbf{r},\omega)$ $\!<$
$\!0$, the expressions under the square roots in Eqs.~(\ref{eq9})
and (\ref{eq10}) are positive. It should be pointed out that the whole
space is assumed to be filled with some (absorbing) media, in which
case the aforementioned conditions for
$\mathrm{Im}\,\varepsilon(\mathbf{r},\omega)$ and 
$\mathrm{Im}\,\mu(\mathbf{r},\omega)$ ensure that the differential
equation (\ref{eq14}) together with the boundary condition (\ref{eq14.1})
presents a well-defined problem. However, as this assumption allows
for both $\varepsilon(\mathbf{r},\omega)$ and
$\mu(\mathbf{r},\omega)$ to be arbitrarily close to unity (i.e., for
arbitrarily dilute matter), it is naturally possible to include 
vacuum regions in the theory, by performing the limit
\mbox{$\varepsilon(\mathbf{r},\omega)$ $\!\to$ $\!1$},
$\mu(\mathbf{r},\omega)$ $\!\to$ $\!1$ in these regions after 
having calculated the desired expectation values of the relevant
quantities as functions of $\varepsilon(\mathbf{r},\omega)$ and 
$\mu(\mathbf{r},\omega)$.

The Green tensor has the following useful properties \cite{Knoll01}:
\begin{equation}
\label{eq15}
     \bm{G}^{\ast}(\mathbf{r},\mathbf{r}',\omega)
     =\bm{G}(\mathbf{r},\mathbf{r}',-\omega^{\ast}),
\end{equation}
\begin{equation}
\label{eq16}
     \bm{G}(\mathbf{r},\mathbf{r}',\omega)
     =\bm{G}^\top(\mathbf{r}',\mathbf{r},\omega),
\end{equation}
\begin{eqnarray}
\label{eq17}
\lefteqn{
    \int \!\mathrm{d}^3 s\, \Bigl\{
    \mathrm{Im}\,\kappa(\mathbf{s},\omega)
    \bigl[
    \bm{G}(\mathbf{r},\mathbf{s},\omega)
    \times\overleftarrow{{\bm{\nabla}}}_{\!\!\mathbf{s}}
    \bigr]
    \bigl[
    {\bm{\nabla}}_{\mathbf{s}} \!\times
           \bm{G}^\ast(\mathbf{s},\mathbf{r}',\omega) \bigr]
}
\nonumber\\&&\hspace{-2ex}
    + \,\frac{\omega^2}{c^2}\, \mathrm{Im}\,
      \varepsilon(\mathbf{s},\omega)
      \,\bm{G}(\mathbf{r},\mathbf{s},\omega)
      \bm{G}^\ast(\mathbf{s},\mathbf{r}',\omega)
      \Bigr\}
      = \mathrm{Im}\,\bm{G}(\mathbf{r},\mathbf{r}',\omega).
\nonumber\\&&
\end{eqnarray}
Combining Eq.~(\ref{eq17}) with Eqs.~(\ref{eq9}) and (\ref{eq10})
yields
\begin{eqnarray}
\label{eq18}
\lefteqn{
 \sum_{\lambda=e,m}\int \mathrm{d}^3 s\,
 G_{\lambda ik}(\mathbf{r},\mathbf{s},\omega)
 G^\ast_{\lambda jk}(\mathbf{r}',\mathbf{s},\omega)
}
\nonumber\\&&\hspace{5ex}
 =\frac{\hbar\mu_0}{\pi}\omega^2\mathrm{Im}\,
 G_{ij}(\mathbf{r},\mathbf{r}',\omega).
\hspace{10ex}
\end{eqnarray}
Note that in Eq.~(\ref{eq18}) and throughout the remaining part of
this paper, summation over repeated vector indices is understood.

The total electric field is given by
\begin{equation}
\label{eq19}
     \hat{\vec{E}}(\mathbf{r}) = \hat{\mathbf{E}}(\mathbf{r})
     - \bm{\nabla} \hat{\varphi}_\mathrm{A}(\mathbf{r}),
      \end{equation}
where
\begin{equation}
\label{eq20}
      \hat{\mathbf{E}}(\mathbf{r})=\int_0^\infty \mathrm{d} \omega\,
      \underline{\hat{\mathbf{E}}}(\mathbf{r},\omega)
      + \mathrm{H.c.},
\end{equation}
with $\underline{\hat{\mathbf{E}}}(\mathbf{r},\omega)$ from
Eq.~(\ref{eq8}). Accordingly, the total induction field reads
\begin{equation}
\label{eq21}
\hat{\vec{B}}(\mathbf{r}) = \hat{\mathbf{B}}(\mathbf{r})
= \int_0^\infty \mathrm{d} \omega\,
      \underline{\hat{\mathbf{B}}}(\mathbf{r},\omega)
      + \mathrm{H.c.},
\end{equation}
where
\begin{equation}
\label{eq22}
\underline{\hat{\mathbf{B}}}(\mathbf{r},\omega) = (i\omega)^{-1}
\bm{\nabla}\times \underline{\hat{\mathbf{E}}}(\mathbf{r},\omega).
\end{equation}
Finally, the displacement and magnetic fields are given by
\begin{eqnarray}
\label{eq23}
     \hat{\vec{D}}(\mathbf{r})
     \hspace{-1ex}&=&\hspace{-1ex}
     \hat{\mathbf{D}}(\mathbf{r})
     - \varepsilon_0\bm{\nabla} 
     \hat{\varphi}_\mathrm{A}(\mathbf{r})\nonumber\\
     \hspace{-1ex} &=&\hspace{-1ex}
     \int_0^\infty \mathrm{d} \omega\,
     \Big[\underline{\hat{\mathbf{D}}}(\mathbf{r},\omega)
     + \mathrm{H.c.}\Big]
     - \varepsilon_0\bm{\nabla} 
     \hat{\varphi}_\mathrm{A}(\mathbf{r}),
      \\
\label{eq24}
\hat{\vec{H}}(\mathbf{r})
     \hspace{-1ex} &=&\hspace{-1ex}
      \hat{\mathbf{H}}(\mathbf{r})
     =\int_0^\infty \mathrm{d} \omega\,
      \underline{\hat{\mathbf{H}}}(\mathbf{r},\omega)
      + \mathrm{H.c.},
\end{eqnarray}
where
\begin{eqnarray}
\label{eq25}
\lefteqn{
\underline{\hat{\mathbf{D}}}(\mathbf{r},\omega)
= \varepsilon_0\varepsilon(\mathbf{r},\omega)
\underline{\hat{\mathbf{E}}}(\mathbf{r},\omega)
}
\nonumber\\&&\hspace{8ex}
+\,i\sqrt{\frac{\hbar\varepsilon_0}{\pi}\,
\mathrm{Im}\,\varepsilon(\mathbf{r},\omega)}\,
\hat{\mathbf{f}}_e(\mathbf{r},\omega),
\\[1ex]
\label{eq26}
\lefteqn{
\underline{\hat{\mathbf{H}}}(\mathbf{r},\omega) 
= \kappa_0\kappa(\mathbf{r},\omega)
\underline{\hat{\mathbf{B}}}(\mathbf{r},\omega)
}
\nonumber\\&&\hspace{8ex}
-\,\sqrt{-\frac{\hbar\kappa_0}{\pi}\,
\mathrm{Im}\,\kappa(\mathbf{r},\omega)}\,
\hat{\mathbf{f}}_m(\mathbf{r},\omega).
\end{eqnarray}

Assuming that the atomic system is sufficiently localized, and
introducing shifted particle coordinates
\begin{equation}
\label{eq27}
      \hat{\bar{\mathbf{r}}}_\alpha=\hat{\mathbf{r}}_\alpha
      -\hat{\mathbf{r}}_\mathrm{A}
\end{equation}
relative to the center of mass
\begin{equation}
\label{eq28}
       \hat{\mathbf{r}}_\mathrm{A} = \sum_\alpha
       \frac{m_\alpha}{m_\mathrm{A}}
       \,\hat{\mathbf{r}}_\alpha
\end{equation}
($m_\mathrm{A}$ $\!=$ $\!\sum_\alpha m_\alpha$), we can apply the
long-wavelength approximation by expanding the fields 
$\hat{\mathbf{A}}(\mathbf{r})$ and $\hat{\varphi}(\mathbf{r})$ around
the center of mass and keeping only the leading nonvanishing terms of
the respective field operators. For a neutral atomic system,
\begin{equation}
\label{eq29}
    q_\mathrm{A}=\sum_\alpha q_\alpha=0,
\end{equation}
this is just the familiar electric dipole approximation, and the
Hamiltonian (\ref{eq1}) simplifies to
\begin{equation}
\label{eq30}
     \hat{H} =  \hat{H}_\mathrm{F} + \hat{H}_\mathrm{A} +
     \hat{H}_\mathrm{AF},
\end{equation}
where
\begin{align}
\label{eq31}
      &\hat{H}_\mathrm{F} \equiv
      \sum_{\lambda=e,m}\int\mathrm{d}^3 {r}
      \int_0^{\infty}\mathrm{d}\omega\,\hbar\omega
      \,\hat{\mathbf{f}}_\lambda^{\dagger}(\mathbf{r},\omega)
      \hat{\mathbf{f}}_\lambda(\mathbf{r},\omega),
\\[1ex]
\label{eq32}
      &\hat{H}_\mathrm{A} \equiv
      \sum_{\alpha}
      \frac{\hat{\mathbf{p}}_{\alpha}^2}{2m_{\alpha}}
      +{\textstyle\frac{1}{2}}\int \mathrm{d}^3 r\,
      \hat{\rho}_\mathrm{A}(\mathbf{r})
      \hat{\varphi}_\mathrm{A}(\mathbf{r}),
\\[1ex]
\label{eq33}
      &\hat{H}_\mathrm{AF} \equiv
      \hat{\mathbf{d}}\bm{\nabla}
      \hat{\varphi}(\mathbf{r})|
      _{\mathbf{r}=\hat{\mathbf{r}}_\mathrm{A}}
      -\sum_{\alpha}\frac{q_{\alpha}}{m_{\alpha}}
      \hat{\mathbf{p}}_{\alpha}
      \hat{\mathbf{A}}({\hat{\mathbf{r}}_\mathrm{A}})
\nonumber\\
      &\hspace{10ex}
      + \sum_{\alpha}\frac{q_{\alpha}^2}{2m_{\alpha}}
      \,\hat{\mathbf{A}}^2({\hat{\mathbf{r}}_\mathrm{A}}),
\end{align}
with
\begin{equation}
\label{eq34}
      \hat{\mathbf{d}}
      = \sum_{\alpha}q_{\alpha}\hat{\mathbf{r}}_{\alpha}
      = \sum_{\alpha}q_{\alpha}\hat{\bar{\mathbf{r}}}_{\alpha}
\end{equation}
being the total electric dipole moment.

%%%%%%%%%%%%%%%%%%%%%%%%%%%%%%%%%%%%%%%%%%%%%%%%%%%%%%%%%%%%%%%%%%%%%%

\subsection{Multipolar coupling}
\label{sec2.2}

Let us turn to the multipolar coupling scheme widely used for studying
the interaction of electromagnetic fields with atoms and molecules.
Just as in standard QED, so in the present formalism
\cite{Knoll01,Ho03}, the multipolar Hamiltonian can be obtained from
the minimal-coupling Hamiltonian by means of a Power-Zienau
transformation,
\begin{equation}
\label{eq35}
      \hat{U} = \exp \left[\frac{i}{\hbar} \int \mathrm{d}^3 r
      \,\hat{\mathbf{P}}_\mathrm{A}(\mathbf{r}) 
      \hat{\mathbf{A}}(\mathbf{r})
      \right],
\end{equation}
where
\begin{equation}
\label{eq36}
     \hat{\mathbf{P}}_\mathrm{A}(\mathbf{r}) =
     \sum_\alpha q_\alpha
     \hat{\bar{\mathbf{r}}}_\alpha
     \int _0^1 \mathrm{d}\lambda
     \,\delta(\mathbf{r} - \hat{\mathbf{r}}_\mathrm{A}
     - \lambda\hat{\bar{\mathbf{r}}}_\alpha ).
\end{equation}
For a neutral atomic system, the multipolar Hamiltonian [which is
obtained by expressing the Hamiltonian (\ref{eq1}) in terms of the
transformed variables] can be given in the form of
(see Appendix~\ref{AppA})
\begin{eqnarray}
\label{eq37}
      \hat{H}&=&
      \sum_{\lambda=e,m}\int\mathrm{d}^3r
      \int_0^{\infty}\mathrm{d}\omega
      \,\hbar\omega\,
      \hat{\mathbf{f}}'_\lambda{\!^\dagger}(\mathbf{r},\omega)
      \hat{\mathbf{f}}_\lambda'(\mathbf{r},\omega)
\nonumber\\&&
      +\, \frac{1}{2\varepsilon_0} \int\mathrm{d}^3r
      \,\hat{\mathbf{P}}^2_\mathrm{A} (\mathbf{r})
      - \int\mathrm{d}^3r\, \hat{\mathbf{P}}_\mathrm{A} (\mathbf{r})
                       \hat{\mathbf{E}}' (\mathbf{r})
\nonumber\\
      &&+\,\sum_{\alpha}\frac{1}{2 m_{\alpha}}
      \left[\hat{\mathbf{p}}'_\alpha
      +  \int\!\mathrm{d}^3 r\, \hat{\bm{\Xi}}_\alpha(\mathbf{r})
      \times \hat{\mathbf{B}}'(\mathbf{r}) \right]^2,\qquad
\end{eqnarray}
where
\begin{equation}
\label{eq38}
        \hat{\bm{\Xi}}_\alpha (\mathbf{r}) =
        q_\alpha \hat{\bm{\Theta}}_\alpha (\mathbf{r})
        - \frac{m_\alpha}{m_\mathrm{A}}
         \sum_\beta q_\beta 
         \hat{\bm{\Theta}}_\beta (\mathbf{r})
        + \frac{m_\alpha}{m_\mathrm{A}} 
        \hat{\mathbf{P}}_\mathrm{A} (\mathbf{r})
\end{equation}
and
\begin{eqnarray}
\label{eq39}
&\displaystyle\hspace{-5ex}
     \hat{\bm{\Theta}}_\alpha(\mathbf{r}) =
     \hat{\bar{\mathbf{r}}}_\alpha
     \int _0^1 \mathrm{d}\lambda\, \lambda
     \,\delta(\mathbf{r} - \hat{\mathbf{r}}_\mathrm{A}
     - \lambda\hat{\bar{\mathbf{r}}}_\alpha).
\end{eqnarray}
Note that due to the unitarity of the transformation (\ref{eq35}), the
transformed variables of the atomic system
$\hat{\mathbf{r}}'_\alpha$ $\!=$ $\!\hat{\mathbf{r}}_\alpha$ and
$\hat{\mathbf{p}}'_\alpha$ and the transformed field variables
$\hat{\mathbf{f}}'_{\lambda}(\mathbf{r},\omega)$ and
$\hat{\mathbf{f}}^{\prime\dagger}_{\lambda}(\mathbf{r},\omega)$
obey the same commutation relations as the untransformed ones.
Needless to say that the transformed fields
$\hat{\mathbf{E}}'(\mathbf{r})$ and $\hat{\mathbf{B}}'(\mathbf{r})$
are related to the transformed fields 
$\hat{\mathbf{f}}_\lambda'(\mathbf{r},\omega)$ and
$\hat{\mathbf{f}}'_\lambda{\!^\dagger}(\mathbf{r},\omega)$ according
to Eq.~(\ref{eq8}) and Eqs.~(\ref{eq20})--(\ref{eq22}), with primed
quantities instead of the unprimed ones. The Hamiltonian
(\ref{eq37}) can be regarded as the generalization of the multipolar
Hamiltonian obtained earlier for moving atoms in vacuum
\cite{Healy77,Baxter93,Lembessis93,Wilkens93,Guillot02} to the case
where dispersing and absorbing magnetodielectric bodies are present.
In particular, it can be used to describe effects specifically due to
the translational motion of an atomic system such as Doppler
and recoil effects.

Applying the long-wavelength approximation to the fields
$\hat{\mathbf{E}}'(\mathbf{r})$ and $\hat{\mathbf{B}}'(\mathbf{r})$
in Eq.~(\ref{eq37}), which is equivalent to approximating
$\delta(\mathbf{r}$ $\!-$ $\!\hat{\mathbf{r}}_\mathrm{A}$
$\!-$ $\!\lambda\hat{\bar{\mathbf{r}}}_\alpha)$ by
$\delta(\mathbf{r}$ $\!-$ $\!\hat{\mathbf{r}}_\mathrm{A})$ in
Eqs.~(\ref{eq36}) and (\ref{eq39}), respectively, i.e.,
\begin{equation}
\label{eq40}
      \hat{\mathbf{P}}_\mathrm{A} (\mathbf{r}) = \hat{\mathbf{d}}
      \,\delta(\mathbf{r}-\hat{\mathbf{r}}_\mathrm{A}),
\end{equation}
\begin{equation}
\label{eq41}
      \hat{\bm{\Theta}}_\alpha (\mathbf{r}) = {\textstyle \frac{1}{2}}
      \,\hat{\bar{\mathbf{r}}}_\alpha
      \delta(\mathbf{r}-\hat{\mathbf{r}}_\mathrm{A}),
\end{equation}
thus
\begin{equation}
\label{eq42}
 \hat{\bm{\Xi}}_\alpha (\mathbf{r})
 ={\textstyle\frac{1}{2}}q_\alpha\hat{\bar{\mathbf{r}}}_\alpha
 \delta(\mathbf{r}-\hat{\mathbf{r}}_\mathrm{A})
 +\frac{m_\alpha}{2m_\mathrm{A}}
 \hat{\mathbf{d}}\delta(\mathbf{r}-\hat{\mathbf{r}}_\mathrm{A}),
\end{equation}
we obtain the multipolar Hamiltonian in long-wavelength approximation,
\begin{equation}
\label{eq43}
     \hat{H} =  \hat{H}'_\mathrm{F} + \hat{H}'_\mathrm{A} +
     \hat{H}'_\mathrm{AF},
\end{equation}
with
\begin{align}
\label{eq44}
      &\hat{H}'_\mathrm{F} \equiv
      \sum_{\lambda=e,m}\int\mathrm{d}^3 {r}
      \int_0^{\infty}\mathrm{d}\omega\,\hbar\omega\,
      \hat{\mathbf{f}}'_\lambda{\!^\dagger}(\mathbf{r},\omega)
      \hat{\mathbf{f}}_\lambda'(\mathbf{r},\omega),
\\[1ex]
\label{eq45}
      &\hat{H}'_\mathrm{A} \equiv
      \sum_{\alpha}
      \frac{\hat{\mathbf{p}}'_{\alpha}{\!^2}}{2m_{\alpha}}
      + \frac{1}{2\varepsilon_0} \int\mathrm{d}^3r\,
      \hat{\mathbf{P}}_\mathrm{A}^2 (\mathbf{r}),
\\[1ex]
\label{eq46}
      &\hat{H}'_\mathrm{AF} \equiv
      -\hat{\mathbf{d}}
      \hat{\mathbf{E}}' (\hat{\mathbf{r}}_\mathrm{A})
       + \sum_\alpha \frac{q_\alpha}{2m_\alpha}\,
        \hat{\bar{\mathbf{p}}}{}'_\alpha
        \hat{\bar{\mathbf{r}}}_\alpha\!\times\!
        \hat{\mathbf{B}}'(\hat{\mathbf{r}}_\mathrm{A})
\nonumber\\&\hspace{5ex}
        + \sum_\alpha \frac{q_\alpha^2}{8m_\alpha}
        \bigl[\hat{\bar{\mathbf{r}}}_\alpha
        \times \hat{\mathbf{B}}'(\hat{\mathbf{r}}_\mathrm{A})
        \bigr]^2
        +\, \frac{3}{8m_\mathrm{A}}
        \bigl[\hat{\mathbf{d}}
        \times \hat{\mathbf{B}}'(\hat{\mathbf{r}}_\mathrm{A})
        \bigr]^2
\nonumber\\&\hspace{5ex}
        +\frac{1}{m_\mathrm{A}}\,
        \hat{\mathbf{p}}'_\mathrm{A}\hat{\mathbf{d}}\!\times\!
        \hat{\mathbf{B}}'(\hat{\mathbf{r}}_\mathrm{A}),
\end{align}
where 
\begin{equation}
\label{eq47}
 \hat{\mathbf{p}}'_\mathrm{A}=\sum_\alpha \hat{\mathbf{p}}'_\alpha
\end{equation}
is the (canonical) momentum of the center of mass, and
\begin{equation}
\label{eq48}
 \hat{\bar{\mathbf{p}}}'_\alpha
 =\hat{\mathbf{p}}'_\alpha-\frac{m_\alpha}{m_\mathrm{A}}\,
 \hat{\mathbf{p}}'_\mathrm{A}
\end{equation}
denote shifted momenta of the particles relative to the center of
mass. The first two terms on the right-hand side of Eq.~(\ref{eq46})
represent electric and magnetic dipole interactions, respectively,
the next two terms describe the (generalized) diamagnetic interaction
of the charged particles with the medium-assisted electromagnetic
fields, while the last term describes the R\"{o}ntgen interaction due
to the translational motion of the center of mass. In particular, in
(generalized) electric dipole approximation, Eq.~(\ref{eq46}) reads
\begin{equation}
\label{eq49}
      \hat{H}'_\mathrm{AF}=
      - \hat{\mathbf{d}}
      \hat{\mathbf{E}}' (\hat{\mathbf{r}}_\mathrm{A})
      +\frac{1}{m_\mathrm{A}}\,
        \hat{\mathbf{p}}'_\mathrm{A}\hat{\mathbf{d}}\!\times\!
        \hat{\mathbf{B}}'(\hat{\mathbf{r}}_\mathrm{A}).
\end{equation}
Recall that the transformed medium-assisted electric field
$\hat{\mathbf{E}}'(\hat{\mathbf{r}})$ is related to the physical one,
$\hat{\mathbf{E}}(\hat{\mathbf{r}})$, according to Eq.~(\ref{A4}).

If the center-of-mass coordinate is treated as a (classical) parameter
($\hat{\mathbf{r}}_\mathrm{A}$ $\!\mapsto$ $\!\mathbf{r}_\mathrm{A}$),
then Eq.~(\ref{eq38}) reduces to
\begin{equation}
\label{eq50}
\hat{\bm{\Xi}}_\alpha (\mathbf{r}) =
        q_\alpha \hat{\bm{\Theta}}_\alpha (\mathbf{r}),
\end{equation}
which corresponds to the limit $m_\alpha/m_\mathrm{A}$ $\!\to 0$.
Hence Eq.~(\ref{eq46}) becomes
\begin{eqnarray}
\label{eq51}
      \hat{H}'_\mathrm{AF}&\hspace{-1 ex}=&\hspace{-1 ex}
      -\hat{\mathbf{d}}
      \hat{\mathbf{E}}' (\mathbf{r}_\mathrm{A})
        +\sum_\alpha \frac{q_\alpha}{2m_\alpha}\,
        \hat{\mathbf{p}}'_\alpha
	\hat{\bar{\mathbf{r}}}_\alpha\!\times\!
         \hat{\mathbf{B}}'(\mathbf{r}_\mathrm{A})
         \nonumber\\
      &\hspace{-1 ex}&\hspace{-1 ex}+ \sum_\alpha \frac{q_\alpha^2}{8m_\alpha}
        \big[\hat{\bar{\mathbf{r}}}_\alpha
              \times \hat{\mathbf{B}}'(\mathbf{r}_\mathrm{A}) \big]^2.
\end{eqnarray}
If the paramagnetic and diamagnetic terms are omitted, the interaction
Hamiltonian simply reduces to the first term on the right-hand side of
Eq.~(\ref{eq51}).

%%%%%%%%%%%%%%%%%%%%%%%%%%%%%%%%%%%%%%%%%%%%%%%%%%%%%%%%%%%%%%%%%%%%%%

\section{Van der Waals potential}
\label{sec3}

According to Casimir's and Polder's pioneering concept
\cite{Casimir48}, the CP force on an atomic system near macroscopic
bodies is commonly regarded as being a conservative force. In
particular, it is assumed that for an atom in an eigenstate
$|l\rangle$ of the atomic Hamiltonian the position-dependent shift of
the corresponding eigenvalue due to the (electric-dipole) interaction
of the atomic system with the body-assisted electromagnetic field is
the potential, also referred to as van der Waals (vdW) potential,
from which the CP force can be derived, where the calculations are
usually performed within the frame of lowest-order perturbation
theory. In this picture, the center-of-mass coordinate is a parameter
rather than a dynamical variable ($\hat{\mathbf{r}}_\mathrm{A}$
$\!\mapsto$ $\!\mathbf{r}_\mathrm{A}$). Following this line, we
first extend previous results \cite{Buhmann03}, and show that minimal and
multipolar coupling schemes yield essentially the same expression for
the force.

%%%%%%%%%%%%%%%%%%%%%%%%%%%%%%%%%%%%%%%%%%%%%%%%%%%%%%%%%%%%%%%%%%%%%%

\subsection{Minimal coupling}
\label{sec3.1}

We start from the minimal-coupling Hamiltonian in electric dipole
approximation as given by Eqs.~(\ref{eq30})--(\ref{eq33}) together
with Eq.~(\ref{eq34})
($\hat{\mathbf{r}}_\mathrm{A}$ $\!\mapsto$ $\!\mathbf{r}_\mathrm{A}$).
Let $|n\rangle$ denote the eigenstates of the multilevel atomic system
and write $\hat{H}_\mathrm{A}$ [Eq.~(\ref{eq32})] as
\begin{equation}
\label{eq52}
      \hat{H}_\mathrm{A}
      =\sum_n E_n |n\rangle\langle n|.
\end{equation}
To calculate the leading-order correction to the unperturbed
eigenvalue of a state $|l\rangle|\{0\}\rangle$ due to the perturbation
Hamiltonian (\ref{eq33}) [$|\{0\}\rangle$, ground state of the
fundamental fields $\hat{\mathbf{f}}_\lambda(\mathbf{r},\omega)$], we
first note that the first two terms have no diagonal elements. Thus
they start to contribute in second order,
\begin{eqnarray}
\label{eq53}
      \Delta_2 E_l
      &=&
      -\frac{1}{\hbar}\sum_k\sum_{\lambda=e,m}\mathcal{P}
      \int_0^{\infty}\!\!
      \frac{\mathrm{d}\omega}{\omega_{kl}+\omega}
      \int\!\mathrm{d}^3{r}
 \nonumber\\&&\times\,
      \Bigl|\langle l|\langle\{0\}|
      \hat{\mathbf{d}}\bm{\nabla}
      \hat{\varphi}(\mathbf{r})|_{\mathbf{r}=\mathbf{r}_\mathrm{A}}
      -\sum_{\alpha}\frac{q_{\alpha}}{m_{\alpha}}
      \hat{\mathbf{p}}_{\alpha}
      \hat{\mathbf{A}}(\mathbf{r}_\mathrm{A})
      \nonumber\\&&\times\,
       |\{\mathbf{1}_\lambda(\mathbf{r},\omega)\}\rangle|k\rangle
     \Bigr
     |^2 \qquad
\end{eqnarray}
($\mathcal{P}$, principal part), whereas the third term starts to
contribute in first order,
\begin{equation}
\label{eq54}
      \Delta_1 E_l =
      \langle l|\langle\{0\}|
      \sum_{\alpha}\frac{q_{\alpha}^2}{2m_{\alpha}}
      \,\hat{\mathbf{A}}^2(\mathbf{r}_\mathrm{A})
      |\{0\}\rangle |l\rangle.
\end{equation}
Here,
$|\{\mathbf{1}_\lambda(\mathbf{r},\omega)\}\rangle$ $\!\equiv$
$\!\hat{\mathbf{f}}^{\dagger}_\lambda(\mathbf{r},\omega)|\{0\}\rangle$
denotes single-quan\-tum Fock states of the fundamental fields, and
\begin{equation}
\label{eq55}
        \omega_{kl} \equiv (E_k-E_l)/\hbar
\end{equation}
are the atomic transition frequencies. Since $\Delta_1 E_l$ and
$\Delta_2 E_l$ are quadratic in the coupling constant [Eqs.~(\ref{B9})
and (\ref{B10}) in Appendix~\ref{AppB}], thus being of the same order
of magnitude, the leading-order correction to the eigenvalue is given
by
\begin{equation}
\label{eq56}
      \Delta E_l = \Delta_1 E_l + \Delta_2 E_l.
\end{equation}
A straightforward but somewhat lengthy calculation yields
(see Appendix~\ref{AppC})
\begin{eqnarray}
\label{eq57}
     \Delta E_l
     &=&
     \frac{\mu_0}{\pi}
     \sum_k \mathcal{P}\int_0^{\infty}
     \frac{\mathrm{d}\omega}{\omega_{kl}+\omega}
     \,\mathbf{d}_{lk}\Bigl\{
     \omega_{kl}\omega
     \nonumber\\&&\times\,
     \bigl[
     \mathrm{Im}\,{\bm G}(\mathbf{r}_\mathrm{A},
     \mathbf{r}_\mathrm{A},\omega)
     -\mathrm{Im}\,^\parallel{\bm G}^\parallel
     (\mathbf{r}_\mathrm{A},\mathbf{r}_\mathrm{A},\omega)
     \bigr]\nonumber\\&&
     -\,\omega^2
     \mathrm{Im}\, ^\parallel{\bm G}^\parallel
     (\mathbf{r}_\mathrm{A},\mathbf{r}_\mathrm{A},\omega)
     \Bigr\} \mathbf{d}_{kl},
\end{eqnarray}
with
\begin{equation}
\label{eq58}
\mathbf{d}_{lk} = \langle l|\hat{\mathbf{d}}|k\rangle
\end{equation}
being the dipole matrix elements.

Since the atomic system should be located in a free-space region, the
Green tensor in this region is a linear superposition of the
(translationally invariant) vacuum Green tensor $\bm{G}^{(0)}$
and the scattering Green tensor $\bm{G}^{(1)}$ that accounts for the
spatial variation of the permittivity and permeability,
\begin{equation}
\label{eq59}
     \bm{G}(\mathbf{r},\mathbf{r}',\omega)
     = \bm{G}^{(0)}(\mathbf{r},\mathbf{r}',\omega)
     + \bm{G}^{(1)}(\mathbf{r},\mathbf{r}',\omega).
\end{equation}
As a consequence, the eigenvalue correction $\Delta E_l$ can be decomposed
into two parts,
\begin{equation}
\label{eq60}
     \Delta E_l = \Delta E_l^{(0)} 
     + \Delta E_l^{(1)}(\mathbf{r}_\mathrm{A}).
\end{equation}
The $\mathbf{r}_\mathrm{A}$-independent term $\Delta E_l^{(0)}$
associated with the vacuum Green tensor gives rise to the vacuum Lamb
shift and is not of interest here. The 
$\mathbf{r}_\mathrm{A}$-dependent term
$\Delta E_l^{(1)}(\mathbf{r}_\mathrm{A})$, associated with the
scattering Green tensor, is just the vdW potential sought,
\begin{equation}
\label{eq61}
        U_l(\mathbf{r}_\mathrm{A}) 
        = \Delta E_l^{(1)}(\mathbf{r}_\mathrm{A})
        = \Delta_1 E_l^{(1)}(\mathbf{r}_\mathrm{A})
        + \Delta_2 E_l^{(1)}(\mathbf{r}_\mathrm{A}).
\end{equation}
Hence from Eq.~(\ref{eq57}) [$\bm{G}(\mathbf{r}_\mathrm{A},
\mathbf{r}_\mathrm{A},\omega)$ $\!\mapsto$
$\!\bm{G}^{(1)}(\mathbf{r}_\mathrm{A},\mathbf{r}_\mathrm{A},\omega)$]
we derive, on recalling Eq.~(\ref{eq15}) and changing the integration
variable from $-\omega$ to $\omega$,
\begin{alignat}{1}
\label{eq62}
     &U_l(\mathbf{r}_\mathrm{A})= \frac{\mu_0}{2i\pi}
     \sum_k \mathbf{d}_{lk}
     \biggl[\mathcal{P}\int_0^{\infty}
     \frac{\mathrm{d} \omega}{\omega_{kl}+\omega}
      \nonumber\\&\quad\times
      \Bigl\{\omega_{kl}\omega
      \big[
     {}\bm{G}^{(1)}(\mathbf{r}_\mathrm{A},
     \mathbf{r}_\mathrm{A},\omega)
     - {}^\parallel\bm{G}^{(1)\parallel}(\mathbf{r}_\mathrm{A},
     \mathbf{r}_\mathrm{A},\omega)\big]\nonumber\\
      &\qquad
      -\,\omega^2
     {}^\parallel\bm{G}^{(1)\parallel}(\mathbf{r}_\mathrm{A},
     \mathbf{r}_\mathrm{A},\omega)\Big\}
   -\mathcal{P}\int_0^{-\infty}\frac{\mathrm{d} \omega}
   {\omega_{kl}-\omega}
       \nonumber\\&\quad\times
      \Big\{\omega_{kl}\omega
      \big[
     {}\bm{G}^{(1)}(\mathbf{r}_\mathrm{A},
     \mathbf{r}_\mathrm{A},\omega)
     - {}^\parallel\bm{G}^{(1)\parallel}(\mathbf{r}_\mathrm{A},
     \mathbf{r}_\mathrm{A},\omega)
      \big]
      \nonumber\\&\qquad
     +\,\omega^2
      {}^\parallel\bm{G}^{(1)\parallel}(\mathbf{r}_\mathrm{A},
      \mathbf{r}_\mathrm{A},\omega)\Big\}
      \biggr] \mathbf{d}_{kl}.
\end{alignat}

This equation can be greatly simplified by using con\-tour-integral
techniques. $\bm{G}^{(1)}(\mathbf{r}_\mathrm{A},
\mathbf{r}_\mathrm{A},\omega)$ is an analytic function in the upper
half of the complex frequency plane, including the real axis (apart
from a possible pole at \mbox{$\omega$ $\!=$ $\!0$}). Furthermore,
knowing the asymptotic behaviour of the Green tensor in the limit
$\omega\rightarrow 0$ (cf. Ref.~\cite{Ho03}), one can verify that all
integrands in Eq.~(\ref{eq62}) remain finite in this limit. We may
therefore apply Cauchy's theorem, and replace the principal value
integral over the positive (negative) real half axis by a contour
integral along the positive imaginary half axis (introducing the
purely imaginary coordinate \mbox{$\omega$ $\!=$ $\!i u$}) and along a
quarter circle with infinitely large radius in the first (second)
quadrant of the complex frequency plane plus, in the case of
\mbox{$\omega_{lk}$ $\!>$ $\!0$}, a contour integral along an
infinitesimally small half circle around $\omega$ $\!=$
$\!\omega_{lk}$ ($\omega$ $\!=$ $\!-\omega_{lk}$) in the first
(second) quadrant of the complex frequency plane. The integrals along
the infinitely large quarter circles vanish due to the asymptotic
property
\begin{equation}
\label{eq63}
 \lim_{|\omega|\rightarrow\infty}\frac{\omega^2}{c^2}
 \,\bm{G}^{(1)}(\mathbf{r},\mathbf{r},\omega)=0
\end{equation}
(cf. Ref.~\cite{Ho03}), so we finally arrive at
\begin{equation}
\label{eq63-1}
U_l(\mathbf{r}_\mathrm{A})=U_l^\mathrm{or}(\mathbf{r}_\mathrm{A})
+U_l^\mathrm{r}(\mathbf{r}_\mathrm{A}),
\end{equation}
where
\begin{eqnarray}
\label{eq64}
     U_l^\mathrm{or}(\mathbf{r}_\mathrm{A})
     &=& \frac{\mu_0}{\pi}
     \sum_k
     \!\int_0^{\infty}\!\!\! \mathrm{d} u\,
     \frac{\omega_{kl} u^2}{\omega_{kl}^2 + u^2}
     \nonumber\\&&\times
     \,\mathbf{d}_{lk} \bm{G}^{(1)}(\mathbf{r}_\mathrm{A},
     \mathbf{r}_\mathrm{A},iu)\mathbf{d}_{kl}
\end{eqnarray}
is the off-resonant part of the vdW potential, and
\begin{eqnarray}
\label{eq65}
U_l^\mathrm{r}(\mathbf{r}_\mathrm{A})&=&
     -\mu_0
     \sum_k \Theta(\omega_{lk})
     \omega_{lk}^2\nonumber\\
     &&\times\,\mathbf{d}_{lk}\, \mathrm{Re}\,
     \bm{G}^{(1)}(\mathbf{r}_\mathrm{A},
     \mathbf{r}_\mathrm{A},\omega_{lk})
     \mathbf{d}_{kl}
\quad
\end{eqnarray}
[$\Theta(z)$, unit step function] is the resonant part arising from
the contribution from the residua at the poles. Note that 
$U_l^\mathrm{r}(\mathbf{r}_\mathrm{A})$ vanishes when the atomic
system is in the ground state. For an atomic system in an
ex\-ci\-ted state, $U_l^\mathrm{r}(\mathbf{r}_\mathrm{A})$ may
dominate $U_l^\mathrm{or}(\mathbf{r}_\mathrm{A})$.

The CP force can be derived from Eq.~(\ref{eq63-1}) according to
\begin{equation}
\label{eq66}
\mathbf{F}_l(\mathbf{r}_\mathrm{A})
= - \bm{\nabla}_{\!\!\mathrm{A}}
U_l(\mathbf{r}_\mathrm{A})
\end{equation}
($\bm{\nabla}_{\!\!\mathrm{A}}$ $\!\equiv$ 
$\!\bm{\nabla}_{\!\!\mathbf{r}_\mathrm{A}}$). A formula of the type of
Eq.~(\ref{eq63-1}) together with Eqs.~(\ref{eq64}) and (\ref{eq65})
was first given in Ref.~\cite{WylieSipe} within the frame of LRT.

To give Eq.~(\ref{eq64}) in a more compact form, we introduce the
generalized atomic polarizability tensor
\begin{eqnarray}
\label{eq67}
      \bm{\alpha}_{mn}(\omega) &=&
      \frac{1}{\hbar} \sum_k
       \biggl[
        \frac{\mathbf{d}_{mk} \otimes \mathbf{d}_{kn}}
       {\tilde{\omega}_{kn} - \omega - i(\Gamma_k+\Gamma_m)/2}
       \nonumber\\
       &&+\,\frac{\mathbf{d}_{kn} \otimes \mathbf{d}_{mk}}
       {\tilde{\omega}_{km} + \omega + i(\Gamma_k+\Gamma_n)/2}
       \biggr],
\end{eqnarray}
where $\tilde{\omega}_{km}$ are the shifted (renormalized) transition
frequencies and $\Gamma_k$ are the excited-state widths. Following
Ref.~\cite{Fain63}, we may regard
\begin{equation}
\label{eq68}
\bm{\alpha}_l(\omega) = \bm{\alpha}_{ll}(\omega)
\end{equation}
as being the ordinary (Kramers-Kronig-con\-sist\-ent) polarizability
tensor of an atom in state $|l\rangle$. Hence we may rewrite
Eq.~(\ref{eq64}) as
\begin{equation}
\label{eq69}
     U_l^\mathrm{or}(\mathbf{r}_\mathrm{A})
     = \frac{\hbar\mu_0}{2\pi}
     \int_0^{\infty} \mathrm{d}u \,u^2 
     \mathrm{Tr}\bigl[\bm{\alpha}_l^{(0)}(iu)
     \,\bm{G}^{(1)}(\mathbf{r}_\mathrm{A},\mathbf{r}_\mathrm{A},iu)
     \bigr],
\end{equation}
where
\begin{equation}
\label{eq70}
     \bm{\alpha}_l^{(0)}(\omega)=
     \lim_{\epsilon\to 0}
     \frac{2}{\hbar}\sum_k \frac{\omega_{kl}}{\omega_{kl}^2-\omega^2
     - i\omega\epsilon}
     \,\mathbf{d}_{lk}\otimes\mathbf{d}_{kl}
\end{equation}
is the polarizability tensor in lowest-order perturbation theory,
which can be obtained from Eq.~(\ref{eq68}) together with
Eq.~(\ref{eq67}) by ignoring both the level shifts and broadenings.
In particular for an atom in a spherically symmetric state, we have
\begin{equation}
\label{eq71}
     \bm{\alpha}_l^{(0)}(\omega) = \alpha_l^{(0)}(\omega)\bm{I}=
     \lim_{\epsilon\to 0}
     \frac{2}{3\hbar}\sum_k
     \frac{\omega_{kl}}{\omega_{kl}^2-\omega^2
     - i\omega\epsilon}
     \,|\mathbf{d}_{lk}|^2\bm{I}
\end{equation}
($\bm{I}$, unit tensor), so that Eq.~(\ref{eq69}) reduces to
\begin{equation}
\label{eq72}
     U_l^\mathrm{or}(\mathbf{r}_\mathrm{A})
     = \frac{\hbar\mu_0}{2\pi}
     \int_0^{\infty} \mathrm{d}u \,u^2 \alpha_l^{(0)}(iu)
     \,\mathrm{Tr}\,
     \bm{G}^{(1)}(\mathbf{r}_\mathrm{A},\mathbf{r}_\mathrm{A},iu),
\end{equation}
and Eq.~(\ref{eq65}) simplifies to
\begin{eqnarray}
\label{eq73}
U_l^\mathrm{r}(\mathbf{r}_\mathrm{A})&=&
-\frac{\mu_0}{3}\sum_k\Theta(\omega_{lk})
     \omega_{lk}^2
     |\mathbf{d}_{lk}|^2
     \nonumber\\
      &&\times\,\mathrm{Tr}\,\bigl[\mathrm{Re}\,
      \bm{G}^{(1)}(\mathbf{r}_\mathrm{A},
      \mathbf{r}_\mathrm{A},\omega_{lk})\bigr].
\end{eqnarray}
Note that
\begin{equation}
\label{eq74}
\bm{\alpha}_l(iu) \simeq \bm{\alpha}_l^{(0)}(iu)
\end{equation}
is typically valid for an atomic system in free space, because of the
smallness of the level shifts and broadenings that result from the
interaction of the atomic system with the vacuum electromagnetic
field.

Equation (\ref{eq63-1}) together with Eqs.~(\ref{eq64}) and (\ref{eq65})
can be regarded as being the natural extension of the QED results obtained
on the basis of the familiar normal-mode formalism, which ignores
material absorption. Moreover, it does not only apply to arbitrary
causal dielectric bodies, but, to our knowledge,  
it first proves applicable to magnetodielectric matter such as
left-handed material, for which standard quantization concepts run
into difficulties. Note that all information about the electric and
magnetic properties of the matter is contained in the scattering
Green tensor.

Finally, let us briefly comment on the ground-state potential as
given by Eq.~(\ref{eq69}) for $l$ $\!=$ $\!0$. In terms of an integral along
the positive frequency axis, it reads 
\begin{eqnarray}
\label{eq148-1}
\lefteqn{
U_0(\mathbf{r}_\mathrm{A})
     =-\frac{\hbar\mu_0}{2\pi}
      \int_0^\infty \mathrm{d}\omega\, \omega^2
}
\nonumber\\
&&\hspace{2ex}\times\;
     \mathrm{Im}\left\{\mathrm{Tr}\,\bigl[\bm{\alpha}_0^{(0)}(\omega)
     \bm{G}^{(1)}
     (\mathbf{r}_\mathrm{A},\mathbf{r}_\mathrm{A},\omega)\bigl]
    \right\}.
     \qquad
\end{eqnarray}
An expression of this type can also be obtained by using the methods
of LRT \cite{WylieSipe,Kryszewski92}. It allows for a simple physical
interpretation for the ground-state CP force as being due to
correlations of the fluctuating electromagnetic field with the
corresponding induced electric dipole of the atomic system plus the
correlations of the fluctuating electric dipole moment with its
induced electric field \cite{Henkel02}.

%%%%%%%%%%%%%%%%%%%%%%%%%%%%%%%%%%%%%%%%%%%%%%%%%%%%%%%%%%%%%%%%%%%%%%
\subsection{Multipolar coupling}
\label{sec3.2}

Let us now consider the multipolar Hamiltonian in long-wavelength
approximation as given by Eqs.~(\ref{eq43})--(\ref{eq45}) together
with Eq.~(\ref{eq51}), and write $\hat{H}'_\mathrm{A}$
[Eq.~(\ref{eq45})] in the form of Eq.~(\ref{eq52}). In contrast to the
electric dipole approximation considered in the minimal coupling
scheme, the present Hamiltonian also includes magnetic interactions.
One might therefore expect that the leading-order corrections to the
unperturbed eigenvalues are given by the second-order corrections due
to the dipole interactions (linear in the field variables) plus the
first-order correction due to the diamagnetic interaction (quadratic
in the field variables), all of these contributions being quadratic in
the coupling constant. However, one can show
[Eqs.~(\ref{B12})--(\ref{B14}) in Appendix~\ref{AppB}] that the
second-order eigenvalue correction due to magnetic dipole interaction
is smaller than that due to the electric dipole interaction by a
factor of $(Z_\mathrm{eff}\alpha_0)^2$, where $Z_\mathrm{eff}$ is the
effective nucleus charge felt by the electrons giving the main 
contribution to the energy shift, and $\alpha_0$ is the fine-structure
constant. The current formalism based on Hamiltonian (\ref{eq1}) only
treats nonrelativistic atomic systems, which are characterized by
$Z_\mathrm{eff}\alpha_0$ $\!\ll$ $\!1$ \cite{Lamoureaux}, so we can
safely neglect the correction arising from the magnetic dipole
interaction. Furthermore, the first-order correction arising from
the diamagnetic term can be shown to be smaller than the second-order
correction due to the electric dipole interaction by the same factor
$(Z_\mathrm{eff}\alpha_0)^2$, so we can disregard it for the same
reason.

In summary, the main contribution to the eigenvalue shift of a state
$|l\rangle |\{0'\}\rangle$ [$|\{0'\}\rangle$, ground state of the
transformed fundamental fields 
$\hat{\mathbf{f}}'_\lambda(\mathbf{r},\omega)$]
is the second-order correction due to the electric dipole interaction
in Eq.~(\ref{eq51}), i.e., 
\begin{alignat}{1}
\label{eq75}
&\Delta E_l = \Delta_2 E_l = -\frac{1}{\hbar}
\sum_k\sum_{\lambda=e,m}\mathcal{P}
      \int_0^{\infty}
      \frac{\mathrm{d}\omega}{\omega_{kl}+\omega}
\nonumber\\&\quad\times\,
      \int\mathrm{d}^3{r}\,
      \big|\langle l|\langle\{0'\}|
              -\hat{\mathbf{d}}\hat{\mathbf{E}}'
              (\mathbf{r}_\mathrm{A})
             |\{\mathbf{1}_\lambda'(\mathbf{r},\omega)\}\rangle
             |k\rangle\big|^2
             \nonumber\\&
\end{alignat}
[$|\{\mathbf{1}'_\lambda(\mathbf{r},\omega)\}\rangle$
$\!\equiv$ $\!\hat{\mathbf{f}}^{\prime\dagger}_\lambda
(\mathbf{r},\omega)|\{0'\}\rangle]$. After some algebra it can be
found that (see Appendix~\ref{AppC})
\begin{eqnarray}
\label{eq76}
     \Delta E_l
      \hspace{-1ex}&=&\hspace{-1ex}
     - \frac{\mu_0}{\pi}\sum_k
     \mathcal{P}\int_0^{\infty}\!\!\mathrm{d}\omega\,
     \frac{\omega^2}{\omega_{kl}+\omega}
\nonumber\\&&\quad \times
     \,\mathbf{d}_{lk}
     \mathrm{Im}\,\bm{G} (\mathbf{r}_\mathrm{A},
     \mathbf{r}_\mathrm{A},\omega)
     \mathbf{d}_{kl}.
\quad
\end{eqnarray}
We now apply the same procedure as in Sec.~\ref{sec3.1}, below
Eq.~(\ref{eq57}). Replacing the Green tensor by its scattering part
and transforming the frequency integral to imaginary frequencies using
contour integral techniques, we arrive at exactly the same form of the
vdW potential as given in Eq.~(\ref{eq63-1}) together with
Eqs.~(\ref{eq64}) and (\ref{eq65}). It is worth noting that the two
schemes lead to equivalent results only if in the minimal-coupling
scheme the $\hat{\mathbf{A}}^2$ coupling term is properly taken into
account.

%%%%%%%%%%%%%%%%%%%%%%%%%%%%%%%%%%%%%%%%%%%%%%%%%%%%%%%%%%%%%%%%%%%%%%

\section{Center-of-mass motion and Lorentz force}
\label{sec4}

Atomic quantities that are related to the atom--field interaction can
drastically change when the atomic system comes close to a macroscopic
body, the spontaneous decay thus becoming purely radiationless, with
decay rates and level shifts being inversely proportional to the
atom-surface separation to the third power \cite{Ho01}. Clearly, in
this case approximations of the type (\ref{eq74}) cannot be made in
general and the perturbative approach to the calculation of the CP
force becomes questionable. Moreover, when the atomic system is not in
the ground state, then dynamical effects can no longer be disregarded.
To go beyond perturbation theory, let us first consider the
center-of-mass Newtonian equation of motion and the Lorentz force
therein.
 
%%%%%%%%%%%%%%%%%%%%%%%%%%%%%%%%%%%%%%%%%%%%%%%%%%%%%%%%%%%%%%%%%%%%%%

\subsection{Minimal coupling}
\label{sec4.1}

As has been shown \cite{Ho03}, the Heisenberg equations of motion
governed by the minimal-coupling Hamiltonian (\ref{eq1}),
\begin{equation}
\label{eq77}
 \ddot{\hat{\mathbf{r}}}_{\alpha}
 = \left(\frac{1}{i\hbar}\right)^2
 \left[\bigl[\hat{\mathbf{r}}_\alpha,\hat{H}\bigr],\hat{H}\right],
\end{equation}
lead to the well-known Newtonian equations of motion for the individual
charged particles,
\begin{equation}
\label{eq78}
           m_{\alpha} \ddot{\hat{\mathbf{r}}}_{\alpha} =
           q_{\alpha}\bigg\{\hat{\vec{E}}(\mathbf{r}_{\alpha})
           + {\textstyle\frac{1}{2}}
           \Big[\dot{\hat{\mathbf{r}}}_{\alpha}\times
           \hat{{\vec{B}}}(\mathbf{r}_{\alpha}) -
           \hat{{\vec{B}}}(\mathbf{r}_{\alpha})\times
           \dot{\hat{\mathbf{r}}}_{\alpha}\Big]\bigg\} .
\end{equation}
Summing Eq.~(\ref{eq78}) over $\alpha$, recalling definition (\ref{eq28}),
and using Eqs.~(\ref{eq19}) and (\ref{eq21}) together with the
relationship
\begin{equation}
\label{eq79}
          \sum_\alpha q_\alpha \bm{\nabla}_\alpha
          \hat{\varphi}_\mathrm{A} (\hat{\mathbf{r}}_\alpha)=0
\end{equation}
($\bm{\nabla}_{\!\!\alpha}$ $\!\equiv$
$\!\bm{\nabla}_{\!\!\hat{\mathbf{r}}_\alpha}$), we derive
\begin{equation}
\label{eq80}
m_\mathrm{A} \ddot{\hat{\mathbf{r}}}_\mathrm{A} =
\hat{\mathbf{F}},
\end{equation}
where the Lorentz force takes the form
\begin{equation}
\label{eq81}
      \hat{\mathbf{F}} =
      \int \mathrm{d}^3 r \big[\hat{\rho}_\mathrm{A}(\mathbf{r})
      \hat{\mathbf{E}}(\mathbf{r})
      + \hat{\mathbf{j}}_\mathrm{A}(\mathbf{r}) \times
      \hat{\mathbf{B}}(\mathbf{r})\big],
\end{equation}
with charge density $\hat{\rho}_\mathrm{A}(\mathbf{r})$ and current
density $\hat{\mathbf{j}}_\mathrm{A}(\mathbf{r})$ being defined by
Eq.~(\ref{eq2}) and
\begin{equation}
\label{eq82}
      \hat{\mathbf{j}}_\mathrm{A}(\mathbf{r}) =
      {\textstyle\frac{1}{2}}\sum_\alpha q_\alpha
      \Bigl[\dot{\hat{\mathbf{r}}}_\alpha
      \delta(\mathbf{r}-\hat{\mathbf{r}}_\alpha)
      +\delta(\mathbf{r}-\hat{\mathbf{r}}_\alpha)
      \dot{\hat{\mathbf{r}}}_\alpha\Bigr],
\end{equation}
respectively. It can be shown \cite{Healy77,Craig84,Knoll01} that for
neutral atoms the atomic charge and current densities can be expressed
in terms of atomic polarization and magnetization according to
\begin{equation}
\label{eq83}
    \hat{\rho}_\mathrm{A}(\mathbf{r}) =
    -\bm{\nabla} \hat{\mathbf{P}}_{\!\mathrm{A}}(\mathbf{r})
\end{equation}
and
\begin{equation}
\label{eq84}
      \hat{\mathbf{j}}_\mathrm{A}(\mathbf{r}) =
     \dot{\hat{\mathbf{P}}}_{\!\mathrm{A}}(\mathbf{r})
     + \bm{\nabla}\times \hat{\mathbf{M}}_\mathrm{A}(\mathbf{r})
     + \bm{\nabla}\times \hat{\mathbf{M}}_\mathrm{R}(\mathbf{r}),
\end{equation}
respectively, where
\begin{eqnarray}
\label{eq85}
      &\displaystyle
      \hat{\mathbf{M}}_\mathrm{A} (\mathbf{r}) =
       {\textstyle\frac{1}{2}} \sum_\alpha  q_\alpha
       \left[
       \hat{\bm{\Theta}}_\alpha (\mathbf{r})
       \times \dot{\hat{\bar{\mathbf{r}}}}_\alpha
       - \dot{\hat{\bar{\mathbf{r}}}}_\alpha
       \times \hat{\bm{\Theta}}_\alpha (\mathbf{r})
        \right],&
\\
\label{eq86}
      &\displaystyle
      \hat{\mathbf{M}}_\mathrm{R} (\mathbf{r}) =
      {\textstyle\frac{1}{2}}
      \left[ \hat{\mathbf{P}}_\mathrm{A}(\mathbf{r})
      \times \dot{\hat{\mathbf{r}}}_\mathrm{A}
      - \dot{\hat{\mathbf{r}}}_\mathrm{A}
      \times \hat{\mathbf{P}}_{\!\mathrm{A}}(\mathbf{r})\right],
\end{eqnarray}
with $\hat{\mathbf{P}}_{\!\mathrm{A}}(\mathbf{r})$ and
$\hat{\bm{\Theta}}_\alpha (\mathbf{r})$ from Eqs.~(\ref{eq36}) and
(\ref{eq39}), respectively. Note that the last term in
Eq.~(\ref{eq84}) represents the so-called R\"{o}ntgen current
\cite{Craig84, Roentgen1888}, which is a feature of the overall
translational motion of any aggregate of charges.

Inspection of Eqs.~(\ref{eq36}), (\ref{eq39}), (\ref{eq85}),
and (\ref{eq86}) shows that the relations
\begin{eqnarray}
 \label{eq88}
 &\displaystyle
 \bm{\nabla}\otimes\hat{\mathbf{P}}_{\!\mathrm{A}}(\mathbf{r}) = -
 \bm{\nabla}_{\!\!\mathrm{A}}
 \otimes\hat{\mathbf{P}}_{\!\mathrm{A}}(\mathbf{r}),&\\
 \label{eq89}
 &\displaystyle
 \bm{\nabla}\otimes\hat{\mathbf{M}}_\mathrm{A(R)}(\mathbf{r}) = -
 \bm{\nabla}_{\!\!\mathrm{A}}
 \otimes\hat{\mathbf{M}}_\mathrm{A(R)}(\mathbf{r})&
\end{eqnarray}
($\bm{\nabla}_{\!\!\mathrm{A}}$ $\!\equiv$
$\!\bm{\nabla}_{\!\!\hat{\mathbf{r}}_\mathrm{A}}$)
are valid. We therefore may write, on recalling Maxwell's equations,
\begin{eqnarray}
\label{eq91}
\lefteqn{
       -\int \mathrm{d}^3 r\,
        \bigl[
        \bm{\nabla}
        \hat{\mathbf{P}}_{\!\mathrm{A}}(\mathbf{r})\bigr]
        \hat{\mathbf{E}}(\mathbf{r})
}
\nonumber\\&&
       =
       \bm{\nabla}_{\!\!\mathrm{A}}
       \int \mathrm{d}^3 r\,
       \bigl[\hat{\mathbf{P}}_{\!\mathrm{A}}(\mathbf{r})
       \hat{\mathbf{E}}(\mathbf{r})\bigr]
       +\int \mathrm{d}^3 r\, 
       \hat{\mathbf{P}}_{\!\mathrm{A}}(\mathbf{r})
       \times \dot{\hat{\mathbf{B}}}(\mathbf{r}),
\qquad
\\[.5ex]
\label{eq92}
\lefteqn{\hspace{6ex}
      \int \mathrm{d}^3 r\,
       \bigl[\bm{\nabla} \times 
       \hat{\mathbf{M}}_\mathrm{A(R)}(\mathbf{r})\bigr]
      \times \hat{\mathbf{B}}(\mathbf{r})
}
\nonumber\\&&\hspace{12ex}
       =
       \bm{\nabla}_{\!\!\mathrm{A}}
       \int \mathrm{d}^3 r\,
       \bigl[\hat{\mathbf{M}}_\mathrm{A(R)}(\mathbf{r})
       \hat{\mathbf{B}}(\mathbf{r})\bigr].
\end{eqnarray}
Substituting Eqs.~(\ref{eq83}) and (\ref{eq84}) into
Eq.~(\ref{eq81}) and using Eqs.~(\ref{eq91}) and (\ref{eq92}),
we may equivalently express the Lorentz force as
\begin{eqnarray}
\label{eq93}
\lefteqn
{
\hat{\mathbf{F}}
      =
      \bm{\nabla}_{\!\!\mathrm{A}}
      \biggl\{
      \int\mathrm{d}^3r\, \hat{\mathbf{P}}_{\!\mathrm{A}}(\mathbf{r})
                       \hat{\mathbf{E}} (\mathbf{r})
}
\nonumber\\&&\hspace{8ex}
+ \int\mathrm{d}^3r\, \big[\hat{\mathbf{M}}_\mathrm{A} (\mathbf{r})
        +\hat{\mathbf{M}}_\mathrm{R} (\mathbf{r})\big]
        \hat{\mathbf{B}}(\mathbf{r})
      \biggr\}
\nonumber\\&&\hspace{3ex}
       +\,\frac{\mathrm{d}}{\mathrm{d}t}
       \int \mathrm{d}^3 r \,
       \hat{\mathbf{P}}_{\!\mathrm{A}}(\mathbf{r}) \times
       \hat{\mathbf{B}}(\mathbf{r}).
\end{eqnarray}

In long-wavelength approximation, Eqs.~(\ref{eq85}) and (\ref{eq86})
simplify to [recall Eqs.~(\ref{eq40}) and (\ref{eq41})]
\begin{eqnarray}
\label{eq94}
 \hat{\mathbf{M}}_\mathrm{A} (\mathbf{r})
       \hspace{-1ex}&=&\hspace{-1ex}
       {\textstyle\frac{1}{4}} \sum_\alpha  q_\alpha
        \bigl[
        \delta(\mathbf{r}-\hat{\mathbf{r}}_\mathrm{A})
        \hat{\bar{\mathbf{r}}}_\alpha
        \times \dot{\hat{\bar{\mathbf{r}}}}_\alpha
\nonumber\\
        &&\qquad\qquad-\,\dot{\hat{\bar{\mathbf{r}}}}_\alpha
        \times\hat{\bar{\mathbf{r}}}_\alpha
        \delta(\mathbf{r}-\hat{\mathbf{r}}_\mathrm{A})
        \bigr]
\end{eqnarray}
and
\begin{eqnarray}
\label{eq95}
 \hat{\mathbf{M}}_\mathrm{R}(\mathbf{r})
       =
       {\textstyle\frac{1}{2}}\bigl[\delta(\mathbf{r}
       -\hat{\mathbf{r}}_\mathrm{A}) \hat{\mathbf{d}}
       \times \dot{\hat{\mathbf{r}}}_\mathrm{A}
       -
       \dot{\hat{\mathbf{r}}}_\mathrm{A}\times\hat{\mathbf{d}}\,
       \delta(\mathbf{r}-\hat{\mathbf{r}}_\mathrm{A})
       \bigr],
&&       
\end{eqnarray}
respectively, so that the Lorentz force (\ref{eq93}) can be written as
\begin{alignat}{1}
\label{eq96}
&\hat{\mathbf{F}}
      =
      \bm{\nabla}_{\!\!\mathrm{A}}
      \biggl\{\hat{\mathbf{d}} 
      \hat{\mathbf{E}}(\hat{\mathbf{r}}_\mathrm{A})
      + {\textstyle\frac{1}{2}} \sum_\alpha q_\alpha
      \dot{\hat{\mathbf{r}}}_\alpha
      \hat{\mathbf{B}}(\hat{\mathbf{r}}_\mathrm{A})
      \times\hat{\bar{\mathbf{r}}}_\alpha
      \nonumber\\&\quad
      +{\textstyle \frac{1}{2}}
      \dot{\hat{\mathbf{r}}}_\mathrm{A}
      \hat{\mathbf{B}}(\hat{\mathbf{r}}_\mathrm{A})
      \times\hat{\mathbf{d}}
      \biggr\}
      +\frac{\mathrm{d}}{\mathrm{d}t}
      \bigl[\hat{\mathbf{d}} \times 
      \hat{\mathbf{B}}(\hat{\mathbf{r}}_\mathrm{A})\bigr].
\end{alignat}
Further, we calculate
\begin{eqnarray}
\label{eq99}
 &&\frac{\mathrm{d}}{\mathrm{d}t}
      \Bigl[\hat{\mathbf{d}} \times 
      \hat{\mathbf{B}}(\hat{\mathbf{r}}_\mathrm{A})\Bigr]
      =\frac{i}{\hbar}\Bigl[\hat{H},\hat{\mathbf{d}}
      \times \hat{\mathbf{B}}(\hat{\mathbf{r}}_\mathrm{A})\Bigr]
 \nonumber\\
      &&=\dot{\hat{\mathbf{d}}}\times 
      \hat{\mathbf{B}}(\hat{\mathbf{r}}_\mathrm{A})
       +\hat{\mathbf{d}}\times
       \dot{\hat{\mathbf{B}}}(\mathbf{r})
       \bigr|_{\mathbf{r}=\hat{\mathbf{r}}_\mathrm{A}}
 \nonumber\\
      &&\quad+\,\hat{\mathbf{d}}\times {\textstyle\frac{1}{2}}\Big[
        \dot{\hat{\mathbf{r}}}_\mathrm{A}\bm{\nabla}_{\!\!\mathrm{A}}
        \otimes\hat{\mathbf{B}}(\hat{\mathbf{r}}_\mathrm{A})
       +\hat{\mathbf{B}}(\hat{\mathbf{r}}_\mathrm{A})
       \otimes\overleftarrow{\bm{\nabla}}_{\!\!\mathrm{A}}
       \dot{\hat{\mathbf{r}}}_\mathrm{A}\Big].\qquad
\end{eqnarray}
Comparing the different terms in Eq.~(\ref{eq96}), one can show
[Eqs.~(\ref{B15}), (\ref{B16}), and (\ref{B18})--(\ref{B20}) in
Appendix~\ref{AppB}] that the second term in curly brackets is
typically smaller than the first one by a factor of 
$v/c+Z_\mathrm{eff}\alpha_0$($v$, velocity of the center of mass),
while the third term is smaller than the first one by a factor of
$v/c$. Similarly, we find [Eqs.~(\ref{B21})--(\ref{B23}) in
Appendix~\ref{AppB}] that the third term in Eq.~(\ref{eq99}) is
smaller than the first two terms by a factor of $v/c$. Thus in the
nonrelativistic limit considered throughout the current work [cf.
Hamiltonian (\ref{eq1})] we can set
\begin{equation}
\label{eq100}
      \hat{\mathbf{F}}
      = \biggl\{\bm{\nabla}
      \bigl[\hat{\mathbf{d}} \hat{\mathbf{E}} (\mathbf{r})\bigr]
      +\frac{\mathrm{d}}{\mathrm{d}t}
      \bigl[\hat{\mathbf{d}} \times \hat{\mathbf{B}}(\mathbf{r})\bigr]
      \biggr\}_{\mathbf{r}=\hat{\mathbf{r}}_\mathrm{A}}.
\end{equation}

In the absense of magnetodielectric bodies, Eq.~(\ref{eq100}) reduces
to earlier results derived within the multipolar coupling scheme for
an atom interacting with the electromagnetic field in free space
\cite{Baxter93,Lembessis93}. However, it should be pointed out that
here the electric and magnetic fields $\hat{\mathbf{E}}(\mathbf{r})$
and $\hat{\mathbf{B}}(\mathbf{r})$, respectively, are the
medium-assisted fields as defined by Eqs.~(\ref{eq20}) and
(\ref{eq21}) [together with Eqs.~(\ref{eq8}) and (\ref{eq22})]. Thus
Eq.~(\ref{eq93}) or, in electric dipole approximation,
Eq.~(\ref{eq100}) determine the force acting on an atomic system in
the very general case of dispersing and absorbing magnetodielectric
bodies being present --- a result that has not yet been derived
elsewhere.

%%%%%%%%%%%%%%%%%%%%%%%%%%%%%%%%%%%%%%%%%%%%%%%%%%%%%%%%%%%%%%%%%%%%%%

\subsection{Multipolar coupling}
\label{sec4.2}

Using the multipolar Hamiltonian (\ref{eq37}), we obtain, on recalling
that \mbox{$\hat{\mathbf{r}}'_\alpha$ $\!=$
$\!\hat{\mathbf{r}}_\alpha$},
\begin{equation}
\label{eq101}
     m_\alpha \dot{\hat{\mathbf{r}}}_\alpha=
     \frac{i}{\hbar} \bigl[\hat{H},m_\alpha 
     \hat{\mathbf{r}}_\alpha\bigr]=
      \hat{\mathbf{p}}'_\alpha
      +  \int\!\mathrm{d}^3 r\, \hat{\bm{\Xi}}_\alpha(\mathbf{r})
      \times \hat{\mathbf{B}}'(\mathbf{r}).
\end{equation}
Summing Eq.~(\ref{eq101}) over $\alpha$ and taking into account
Eqs.~(\ref{eq28}) and (\ref{eq47}) yields
\begin{equation}
\label{eq102}
     m_\mathrm{A} \dot{\hat{\mathbf{r}}}_\mathrm{A}=
       \hat{\mathbf{p}}'_\mathrm{A}
       + \int \mathrm{d}^3 r\,
       \hat{\mathbf{P}}_{\!\mathrm{A}} (\mathbf{r}) 
       \times \hat{\mathbf{B}}'(\mathbf{r}).
\end{equation}
Equation (\ref{eq102}) leads to
\begin{eqnarray}
\label{eq103}
\lefteqn{
      m_\mathrm{A} \ddot{\hat{\mathbf{r}}}_\mathrm{A} =
      \hat{\mathbf{F}}
      = \frac{i}{\hbar} \biggl[\hat{H},
      \hat{\mathbf{p}}'_\mathrm{A}
       + \int \mathrm{d}^3 r\,
       \hat{\mathbf{P}}_\mathrm{A} (\mathbf{r})
       \times \hat{\mathbf{B}}'(\mathbf{r})\biggr]
}
\nonumber\\&&\hspace{4ex}
      = 
      \frac{i}{\hbar}\bigl[\hat{H},
      \hat{\mathbf{p}}'_\mathrm{A}\bigr]
      +\frac{\mathrm{d}}{\mathrm{d}t}
      \int\mathrm{d}^3 r\,
      \hat{\mathbf{P}}_{\!\mathrm{A}} (\mathbf{r})
      \times \hat{\mathbf{B}}'(\mathbf{r})
.\qquad
\end{eqnarray}
To evaluate the different contributions to the first term in
Eq.~(\ref{eq103}), we first recall Eq.~(\ref{eq88}) and note that
\begin{equation}
\label{eq104}
 \frac{i}{\hbar}\left[
 \frac{1}{2\varepsilon_0}\! \int\! \mathrm{d}^3r\,
 \hat{\mathbf{P}}^2_{\!\mathrm{A}} (\mathbf{r})
 ,\hat{\mathbf{p}}'_\mathrm{A}\right]
 =
 \frac{1}{2\varepsilon_0}\! \int\!\mathrm{d}^3r\,
 \bm{\nabla}
 \hat{\mathbf{P}}^2_{\!\mathrm{A}}(\mathbf{r})
 =0.
\end{equation}
Further, we derive, on recalling Eq.~(\ref{eq101}),
\begin{alignat}{1}
\label{eq106}
 &\frac{i}{\hbar}\left[
 \sum_{\alpha}\frac{1}{2 m_{\alpha}}
      \left(\hat{\mathbf{p}}'_\alpha
      +  \int\!\mathrm{d}^3 r\, \hat{\bm{\Xi}}_\alpha(\mathbf{r})
      \times \hat{\mathbf{B}}'(\mathbf{r}) \right)^2
      \!\!,\hat{\mathbf{p}}'_\mathrm{A}\right]
\nonumber\\&\quad
= -
\bm{\nabla}_{\!\!\mathrm{A}}
\int\mathrm{d}^3r\, {\textstyle \frac{1}{2}}\sum_\alpha
\left[
\dot{\hat{\mathbf{r}}}_\alpha\times\hat{\bm{\Xi}}_\alpha
(\mathbf{r})-\hat{\bm{\Xi}}_\alpha(\mathbf{r})
\times\dot{\hat{\mathbf{r}}}_\alpha
\right]
\hat{\mathbf{B}}'(\mathbf{r}).
\end{alignat}
Substituting Eqs.~(\ref{eq104}) and (\ref{eq106}) into
Eq.~(\ref{eq103}), with $\hat{H}$ as given in Eq.~(\ref{eq37}), we
eventually obtain
\begin{eqnarray}
\label{eq107}
&&
\hat{\mathbf{F}}
      =
      \bm{\nabla}_{\!\!\mathrm{A}}
      \biggl\{
      \int\mathrm{d}^3r\, \hat{\mathbf{P}}_{\!\mathrm{A}} (\mathbf{r})
                       \hat{\mathbf{E}}' (\mathbf{r})
\nonumber\\&&\quad
      + \,{\textstyle \frac{1}{2}} \int\mathrm{d}^3r\, \sum_\alpha
      \left[
  \hat{\bm{\Xi}}_\alpha (\mathbf{r})\times
  \dot{\hat{\mathbf{r}}}_\alpha
 -\dot{\hat{\mathbf{r}}}_\alpha\times\hat{\bm{\Xi}}_\alpha(\mathbf{r})
     \right]
     \hat{\mathbf{B}}'(\mathbf{r})
      \biggr\}
\nonumber\\&&\quad
      +\,\frac{\mathrm{d}}{\mathrm{d}t}
      \int \mathrm{d}^3 r\,
       \hat{\mathbf{P}}_{\!\mathrm{A}} (\mathbf{r}) \times
       \hat{\mathbf{B}}'(\mathbf{r}).
\end{eqnarray}
It can be shown (see Appendix~\ref{AppD}) that Eq.~(\ref{eq107}) is
identical to Eq.~(\ref{eq93}).

It is not difficult to see [recall Eqs.~(\ref{eq40}) and (\ref{eq42})]
that in long-wavelength approximation Eq.~(\ref{eq107}) takes the form
of Eq.~(\ref{eq96}), but with 
$\hat{\mathbf{E}}'(\hat{\mathbf{r}}_\mathrm{A})$
and $\hat{\mathbf{B}}'(\hat{\mathbf{r}}_\mathrm{A})$ in place of
$\hat{\mathbf{E}}(\hat{\mathbf{r}}_\mathrm{A})$ and 
$\hat{\mathbf{B}}(\hat{\mathbf{r}}_\mathrm{A})$, respectively. The
time derivative \mbox{$\mathrm{d}[\hat{\mathbf{d}} \times
\hat{\mathbf{B}}'(\hat{\mathbf{r}}_\mathrm{A})]/\mathrm{d}t$} can then
be calculated to give an expression of the form of Eq.~(\ref{eq99})
with $\hat{\mathbf{B}}(\hat{\mathbf{r}}_\mathrm{A})$ replaced by 
$\hat{\mathbf{B}}'(\hat{\mathbf{r}}_\mathrm{A})$. Obviously, in the
nonrelativistic limit we are left with an expression similar to
Eq.~(\ref{eq100}). It should be pointed out that Eqs.~(\ref{eq96}) and
(\ref{eq100}) with $\hat{\mathbf{E}}(\hat{\mathbf{r}}_\mathrm{A})$ and
$\hat{\mathbf{B}}(\hat{\mathbf{r}}_\mathrm{A})$ replaced by 
$\hat{\mathbf{E}}'(\hat{\mathbf{r}}_\mathrm{A})$ and
$\hat{\mathbf{B}}'(\hat{\mathbf{r}}_\mathrm{A})$, respectively, yield
exactly the same force as the equations with the unprimed quantities,
although the physical meaning of 
$\hat{\mathbf{E}}'(\hat{\mathbf{r}}_\mathrm{A})$ is
different from that of $\hat{\mathbf{E}}(\hat{\mathbf{r}}_\mathrm{A})$
[recall that $\hat{\mathbf{B}}'(\hat{\mathbf{r}}_\mathrm{A})$ $\!=$
$\!\hat{\mathbf{B}}(\hat{\mathbf{r}}_\mathrm{A})$].

It is worth noting that the results of this section can serve as an
example to illustrate that the electric dipole approximation has to be
employed with great care. If in electric dipole approximation the
R\"{o}ntgen interaction primarily related to the induction field had
been disregarded and Eq.~(\ref{eq49}) without the second term on the
right-hand side had been used, then in the resulting expression for
the force the time-derivative term, i.e., the magnetic part of the
force, would have been lost. Note that the pressure exerted by
external laser fields on macroscopic bodies can be dominated by this
magnetic force \cite{Gordon73,Loudon03}, which contrasts with
arguments \cite{Baxter93, Stenholm86} that the contribution of this
term to the radiation force on atoms can be neglected.

%%%%%%%%%%%%%%%%%%%%%%%%%%%%%%%%%%%%%%%%%%%%%%%%%%%%%%%%%%%%%%%%%%%%%%

\section{Average Lorentz force}
\label{sec5}

Let us now turn to the problem of determining the electromagnetic
force acting on an atomic system that is initially prepared in an
arbitrary internal (electronic) quantum state. For convenience, we
shall employ the multipolar formalism. On recalling Eqs.~(\ref{eq20})
and (\ref{eq21}) together with Eq.~(\ref{eq22}), we find that
Eq.~(\ref{eq100}) [with 
$\hat{\mathbf{E}}(\hat{\mathbf{r}}_\mathrm{A})$ and
$\hat{\mathbf{B}}(\hat{\mathbf{r}}_\mathrm{A})$ replaced by
$\hat{\mathbf{E}}'(\hat{\mathbf{r}}_\mathrm{A})$ and
$\hat{\mathbf{B}}'(\hat{\mathbf{r}}_\mathrm{A})$, respectively] can be
rewritten as
\begin{eqnarray}
\label{eq108}
\lefteqn{
      \hat{\mathbf{F}}
      = \bigg\{\int_0^\infty \mathrm{d}\omega\, \bm{\nabla}
      \left[\hat{\mathbf{d}}
      \underline{\hat{\mathbf{E}}}{'} (\mathbf{r},\omega)\right]
}
\nonumber\\&&\hspace{3ex}
      +\frac{1}{i\omega}\,\frac{\mathrm{d}}{\mathrm{d}t}
      \hat{\mathbf{d}}
       \times\! \left[ \bm{\nabla} \times
        \underline{\hat{\mathbf{E}}}{'} (\mathbf{r},\omega)\right]
      \bigg\}_{\mathbf{r}=\hat{\mathbf{r}}_\mathrm{A}}
      +\mathrm{H.c.},
\qquad
\end{eqnarray}
where $\underline{\hat{\mathbf{E}}}{'} (\mathbf{r},\omega)$ is defined
according to Eq.~(\ref{eq8}). Decomposing $\hat{\mathbf{F}}$ into an
average component $\langle\hat{\mathbf{F}}\rangle$ (where the
expectation value $\langle\ldots\rangle$ is taken with respect to the
internal atomic motion and the medium-assisted electromagnetic field
only) and a fluctuating component
\begin{equation}
\label{eq109}
\Delta\hat{\mathbf{F}} = \hat{\mathbf{F}} -
\bigl\langle\hat{\mathbf{F}}\bigr\rangle,
\end{equation}
we may write
\begin{equation}
\label{eq110}
\hat{\mathbf{F}} =
\bigl\langle\hat{\mathbf{F}}\bigr\rangle
+ \Delta\hat{\mathbf{F}}.
\end{equation}
In the following, we will only consider the average force
$\langle\hat{\mathbf{F}}\rangle$ (for a discussion of the force
fluctuation $\langle\Delta\hat{\mathbf{F}}^2\rangle$, see, e.g.,
Ref.~\cite{Wu}). Note that we are free to choose a convenient operator
ordering in Eq.~(\ref{eq108}), because 
$\underline{\hat{\mathbf{E}}}{'}(\mathbf{r},\omega)$ commutes with 
$\hat{\mathbf{d}}$.

%%%%%%%%%%%%%%%%%%%%%%%%%%%%%%%%%%%%%%%%%%%%%%%%%%%%%%%%%%%%%%%%%%%%%%

\subsection{General case}
\label{sec5.1}

In order to calculate the average force as a function of time, we
first formally integrate the Heisenberg equations of motion for the
fundamental fields $\hat{\mathbf{f}}'_\lambda(\mathbf{r},\omega,t)$ to
obtain the source-quantity representation of 
$\underline{\hat{\mathbf{E}}}{'}(\mathbf{r},\omega,t)$. The result
reads (see Appendix~\ref{AppE})
\begin{equation}
\label{eq111}
\underline{\hat{\mathbf{E}}}{}'(\mathbf{r},\omega,t)
= \underline{\hat{\mathbf{E}}}{}'_\mathrm{free}(\mathbf{r},\omega,t)
+ \underline{\hat{\mathbf{E}}}{}'_\mathrm{source}
(\mathbf{r},\omega,t),
\end{equation}
where
\begin{equation}
\label{eq112}
      \underline{\hat{\mathbf{E}}}{}'_\mathrm{free}
      (\mathbf{r},\omega,t)
      = \underline{\hat{\mathbf{E}}}{}'(\mathbf{r},\omega)
      e^{-i\omega t}
\end{equation}
and
\begin{eqnarray}
\label{eq113}
\lefteqn{
      \underline{\hat{\mathbf{E}}}{}'_\mathrm{source}
      (\mathbf{r},\omega,t)
}
\nonumber\\&&
      = \frac{i\mu_0}{\pi} \omega^2
      \!\int_0^t\!\! \mathrm{d}t'\,
      e^{-i\omega(t-t')}\mathrm{Im}\,\bm{G}[\mathbf{r},
      \hat{\mathbf{r}}_\mathrm{A}(t'),\omega]
      \hat{\mathbf{d}}(t').
\qquad
\end{eqnarray}
Substituting Eq.~(\ref{eq111}) together with Eqs.~(\ref{eq112}) and
(\ref{eq113}) into Eq.~(\ref{eq108}), we arrive at
\begin{equation}
\label{eq114}
\bigl\langle\hat{\mathbf{F}}(t)\bigr\rangle
= \bigl\langle\hat{\mathbf{F}}_\mathrm{free}(t)\bigr\rangle
+ \bigl\langle\hat{\mathbf{F}}_\mathrm{source}(t)\bigr\rangle,
\end{equation}
where
\begin{alignat}{1}
\label{eq115}
&
\bigl\langle\hat{\mathbf{F}}_\mathrm{free}(t)\bigr\rangle
 =\bigg\{\int_0^\infty \mathrm{d}\omega\, \bm{\nabla}
      \bigl\langle\hat{\mathbf{d}}(t)
      \underline{\hat{\mathbf{E}}}{}'_\mathrm{free}
      (\mathbf{r},\omega,t)
      \bigr\rangle
\nonumber\\&
\quad+\frac{1}{i\omega}\,\frac{\mathrm{d}}{\mathrm{d}t}
       \bigl\langle\hat{\mathbf{d}}(t)
       \times\! \big[ \bm{\nabla} \times
        \underline{\hat{\mathbf{E}}}{}'_\mathrm{free} 
      (\mathbf{r},\omega,t)\big]\bigr\rangle
      \bigg\}
      _{\mathbf{r}=\hat{\mathbf{r}}_\mathrm{A}(t)}
      +\mathrm{H.c.}\nonumber\\&
\end{alignat}
and
\begin{equation}
\label{eq116}
\bigl\langle\hat{\mathbf{F}}_\mathrm{source}(t)\bigr\rangle=
\bigl\langle\hat{\mathbf{F}}_\mathrm{source}^\mathrm{el}(t)
\bigr\rangle
+\bigl\langle\hat{\mathbf{F}}_\mathrm{source}^\mathrm{mag}(t)
\bigr\rangle.
\end{equation}
Here,
\begin{eqnarray}
\label{eq117}
\lefteqn{
\bigl\langle\hat{\mathbf{F}}_\mathrm{source}^\mathrm{el}(t)
\bigr\rangle
= \biggl\{\frac{i\mu_0}{\pi}\int_0^\infty \mathrm{d}\omega\,\omega^2
 \int_0^t\! \mathrm{d}t'\,
      e^{-i\omega(t-t')}
}
      \nonumber\\&&\times
      \bm{\nabla}
      \bigl\langle\hat{\mathbf{d}}(t)
      \mathrm{Im}\,\bm{G} [\mathbf{r},
      \hat{\mathbf{r}}_\mathrm{A}(t'),\omega]
      \hat{\mathbf{d}}(t')\bigr\rangle
      \biggr\}
      _{\mathbf{r}=\hat{\mathbf{r}}_\mathrm{A}(t)}
      +\mathrm{H.c.}\nonumber\\
\end{eqnarray}
is the electric part of the average force associated with the
source-field part of the medium-assisted electromagnetic field, and
\begin{eqnarray}
\label{eq118}
\lefteqn{
\bigl\langle\hat{\mathbf{F}}_\mathrm{source}^\mathrm{mag}(t)
\bigr\rangle
= \biggl\{\frac{\mu_0}{\pi}\int_0^\infty \mathrm{d}\omega\,\omega
 \frac{\mathrm{d}}{\mathrm{d}t}\int_0^t\! \mathrm{d}t'\,
      e^{-i\omega(t-t')}
}
      \nonumber\\&&\times
      \bigl\langle\hat{\mathbf{d}}(t)\!\times\!\big\{
      \bm{\nabla}\!\times\!
   \mathrm{Im}\,\bm{G} [\mathbf{r},
   \hat{\mathbf{r}}_\mathrm{A}(t'),\omega]
      \big\}
      \hat{\mathbf{d}}(t')\bigr\rangle
      \biggr\}
      _{\mathbf{r}=\hat{\mathbf{r}}_\mathrm{A}(t)}
      \!\!\!+\mathrm{H.c.}\nonumber\\&&
\end{eqnarray}
is the respective magnetic part. Equations
(\ref{eq114})--(\ref{eq118}) are still general in the sense that they
apply to both driven and nondriven atomic systems and to both weak
and strong atom-field coupling.

%%%%%%%%%%%%%%%%%%%%%%%%%%%%%%%%%%%%%%%%%%%%%%%%%%%%%%%%%%%%%%%%%%%%%%

\subsection{Nondriven atom in the weak-coupling regime}
\label{sec5.2}

When the atomic system is not driven, i.e.,
\begin{equation}
\label{eq119}
\bigl\langle\ldots
\underline{\hat{\mathbf{E}}}{}'_\mathrm{free}
[\hat{\mathbf{r}}_\mathrm{A}(t),\omega,t]
\bigr\rangle
=\bigl\langle
\underline{\hat{\mathbf{E}}}{}'^\dagger_\mathrm{free}
[\hat{\mathbf{r}}_\mathrm{A}(t),\omega,t]
\ldots\bigr\rangle
=0,
\end{equation}
then $\langle\hat{\mathbf{F}}_\mathrm{free}(t)\rangle$ $\!=$ $\!0$.
Consequently, the average force, referred to as CP force, is
determined by the source-field part only,
\begin{equation}
\label{eq120}
\bigl\langle\hat{\mathbf{F}}(t)\bigr\rangle
= \bigl\langle\hat{\mathbf{F}}_\mathrm{source}(t)\bigr\rangle.
\end{equation}
Even more specifically, we assume that the density operator of the
initial quantum state of the field and the internal (electronic)
motion of the atomic system reads 
\begin{equation}
\label{eq121}
\hat{\varrho}=|\{0'\}\rangle\langle\{0'\}|\otimes
\hat{\sigma},
\end{equation}
where the density operator of the internal motion of the atomic system
$\hat{\sigma}$ can be written as
\begin{eqnarray}
\label{eq122}
\hat{\sigma}
  = \sum_{m,n} \sigma_{mn}
  \hat{A}_{mn}
\end{eqnarray}
($\hat{A}_{mn}$ $\!=$ $\!|m\rangle\langle n|$, with
$|n\rangle,\,|m\rangle$ being the internal atomic energy eigenstates).
In order to calculate the dipole-dipole correlation function appearing
in Eqs.~(\ref{eq117}) and (\ref{eq118}), we make use of the expansion
\begin{equation}
\label{eq123}
      \hat{\mathbf{d}}(t)
     = \sum_{m,n}
     \mathbf{d}_{mn} \hat{A}_{mn}(t)
\end{equation}
and write
\begin{eqnarray}
\label{eq124}
&&\bigl\langle\hat{\mathbf{d}}(t)\otimes
\hat{\mathbf{d}}(t')\bigr\rangle
\nonumber\\&&\quad
= \sum_{m,n}\sum_{m',n'} \mathbf{d}_{mn}\otimes \mathbf{d}_{m'n'}
\bigl\langle\hat{A}_{mn}(t)\hat{A}_{m'n'}(t')\bigr\rangle.
\qquad
\end{eqnarray}

In the weak-coupling regime, the Markov approximation can be exploited
and the correlation functions
$\langle\hat{A}_{mn}(t)\hat{A}_{m'n'}(t')\rangle$
can be calculated by means of the quantum regression theorem (see,
e.g., Ref.~\cite{Vogel01}). For this purpose, the (intra-atomic)
master equation has to be solved for arbitray initial conditions,
which in general requires knowledge of the specific level structure of
the atomic system under consideration. Only if the relevant atomic
transition frequencies are well separated from each other, one can go
a step forward constructing a general solution. In this case, the
off-diagonal density-matrix elements can be regarded as being
decoupled from each other and from the diagonal elements. We find
(see Appendix~\ref{AppF})
\begin{alignat}{1}
\label{eq125}
       &\bigl\langle \hat{A}_{mn}(t)
       \hat{A}_{m'n'}(t')\bigr\rangle
       = \delta_{nm'}
       \bigl\langle \hat{A}_{mn'}(t')\bigr\rangle
\nonumber\\&\quad\times
        e^{\{i\tilde{\omega}_{mn}(\hat{\mathbf{r}}_\mathrm{A})
       -[\Gamma_m(\hat{\mathbf{r}}_\mathrm{A})
       +\Gamma_n(\hat{\mathbf{r}}_\mathrm{A})]/2\}(t-t')}
\end{alignat}
($t$ $\!\ge$ $\!t'$, $m$ $\!\neq$ $n$). Here,
\begin{eqnarray}
\label{eq126}
         \tilde{\omega}_{mn}(\hat{\mathbf{r}}_\mathrm{A})
         =
         \omega_{mn}
         +\delta\omega_m(\hat{\mathbf{r}}_\mathrm{A})
         -\delta\omega_n(\hat{\mathbf{r}}_\mathrm{A})
\end{eqnarray}
are the body-induced position-dependent shifted transition frequencies
[$\hat{\mathbf{r}}_\mathrm{A}$ $\!=$
$\!\hat{\mathbf{r}}_\mathrm{A}(t)$], where
\begin{equation}
\label{eq127}
   \delta\omega_m(\hat{\mathbf{r}}_\mathrm{A})
   =\sum_k \delta\omega_m^k(\hat{\mathbf{r}}_\mathrm{A}),
\end{equation}
with
\begin{equation}
\label{eq128}
        \delta\omega_m^k(\hat{\mathbf{r}}_\mathrm{A})
         = \frac{\mu_0}{\pi\hbar}
        {\cal P}\int_0^\infty \!\!\mathrm{d}\omega\, \omega^2
        \frac{\mathbf{d}_{km}
        \mathrm{Im}\bm{G}^{(1)}\,
        (\hat{\mathbf{r}}_\mathrm{A},
        \hat{\mathbf{r}}_\mathrm{A},\omega)
        \mathbf{d}_{mk}}{\tilde{\omega}_{mk}
	(\hat{\mathbf{r}}_\mathrm{A})-\omega}\,,
\end{equation}
and
\begin{equation}
\label{eq129}
         \Gamma_m(\hat{\mathbf{r}}_\mathrm{A})
         = \sum_k \Gamma_m^k(\hat{\mathbf{r}}_\mathrm{A})
\end{equation}
are the position-dependent level widths, with
\begin{eqnarray}
\label{eq130}
        \Gamma_m^k(\hat{\mathbf{r}}_\mathrm{A})
        &=& \frac{2\mu_0 }{\hbar}\,
        \Theta[\tilde{\omega}_{mk}(\hat{\mathbf{r}}_\mathrm{A})]
        [\tilde{\omega}_{mk}
        (\hat{\mathbf{r}}_\mathrm{A})]^2\nonumber\\
        &&\times\mathbf{d}_{km}
        \mathrm{Im}\bm{G}\,
        [\hat{\mathbf{r}}_\mathrm{A},
        \hat{\mathbf{r}}_\mathrm{A},
	\tilde{\omega}_{mk}(\hat{\mathbf{r}}_\mathrm{A})]
        \mathbf{d}_{mk}.
\end{eqnarray}
One should point out that the position-independent (infinite)
Lamb-shift terms resulting from
$\bm{G}^{(0)}\,(\hat{\mathbf{r}}_\mathrm{A},
\hat{\mathbf{r}}_\mathrm{A},\omega)$
[recall Eq.~(\ref{eq59})] have been thought to be absorbed in the
transitions frequencies $\omega_{mn}$. Equation (\ref{eq128}) can be
rewritten by changing to imaginary frequencies [cf. the discussion
below Eq.~(\ref{eq62})], resulting in
\begin{alignat}{1}
\label{eq131}
       &\delta\omega_m^k(\hat{\mathbf{r}}_\mathrm{A})=
        -\frac{\mu_0}{\hbar}
        \Theta[\tilde{\omega}_{mk}(\hat{\mathbf{r}}_\mathrm{A})]
        [\tilde{\omega}_{mk}(\hat{\mathbf{r}}_\mathrm{A})]^2
       \nonumber\\
       &\quad\times\mathbf{d}_{km}\mathrm{Re}\,\bm{G}^{(1)}\,
        [\hat{\mathbf{r}}_\mathrm{A},
	\hat{\mathbf{r}}_\mathrm{A},
        \tilde{\omega}_{mk}(\hat{\mathbf{r}}_\mathrm{A})]
        \mathbf{d}_{mk}
        \nonumber\\
       &\quad+\frac{\mu_0}{\pi\hbar}
        \int_0^\infty \mathrm{d}u\, u^2
        \tilde{\omega}_{km}(\hat{\mathbf{r}}_\mathrm{A})
        \frac{\mathbf{d}_{km}
        \bm{G}^{(1)}
        (\hat{\mathbf{r}}_\mathrm{A},
        \hat{\mathbf{r}}_\mathrm{A},iu)
        \mathbf{d}_{mk}}{[\tilde{\omega}_{km}
        (\hat{\mathbf{r}}_\mathrm{A})]^2+u^2}\,.\quad
\end{alignat}
Recall that in the perturbative treatment the vdW potential of an
atomic system in a state $|m\rangle$ is identified with the energy
shift $\hbar\delta\omega_m$, so it is not surprising that
Eq.~(\ref{eq127}) together with Eq.~(\ref{eq131}) corresponds to
Eq.~(\ref{eq63-1}) together with Eqs.~(\ref{eq64}) and (\ref{eq65}),
if in Eq.~(\ref{eq131}) the $\tilde{\omega}_{mk}$ are replaced with
$\omega_{mk}$. The calculation of
\begin{equation}
\label{eq132}
\big\langle\hat{A}_{mn}(t)\big\rangle
=\sigma_{nm}(t)
\end{equation}
[$\sigma_{nm}(0)$ $\!=$ $\!\sigma_{nm}$] then leads (under the
assumptions made) to
\begin{equation}
\label{eq133}
       \sigma_{nm}(t)
       = e^{\{i\tilde{\omega}_{mn}(\hat{\mathbf{r}}_\mathrm{A})
       -[\Gamma_m(\hat{\mathbf{r}}_\mathrm{A})
       +\Gamma_n(\hat{\mathbf{r}}_\mathrm{A})]/2\}t}
       \sigma_{nm}
\end{equation}
for $m$ $\!\neq$ $\!n$ [cf. Eq.~(\ref{eq125})], so the remaining task
consists in solving the balance equations
\begin{equation}
\label{eq134}
       \dot{\sigma}_{mm}(t)
       = -\Gamma_m(\hat{\mathbf{r}}_\mathrm{A})
       \sigma_{mm}(t)
       + \sum_n \Gamma_n^m(\hat{\mathbf{r}}_\mathrm{A})
              \sigma_{nn}(t).
\end{equation}

With these preparations at hand, the CP force can be calculated in the
following steps. We first substitute Eq.~(\ref{eq124}) together with
Eqs.~(\ref{eq125}) and (\ref{eq132}) into Eqs.~(\ref{eq117}) and
(\ref{eq118}) and perform the time derivative in Eq.~(\ref{eq118}).
Introducing slowly varying density-matrix elements
\mbox{$\tilde{\sigma}_{nm}(t)$
$\!=$ $\!e^{i\tilde{\omega}_{nm}t}\sigma_{nm}(t)$}, we then perform
the time integrals in the spirit of the Markov approximation, by
making the replacements $\tilde{\sigma}_{nm}(t')\mapsto 
\tilde{\sigma}_{nm}(t)$ as well as 
$\hat{\mathbf{r}}_\mathrm{A}(t')
\mapsto\hat{\mathbf{r}}_\mathrm{A}(t)$
and letting the upper limit of integration tend to infinity. Recalling
Eq.~(\ref{eq120}) together with Eq.~(\ref{eq116}), we derive
\begin{equation}
\label{eq135}
\bigl\langle\hat{\mathbf{F}}(t)\bigr\rangle
=\sum_{m,n} \sigma_{nm}(t)\mathbf{F}_{mn}
(\hat{\mathbf{r}}_\mathrm{A}),
\end{equation}
\begin{equation}
\label{eq136}
\mathbf{F}_{mn}(\hat{\mathbf{r}}_\mathrm{A})
=\mathbf{F}_{mn}^\mathrm{el}(\hat{\mathbf{r}}_\mathrm{A})
+\mathbf{F}_{mn}^\mathrm{mag}(\hat{\mathbf{r}}_\mathrm{A}),
\end{equation}
where
\begin{eqnarray}
\label{eq137}
\lefteqn{
      \mathbf{F}_{mn}^\mathrm{el}(\hat{\mathbf{r}}_\mathrm{A}) =
      \Biggl\{
      \frac{\mu_0}{\pi} \sum_{k}
      \int_0^\infty \mathrm{d}\omega\,
      \omega^2
}
\nonumber\\&&\times\,
      \frac{\bm{\nabla}\otimes
      \mathbf{d}_{mk}
      \mathrm{Im}\,\bm{G}^{(1)} (\mathbf{r},
      \hat{\mathbf{r}}_\mathrm{A},\omega) \mathbf{d}_{kn} }
      {\omega+\tilde{\omega}_{kn}(\hat{\mathbf{r}}_\mathrm{A})
      -i[\Gamma_k(\hat{\mathbf{r}}_\mathrm{A})
      +\Gamma_m(\hat{\mathbf{r}}_\mathrm{A})]/2}
      \Biggr\}_{\mathbf{r}=\hat{\mathbf{r}}_\mathrm{A}}
      \!\!+\, \mathrm{H.c.},
\nonumber\\&&
\end{eqnarray}
and
\begin{eqnarray}
\label{eq138}
\lefteqn{
     \mathbf{F}^\mathrm{mag}_{mn}(\hat{\mathbf{r}}_\mathrm{A})=
     \Biggl\{\frac{\mu_0}{\pi}\sum_k
     \int_0^\infty \mathrm{d}\omega\,\omega
     \tilde{\omega}_{mn}(\hat{\mathbf{r}}_\mathrm{A})
}
\nonumber\\&&\times\,
     \frac{\mathbf{d}_{mk}\times\bigl[\bm{\nabla}\times
     \mathrm{Im}\,\bm{G}^{(1)} (\mathbf{r},
     \hat{\mathbf{r}}_\mathrm{A},\omega)\bigr]
     \mathbf{d}_{kn}}{\omega
     +\tilde{\omega}_{kn}(\hat{\mathbf{r}}_\mathrm{A})
     -i[\Gamma_k(\hat{\mathbf{r}}_\mathrm{A})
     +\Gamma_m(\hat{\mathbf{r}}_\mathrm{A})]/2}
     \Biggr\}_{\mathbf{r}=\hat{\mathbf{r}}_\mathrm{A}}
     \!\!+\,\mathrm{H.c.}
\nonumber\\&&
\end{eqnarray}

This result requires two comments. First, in Eqs.~(\ref{eq137}) and
(\ref{eq138}) the replacement 
$\bm{G}(\mathbf{r},\hat{\mathbf{r}}_\mathrm{A},\omega)
\mapsto\bm{G}^{(1)}(\mathbf{r},\hat{\mathbf{r}}_\mathrm{A},\omega)$
has again been made,
which can be justified by similar arguments as in Sec.~\ref{sec3}
[cf. the discussion preceding Eq.~(\ref{eq61})]. Second, from the
derivation of Eqs.~(\ref{eq135})--(\ref{eq138}) it is clear that these
equations are valid provided that the center-of-mass motion can be
regarded as being sufficiently slow. More precisely, they hold if the
condition
\begin{equation}
\label{eq139}
\bm{G} [\mathbf{r},
\hat{\mathbf{r}}_\mathrm{A}(t\!+\!\Delta t),\omega]\approx
\bm{G} [\mathbf{r},\hat{\mathbf{r}}_\mathrm{A}(t),\omega] \
\mathrm{for}\ \Delta t \le \Gamma_\mathrm{C}^{-1}
\end{equation}
is satisfied, where $\Gamma_\mathrm{C}$ is a characteristic
intra-atomic decay rate. Under this condition, the internal
(electronic) and external (center-of-mass) motion of the atomic system
decouple in the spirit of a Born-Oppenheimer approximation. As a
result, $\hat{\mathbf{r}}_\mathrm{A}$ effectively enters the equations
as a parameter, so that the caret will be removed in the following
($\hat{\mathbf{r}}_\mathrm{A}$ $\!\mapsto$ $\!\mathbf{r}_\mathrm{A}$).

We finally rewrite Eqs.~(\ref{eq137}) and (\ref{eq138}), by using
contour integration and going over to imaginary frequencies [cf. the
discussion below Eq.~(\ref{eq62})]. Recalling the definition of
$\bm{\alpha}_{mn}(\omega)$ $\!=$
$\!\bm{\alpha}_{mn}(\mathbf{r}_\mathrm{A},\omega)$ as given in
Eq.~(\ref{eq67}) and introducing the abbreviating notation
\begin{equation}
\label{eq140}
\Omega_{mnk}(\mathbf{r}_\mathrm{A})
=\tilde{\omega}_{nk}(\mathbf{r}_\mathrm{A})
+i[\Gamma_m(\mathbf{r}_\mathrm{A})
+\Gamma_k(\mathbf{r}_\mathrm{A})]/2,
\end{equation}
we derive
\begin{eqnarray}
\label{eq140-1}
\mathbf{F}_{mn}^\mathrm{el}(\mathbf{r}_\mathrm{A})
&=&\mathbf{F}_{mn}^\mathrm{el, or}(\mathbf{r}_\mathrm{A})
+\mathbf{F}_{mn}^\mathrm{el, r}(\mathbf{r}_\mathrm{A}),\\
\label{eq140-2}
\mathbf{F}_{mn}^\mathrm{mag}(\mathbf{r}_\mathrm{A})
&=&\mathbf{F}_{mn}^\mathrm{mag, or}(\mathbf{r}_\mathrm{A})
+\mathbf{F}_{mn}^\mathrm{mag, r}(\mathbf{r}_\mathrm{A}),
\end{eqnarray}
where
\begin{widetext}
\begin{equation}
\label{eq141}
     \mathbf{F}_{mn}^\mathrm{el, or}(\mathbf{r}_\mathrm{A})=-\Biggl\{
     \frac{\hbar\mu_0}{2\pi}
     \int_0^\infty \mathrm{d}u\, u^2
     \bigl[(\alpha_{mn})_{ij}(\mathbf{r}_\mathrm{A},iu)
     +(\alpha_{mn})_{ij}(\mathbf{r}_\mathrm{A},-iu)\bigr]
     \bm{\nabla}G^{(1)}_{ij}(\mathbf{r},\mathbf{r}_\mathrm{A},iu)
     \Biggr\}_{\mathbf{r}=\mathbf{r}_\mathrm{A}},
\end{equation}
\begin{eqnarray}
\label{eq142}
\mathbf{F}_{mn}^\mathrm{el,r}(\mathbf{r}_\mathrm{A})=
     \Biggl\{\mu_0\sum_{k}
     \Theta({\tilde{\omega}_{nk}})
     \Omega^2_{mnk}(\mathbf{r}_\mathrm{A})
     \bm{\nabla}\otimes\mathbf{d}_{mk}\bm{G}^{(1)}
     [\mathbf{r},\mathbf{r}_\mathrm{A},
     \Omega_{mnk}(\mathbf{r}_\mathrm{A})]
     \mathbf{d}_{kn}
     \Biggr\}_{\mathbf{r}=\mathbf{r}_\mathrm{A}}+\mathrm{H.c.}
\end{eqnarray}
and
\begin{equation}
\label{eq143}
     \mathbf{F}_{mn}^\mathrm{mag, or}(\mathbf{r}_\mathrm{A})
     =
     \Biggl\{
     \frac{\hbar\mu_0}{2\pi}
      \int_0^\infty \mathrm{d}u\, u^2 \mathrm{Tr}
      \biggl(\left[
      \frac{\tilde{\omega}_{mn}(\mathbf{r}_\mathrm{A})}{iu}\,
      \bm{\alpha}_{mn}^\top(\mathbf{r}_\mathrm{A},iu)-
      \frac{\tilde{\omega}_{mn}(\mathbf{r}_\mathrm{A})}{iu}\,
      \bm{\alpha}_{mn}^\top(\mathbf{r}_\mathrm{A},-iu)
      \right]\times
      \big[\bm{\nabla}\times\bm{G}^{(1)}
      (\mathbf{r},\mathbf{r}_\mathrm{A},iu)\big]
      \biggr)\Biggr\}_{\mathbf{r}=\mathbf{r}_\mathrm{A}},
\end{equation}
\begin{equation}
\label{eq144}
    \mathbf{F}_{mn}^\mathrm{mag,r}(\mathbf{r}_\mathrm{A})=
     \Biggl\{\mu_0\sum_{k}
     \Theta({\tilde{\omega}_{nk}})
     \tilde{\omega}_{mn}(\mathbf{r}_\mathrm{A})
     \Omega_{mnk}(\mathbf{r}_\mathrm{A})
     \mathbf{d}_{mk}\times\left(\bm{\nabla}
     \times\bm{G}^{(1)}[\mathbf{r},\mathbf{r}_\mathrm{A},
     \Omega_{mnk}(\mathbf{r}_\mathrm{A})]
     \mathbf{d}_{kn}
     \right)
     \Biggr\}_{\mathbf{r}=\mathbf{r}_\mathrm{A}}
     +\mathrm{H.c.}
\end{equation}
\end{widetext}
[$(\mathrm{Tr}\,\bm{T})_j$ $\!=$ $\!T_{ljl}$].
Equation (\ref{eq135}) together with Eq.~(\ref{eq136}) and
Eqs.~(\ref{eq140-1})--(\ref{eq144}) is the natural generalization of
Eq.~(\ref{eq66}) together with Eqs.~(\ref{eq63-1}), (\ref{eq65}), and
(\ref{eq69}). The above result is the first nonperturbative
expression for the CP force that incorporates its time dependence 
in case of excited atoms and correctly accounts for body-induced
shifting and broadening of atomic transition lines. 

In the short-time limit,
\mbox{$\Gamma_C t$ $\!\ll$ $\!1$}, Eq.~(\ref{eq135}) reads
\begin{equation}
\label{eq145}
\bigl\langle\hat{\mathbf{F}}(t)\bigr\rangle
\simeq \bigl\langle\hat{\mathbf{F}}(0)\bigr\rangle
=\sum_{m,n}
\sigma_{nm}(0)\mathbf{F}_{mn}(\mathbf{r}_\mathrm{A}),
\end{equation}
which for $\sigma_{nm}(0)$ $\!=$ $\!\delta_{nl}\delta_{ml}$
reduces to
\begin{equation}
\label{eq146}
\bigl\langle\hat{\mathbf{F}}(t)\bigr\rangle
\simeq \bigl\langle\hat{\mathbf{F}}(0)\bigr\rangle
=
{\mathbf{F}}_{ll}^\mathrm{el}(\mathbf{r}_\mathrm{A}).
\end{equation}
For the nonrelativistic Hamiltonian (\ref{eq45}), we can always
choose real dipole matrix elements
(\mbox{$\mathbf{d}_{mn}$ $\!=$ $\!\mathbf{d}_{nm}$}), revealing that
$\mathbf{d}_{mn}\otimes\mathbf{d}_{nm}$ is a symmetric tensor so that,
recalling Eq.~(\ref{eq16}), we may exploit the rule
\begin{equation}
\label{eq147}
S_{ij}\bm{\nabla}G^{(1)}_{ij}(\mathbf{r},\mathbf{r},\omega)
= 2S_{ij}\bm{\nabla}_{\!\!\!\mathbf{s}}
G^{(1)}_{ij}(\mathbf{s},\mathbf{r},\omega)|_{\mathbf{s}=\mathbf{r}},
\end{equation}
which is valid for any symmetric tensor $\bm{S}$. Hence,
Eqs.~(\ref{eq141}) and (\ref{eq142}) [together with Eq.~(\ref{eq68})]
lead to
\begin{eqnarray}
\label{eq148}
\lefteqn{
{\mathbf{F}}_{ll}^\mathrm{el, or}(\mathbf{r}_\mathrm{A})
     =     -\frac{\hbar\mu_0}{4\pi}
      \int_0^\infty \mathrm{d}u u^2\big[(\alpha_l)_{ij}
      (\mathbf{r}_\mathrm{A},iu)
}
      \nonumber\\&&\hspace{-1ex}
      +\,(\alpha_l)_{ij}(\mathbf{r}_\mathrm{A},-iu)\big]
     \bm{\nabla}_{\!\!\mathrm{A}}G^{(1)}_{ij}
     (\mathbf{r}_\mathrm{A},\mathbf{r}_\mathrm{A},iu)\qquad
\end{eqnarray}
and
\begin{alignat}{1}
\label{eq149}
&{\mathbf{F}}_{ll}^\mathrm{el,r}(\mathbf{r}_\mathrm{A})
     =\frac{\mu_0}{2}\sum_{k}
      \Theta({\tilde{\omega}_{lk}})
     \Omega_{lk}^2(\mathbf{r}_\mathrm{A})\Bigl\{
     \bm{\nabla}
\nonumber\\&\quad     
     \otimes\mathbf{d}_{lk}\bm{G}^{(1)}[\mathbf{r},
     \mathbf{r},\Omega_{lk}(\mathbf{r}_\mathrm{A})]
     \mathbf{d}_{kl}
     \Bigr\}_{\mathbf{r}=\mathbf{r}_\mathrm{A}}
     \!\!+\mathrm{H.c.}
 \end{alignat}
[$\Omega_{lk}(\mathbf{r}_\mathrm{A})$ $\!\equiv$
$\!\Omega_{llk}(\mathbf{r}_\mathrm{A})$].
Ignoring the position-dependent shifts and broadenings of the atomic
energy levels, i.e., disregarding the position dependence of the
atomic polarizability [$\bm{\alpha}_l(\mathbf{r}_\mathrm{A},iu)$
$\!\mapsto$
$\!\bm{\alpha}^{(0)}_l(iu)$], Eqs.~(\ref{eq148}) and (\ref{eq149})
reduce to the perturbative result in Eq.~(\ref{eq66}) together with
Eqs.~(\ref{eq63-1}), (\ref{eq65}), and (\ref{eq69})
[${\mathbf{F}}_{ll}^\mathrm{el}(\mathbf{r}_\mathrm{A})$
$\!\mapsto$ $\!{\mathbf{F}}_{l}({\mathbf{r}}_\mathrm{A})$].
Note that this result can be obtained without choosing real dipole
matrix elements [$\bm{\alpha}_l(iu)+\bm{\alpha}_l(-iu)$ being
symmetric in this case]. In the long-time limit,
$\Gamma_C t$ $\!\gg$ $\!1$, Eq.~(\ref{eq135}) obviously reduces to
ground-state force
\begin{equation}
\label{eq150}
\bigl\langle\hat{\mathbf{F}}(t)\bigr\rangle
\simeq \sum_{m,n} \sigma_{nm}(\infty)
\mathbf{F}_{mn}(\mathbf{r}_\mathrm{A})
= \mathbf{F}^\mathrm{el, or}_{00}(\mathbf{r}_\mathrm{A})
\end{equation}
[$\mathbf{F}_{00}^\mathrm{el,r}(\mathbf{r}_\mathrm{A})$ $\!=$ $\!0$],
because of $\sigma_{nm}(\infty)$ $\!=$ $\!\delta_{n0}\delta_{m0}$.

As already mentioned, the expression for the ground-state CP force
$\mathbf{F}_{00}(\mathbf{r}_\mathrm{A})$ obtained in lowest-order
perturbation theory, Eq.~(\ref{eq148-1}), agrees with the expression
obtained from LRT. However, its naive extrapolation in the sense of
the replacement $\bm{\alpha}_0^{(0)}(\omega)$ $\!\mapsto$
$\!\bm{\alpha}_0(\mathbf{r}_\mathrm{A},\omega)$ in Eq.~(\ref{eq148-1})
\cite{Kryszewski92} is wrong, because it results in Eq.~(\ref{eq148})
with $2(\alpha_0)_{ij}(\mathbf{r}_\mathrm{A},iu)$ instead
of $(\alpha_0)_{ij}(\mathbf{r}_\mathrm{A},iu)$
$\!+$ $\!(\alpha_0)_{ij}(\mathbf{r}_\mathrm{A},-iu)$.
As a result, a noticeable influence of the level broadening
on the off-resonant part of the CP force is 
erroneously predicted in Ref.~\cite{Kryszewski92} (cf.
Sec.~\ref{sec5.3}), thus demonstrating that body-induced level
broadening is a nonperturbative effect which lies beyond the scope
of the LRT approach to the problem.

Equation (\ref{eq148}) reveals that even the ground-state CP force
cannot be derived from a potential in the usual way, because of the
position dependence of the atomic polarizability. Nevertheless, it is
a potential force, provided that it is an irrotational vector, i.e.,
\begin{eqnarray}
\label{eq148-2}
\lefteqn{
\bm{\nabla}_{\!\!\mathrm{A}}\times\mathbf{F}_{00}
(\mathbf{r}_\mathrm{A})
     = \int_0^\infty \mathrm{d}u u^2
     \sum_k \bigg\{\bigl[\bm{\nabla}_{\!\!\mathrm{A}}
     \tilde{\omega}_{k0}(\mathbf{r}_\mathrm{A})\bigr]
     \frac{\partial}{\partial \tilde{\omega}_{k0}}
}
\nonumber\\&&
     +\,\bigl[\bm{\nabla}_{\!\!\mathrm{A}}
     \Gamma_k(\mathbf{r}_\mathrm{A})\bigr]
     \frac{\partial}{\partial \Gamma_k}\bigg\}
     \bigl[(\alpha_0)_{ij}
      (\mathbf{r}_\mathrm{A},iu)
\nonumber\\&&
     +\,(\alpha_0)_{ij}(\mathbf{r}_\mathrm{A},-iu)\bigr]\times
     \bm{\nabla}_{\!\!\mathrm{A}}G^{(1)}_{ij}
     (\mathbf{r}_\mathrm{A},\mathbf{r}_\mathrm{A},iu)
     = 0.
     \qquad
\end{eqnarray}
While for effectively one-dimensional problems (e.g., for an atom in the
presence of planarly, spherically, or cylindrically multilayered
media) this condition is satisfied, there are of course situations
where it is violated, implying that Eq.~(\ref{eq148}) is inaccessible
to perturbative methods in principle.

When the atomic system is initially prepared in a coherent
superposition of states such that $\sigma_{nm}(0)$ $\!\neq$ $\!0$
is valid for certain values $n$ and $m$ with $n$ $\!\neq$ $\!m$, then
--- according to Eq.~(\ref{eq135}) --- the corresponding off-diagonal
force components
$\sigma_{nm}(t)\mathbf{F}_{mn}(\mathbf{r}_\mathrm{A})$
can also contribute to the total force acting on the atomic system.
Interestingly, such transient off-diagonal force components contain
contributions not only from the electric part of the Lorentz force but
also from the magnetic part, as can be easily seen from inspection of
Eqs.~(\ref{eq143}) and (\ref{eq144}). Thus an atomic qubit
\mbox{$|\psi\rangle$ $\!=$ $\!(|0\rangle$ $\!+$
$\!|1\rangle)/\sqrt{2}$} (cf., e.g., Ref.~\cite{Zoller01})
near a body feels, in electric dipole approximation, both an electric
and a magnetic force in general.

Let us briefly comment on atomic systems displaying
(qua\-s\mbox{i)d}e\-ge\-neracies, i.e., systems exhibiting
transitions with $\omega_{mn}$ $\!\simeq$ $\!\omega_{m'n'}$
(\mbox{$m$ $\!\neq$ $\!m'$} and/or $n$ $\!\neq$ $\!n'$).
In such a case, the assumption that the (relevant) off-diagonal
density-matrix elements decouple from each other as well as from the
diagonal ones can no longer be made. Let us assume that the degenerate
sublevels are not connected via electric dipole transitions 
($\mathbf{d}_{mm'}$ $\!=$ $\!0$ if $\omega_{mm'}$ $\!\simeq$ $\!0$).
The degeneracy related to the different possible projections of the
angular momentum of an atom (in free space) onto a chosen direction
is a typical example. Taking into account that the degeneracy is
removed when the atom is close to a body, it may be advantageous
to change the basis within each degenerate sublevel accordingly and
consider the master equation in the new basis. An equation of the form
of Eq.~(\ref{eq125}) is then valid in the new basis. Note that the new
basis will in general depend on the position of the atom, thus
introducing an additional position dependence of the CP force.
While Eq.~(\ref{eq133}) also remains valid in the new basis for
$\omega_{mn}$ $\!\neq$ $\!0$, this is not in general true for
the temporal evolution of the density-matrix elements with
$\omega_{mm'}$ $\!\simeq$ $\!0$ so that, instead of the balance
equations (\ref{eq134}), a system of equations has to be solved in
which diagonal density-matrix elements and off-diagonal elements
with $\omega_{mm'}$ $\!\simeq$ $\!0$ are coupled to each other.

%%%%%%%%%%%%%%%%%%%%%%%%%%%%%%%%%%%%%%%%%%%%%%%%%%%%%%%%%%%%%%%%%%%%%%

\subsection{Example: Excited atom near an interface}
\label{sec5.3}

\begin{figure}[!t!]
\noindent
\begin{center}
\includegraphics[width=\linewidth]{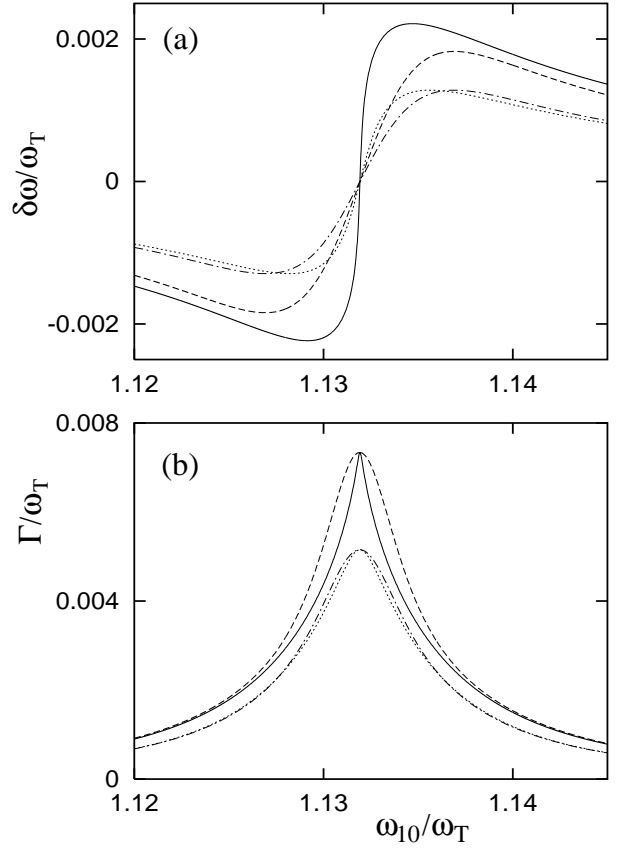}
\end{center}
\caption{
\label{Fig1}
(a) Transition frequency shift (solid and dotted lines) and
(b) decay rate (solid and dotted lines) versus bare transition
frequency for a two-level atom that is situated at distance
$z_\mathrm{A}$ from a semi-infinite half space medium of complex
permittivity according to Eq.~(\ref{eq155}) and whose transition
dipole moment is perpendicular to the interface
[$\omega_\mathrm{P}/\omega_\mathrm{T}$ $\!=$ $\!0.75$,
$\gamma/\omega_\mathrm{T}$ $\!=$ $\!0.01$; ${\omega}_\mathrm{T}^2
d_\mathrm{A}^2/(3\pi\hbar\varepsilon_0c^3)$ $\!=$ $\!10^{-7}$;
$z_\mathrm{A}/\lambda_\mathrm{T}$ $\!=$ $\!0.0075$ (solid and dashed
lines), $z_\mathrm{A}/\lambda_\mathrm{T}$ $\!=$ $\!0.009$ (dotted and
dot-dashed lines)]. For comparison, the approximate results obtained
by using the bare frequencies in Eqs.~(\ref{eq157}) and (\ref{eq160})
are also displayed (dashed and dot-dashed lines).}
\end{figure}

To illustrate the effects of body-induced level shifting and
broadening, let us consider a two-level atom with (real) transition
dipole matrix element
$\mathbf{d}_\mathrm{A}$ $\!\equiv$ $\!\mathbf{d}_{10}$ $\!=$ 
$\!d_\mathrm{A} (\cos\phi\sin\theta\,\mathbf{e}_x$ $\!+$
$\!\sin\phi\sin\theta\,\mathbf{e}_y$ $\!+$
$\!\cos\theta\,\mathbf{e}_z)$
($\mathbf{d}_{00}$ $\!=$ $\!\mathbf{d}_{11}$ $\!=$ $\!0$),
which is situated at position $z_\mathrm{A}$ very close above 
($z$ $\!>$ $\!0$) a semi-infinite half space \mbox{($z$ $\!<$ $\!0$)}
containing a homogeneous dispersing and absorbing magnetodielectric
medium. Let $\delta\omega$ $\!=$ $\!\delta\omega_1$
$\!-$ $\!\delta\omega_0$ denote the (position-dependent) shift of the
transition frequency. Using the Green tensor in the short-distance
limit, from Eqs.~(\ref{eq126}), (\ref{eq127}), and (\ref{eq131}) we
derive (see Appendix~\ref{AppG})
\begin{equation}
\label{eq151-1}
\delta\omega(z_\mathrm{A}) = \delta\omega_\mathrm{r}(z_\mathrm{A})
+ \delta\omega_\mathrm{or}(z_\mathrm{A}),
\end{equation}
\begin{equation}
\label{eq152}
\delta\omega_\mathrm{r}(z_\mathrm{A})
= - \frac{C}{\hbar z_\mathrm{A}^3}
\frac{|\varepsilon[\tilde{\omega}_{10}(z_\mathrm{A})]|^2-1}
{|\varepsilon[\tilde{\omega}_{10}(z_\mathrm{A})]+1|^2}\,,
\end{equation}
\begin{equation}
\label{eq152-1}
\delta\omega_\mathrm{or}(z_\mathrm{A})
= \frac{2C\tilde{\omega}_{10}(z_\mathrm{A})}{\hbar \pi z_\mathrm{A}^3}
\int_0^\infty\!\!\frac{\mathrm{d}u}
{\tilde{\omega}_{10}^2(z_\mathrm{A})+u^2}
\frac{\varepsilon(iu)-1}{\varepsilon(iu)+1}\,,
\end{equation}
where
\begin{equation}
\label{eq153}
C_{}=\frac{d_\mathrm{A}^2(1+\cos^2\theta)}{32\pi\varepsilon_0}\,.
\end{equation}
Note that in the short-distance limit the medium effectively acts like
a dielectric one. Since the relation $\tilde{\omega}_{10}$ $\!=$
$\!\omega_{10}$ $\!+$ $\!\delta\omega$ is valid, Eq.~(\ref{eq151-1})
together with Eqs.~(\ref{eq152}) and (\ref{eq152-1}) is a highly
transcendental equation for the determination of $\delta\omega$. To
solve it, we first note that the off-resonant term
$\delta\omega_\mathrm{or}$ may be neglected in most practical
situations. For example, for a single-resonance medium of
Drude-Lorentz type, 
\begin{equation}
\label{eq155}
\varepsilon(\omega)=1
+\frac{\omega_\mathrm{P}^2}
{\omega_\mathrm{T}^2-\omega^2-i\gamma\omega}\,,
\end{equation}
and the parameter values in Fig.~\ref{Fig1}, one can easily verify
the inequality
\begin{eqnarray}
\label{eq156}
\frac{\delta\omega_\mathrm{or}(z_\mathrm{A})}
{\tilde{\omega}_{10}(z_\mathrm{A})}
\le \frac{C\omega_\mathrm{P}^2}
{2\hbar z_\mathrm{A}^3
\omega_\mathrm{T}^2\tilde{\omega}_{10}(z_\mathrm{A})}
\lesssim 10^{-4}.
\end{eqnarray}
Thus, keeping only the resonant part of the frequency shift, we may
set
\begin{equation}
\label{eq157}
\delta\omega(z_\mathrm{A})
=-\frac{C}{\hbar z_\mathrm{A}^3}
\frac{|\varepsilon[\tilde{\omega}_{10}(z_\mathrm{A})]|^2-1}
{|\varepsilon[\tilde{\omega}_{10}(z_\mathrm{A})]+1|^2}\,.
\end{equation}
%
%%%%%%%%%%%%%%%  F I G U R E %%%%%%%%%%%%%%%%%%%%%%%%%%%%%%%%%%%%%%%%%
%%%%%%%%%%%%%%%%%%%%%%%%%%%%%%%%%%%%%%%%%%%%%%%%%%%%%%%%%%%%%%%%%%%%%%
%
For $\varepsilon(\tilde{\omega}_{10})$ from Eq.~(\ref{eq155}),
Eq.~(\ref{eq157}) is a fifth-order polynomial conditional equation for
$\delta\omega$, which may be solved numerically. Having calculated
$\delta\omega$, we may calculate the (position-dependent) decay rate
$\Gamma$ $\!\equiv$ $\!\Gamma_1$. Neglecting the small free-space
decay rate, we replace the Green tensor by its scattering part
as given by Eq.~(\ref{G4}), hence from Eqs.~(\ref{eq129}) and
(\ref{eq130}) we obtain
\begin{equation}
\label{eq160}
\Gamma(z_\mathrm{A})=
\frac{4C_{}}{\hbar z_\mathrm{A}^3}
\frac{\mathrm{Im}\,\varepsilon[\tilde{\omega}_{10}(z_\mathrm{A})]}
{|\varepsilon[\tilde{\omega}_{10}(z_\mathrm{A})]+1|^2}\,.
\end{equation}

%%%%%%%%%%%%%%%  F I G U R E %%%%%%%%%%%%%%%%%%%%%%%%%%%%%%%%%%%%%%%%%
\begin{figure}[!b!]
\noindent
\begin{center}
\includegraphics[width=\linewidth]{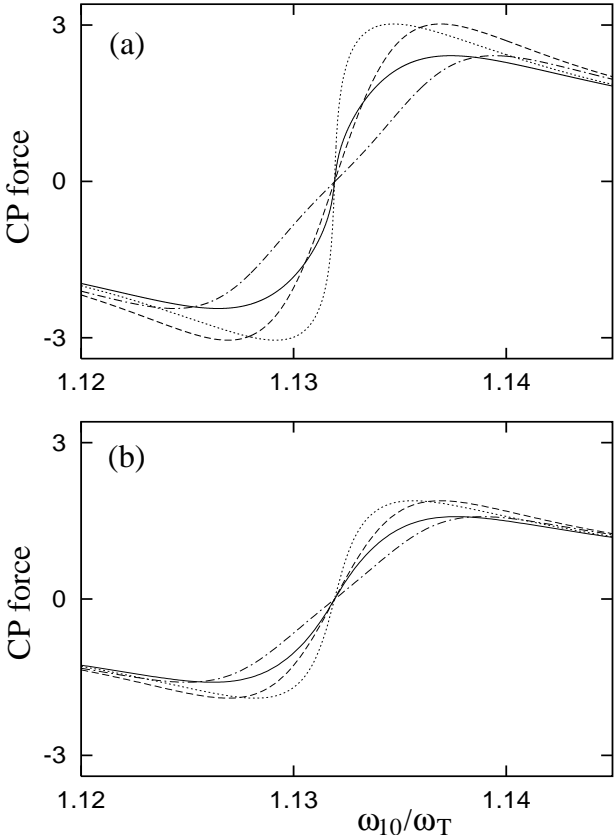}
\end{center}
\caption{
\label{Fig2}
The resonant part of the CP force 
$F_{11}^\mathrm{r}\lambda_\mathrm{T}^4\times
10^{-9}/(3C)$ on a two-level atom that is situated at distance
(a) $z_\mathrm{A}/\lambda_\mathrm{T}$ $\!=$ $\!0.0075$ and
(b) $z_\mathrm{A}/\lambda_\mathrm{T}$ $\!=$ $\!0.009$
of a semi-infinite half space medium of complex permittivity according
to Eq.~(\ref{eq155}) and whose transition dipole moment is
perpendicular to the interface (solid lines). The parameters are the
same as in Fig.~\ref{Fig1}. For comparison, both the perturbative
result (dashed lines) and the separate effects of level shifting
(dotted lines) and level broadening (dash-dotted lines) are shown.
}
\end{figure}
%%%%%%%%%%%%%%%%%%%%%%%%%%%%%%%%%%%%%%%%%%%%%%%%%%%%%%%%%%%%%%%%%%%%%%

The resonant part of the CP force on the excited atom in the
short-distance limit can be found by taking the derivative of
the scattering part of the Green tensor [Eq.~(\ref{G4})] with respect
to $z_\mathrm{A}$ and substituting the result into Eq.~(\ref{eq149})
($l$ $\!=$ $\!1$). We derive (${\mathbf{F}}_{11}^\mathrm{r}$ $\!=$
$\!F_{11}^\mathrm{r}\mathbf{e}_z$)
\begin{equation}
\label{eq163}
F_{11}^\mathrm{r}(z_\mathrm{A})
=
-\frac{3C_{}}{z_\mathrm{A}^4}
\frac{|\varepsilon[\Omega_{10}(z_\mathrm{A})]|^2-1}
{|\varepsilon[\Omega_{10}(z_\mathrm{A})]+1|^2}\,,
\end{equation}
where, according to Eq.~(\ref{eq140}),
\begin{equation}
\label{eq164}
 \Omega_{10}(z_\mathrm{A})=\tilde{\omega}_{10}(z_\mathrm{A})
 +i\Gamma(z_\mathrm{A})/2.
\end{equation}
Using Eq.~(\ref{eq155}), we see that 
($\gamma,\Gamma$ $\!\ll$ $\!\omega_\mathrm{T}$)
\begin{equation}
\label{eq165}
\varepsilon[\Omega_{10}(z_\mathrm{A})]=1
+\frac{\omega_\mathrm{P}^2}{\omega_\mathrm{T}^2
-\tilde{\omega}_{10}^2(z_\mathrm{A})
-i[\Gamma(z_\mathrm{A})+\gamma]
\tilde{\omega}_{10}(z_\mathrm{A})}\,.
\end{equation}
Equation (\ref{eq163}) differs from the perturbative result in two
respects. First, the bare atomic transition frequency $\omega_{10}$
is replaced with the (position-dependent) shifted frequency
$\tilde{\omega}_{10}$. Second, the absorption parameter $\gamma$ of
the medium is replaced with the sum of $\gamma$ and the
(position-dependent) atomic decay rate $\Gamma$. The sum 
\mbox{$\gamma$ $\!+$ $\!\Gamma$} obviously plays the role of the total
absorption parameter.

The dependence of $\delta\omega$ and $\Gamma$ on $\omega_{10}$
in the short-distance limit is shown in Figs.~\ref{Fig1}(a) and (b),
respectively, and Fig.~\ref{Fig2} displays the resonant part of the
CP force as a function of $\omega_{10}$. From Fig.~\ref{Fig2} it
is seen that in the vicinity of the (surface-plasmon induced)
frequency $\omega_\mathrm{S}$ $\!=$ $\!\sqrt{\omega_\mathrm{T}^2+
\omega_\mathrm{P}^2/2}$ an enhanced force is observed, which is
attractive (repulsive) for red (blue) detuned atomic transition
frequencies $\omega_{10}$ $\!<$ $\!\omega_\mathrm{S}$
($\omega_{10}$ $\!>$ $\!\omega_\mathrm{S}$) --- a result already known
from perturbation theory (dashed curves in the figure).
However, it is also seen that due to body-induced level shifting and
broadening the absolute value of the force can be noticeably reduced
(solid curves in the figure). Interestingly, the positions of the
extrema of the force remain nearly unchanged, because level shifting
and broadening give rise to competing effects that almost cancel.

In order to calculate the off-resonant part of the CP force on the
excited atom in the short-distance limit, we first note that,
according to Eq.~(\ref{eq67}),
\begin{eqnarray}
\label{eq166}
      &\!&\!\bm{\alpha}_{1}(z_\mathrm{A},iu)
      +\bm{\alpha}_{1}(z_\mathrm{A},-iu) =
      -\frac{4\mathbf{d}_\mathrm{A} \otimes 
      \mathbf{d}_\mathrm{A}}{\hbar}
      \nonumber\\
      &\!&\!\quad\times\frac{\tilde{\omega}_{10}(z_\mathrm{A})}
      {\tilde{\omega}_{10}^2(z_\mathrm{A})
        +[u+\Gamma(z_\mathrm{A})/2]^2}
      \nonumber\\
      &\!&\!\quad\times
        \frac{
        \tilde{\omega}_{10}^2(z_\mathrm{A})
        +u^2+\Gamma^2(z_\mathrm{A})/4}
        {\tilde{\omega}_{10}^2(z_\mathrm{A})
        +[u-\Gamma(z_\mathrm{A})/2]^2}\,.
\end{eqnarray}
Substituing Eq.~(\ref{eq166}) into Eq.~(\ref{eq148}) and making use
of Eq.~(\ref{G6}) [where $f(u)$ is given by $u^2$ times
Eq.~(\ref{eq166})], we derive (${\mathbf{F}}_{11}^\mathrm{or}$ $\!=$
$\!F_{11}^\mathrm{or}\mathbf{e}_z$)
\begin{eqnarray}
\label{eq167}
      F_{11}^\mathrm{or}(z_\mathrm{A})
      &\!=&\!\frac{3C}{\pi z_\mathrm{A}^4}
      \int_0^\infty\mathrm{d} u\,
      \frac{\varepsilon(iu)-1}{\varepsilon(iu)+1}\nonumber\\
      &\!&\!\times\frac{\tilde{\omega}_{10}(z_\mathrm{A})}
      {\tilde{\omega}_{10}^2(z_\mathrm{A})
        +[u+\Gamma(z_\mathrm{A})/2]^2}
      \nonumber\\
      &\!&\!\times\frac{\tilde{\omega}_{10}^2(z_\mathrm{A})+u^2
        +\Gamma^2(z_\mathrm{A})/4}
        {\tilde{\omega}_{10}^2(z_\mathrm{A})
        +[u-\Gamma(z_\mathrm{A})/2]^2}\,.
\end{eqnarray}
Note that for a two-level atom the relation
\begin{equation}
\label{eq168}
       F_{00}^\mathrm{or}(z_\mathrm{A})
       =- F_{11}^\mathrm{or}(z_\mathrm{A})
\end{equation}
is valid.

%%%%%%%%%%%%%%%  F I G U R E %%%%%%%%%%%%%%%%%%%%%%%%%%%%%%%%%%%%%%%%%
\begin{figure}[!t!]
\noindent
\begin{center}
\includegraphics[width=\linewidth]{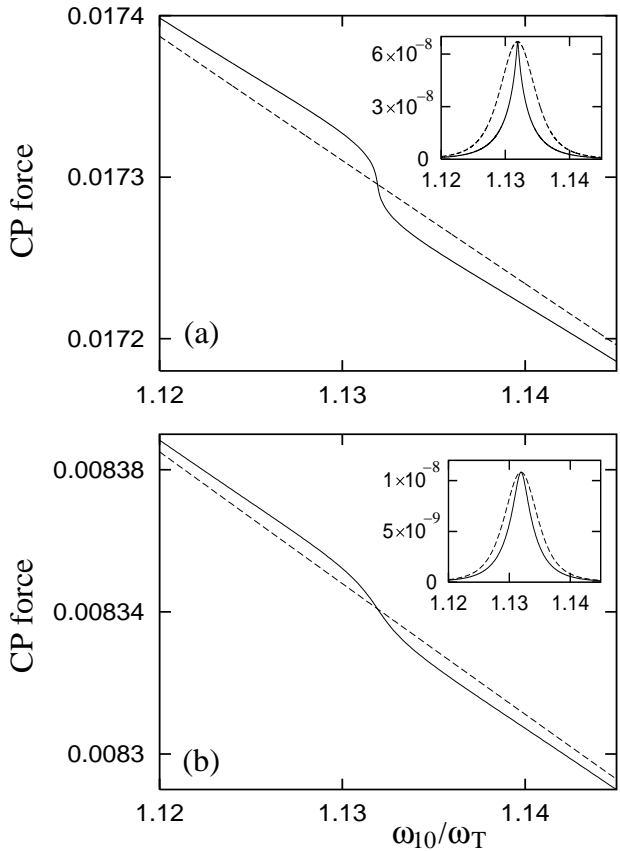}
\end{center}
\caption{
\label{Fig3}
The off-resonant part of the CP force
\mbox{$F_{11}^\mathrm{r}\lambda_\mathrm{T}^4\times 10^{-9}$}
\mbox{$/(3C)$} on a two-level atom that is situated at distance
(a) $z_\mathrm{A}/\lambda_\mathrm{T}$ $\!=$ $\!0.0075$ and
(b) $z_\mathrm{A}/\lambda_\mathrm{T}$ $\!=$ $\!0.009$
of a semi-infinite half space medium of complex permittivity according
to Eq.~(\ref{eq155}) and whose transition dipole moment is
perpendicular to the interface (solid lines). The parameters are the
same as in Fig.~\ref{Fig1}. For comparison, the perturbative
result (dashed lines) is shown. The insets display the difference
between the force with and without consideration of the level
broadening (solid lines). For comparison, we show this difference in
the case where the level shifts are ignored (dashed lines).
}
\end{figure}
%%%%%%%%%%%%%%%%%%%%%%%%%%%%%%%%%%%%%%%%%%%%%%%%%%%%%%%%%%%%%%%%%%%%%%

In Fig.~\ref{Fig3}, the off-resonant part of the CP force is shown
as a function of the bare atomic transition frequency. Obviously, the
shift of the transition frequency has the effect of raising and
lowering the perturbative values of the force (dashed curves) for
\mbox{$\omega_{10}$ $\!<$ $\!\omega_\mathrm{S}$} and
\mbox{$\omega_{10}$ $\!>$ $\!\omega_\mathrm{S}$}, respectively,
which is in full agreement with the frequency response of the
frequency shift shown in Fig.~\ref{Fig1}(a). The influence of the
decay rate on the CP force is extremely weak, as it can be seen from
the insets in the figure. This may be understood by the fact that in
contrast to the case of the resonant part of the CP force, where the
decay rate enters directly via the Green tensor, the influence on the
off-resonant part is more indirect via the atomic polarizability. Due
to the specific dependence on the atomic polarizability, the
leading-order dependence is quadratic in $\Gamma$ and not linear in
$\Gamma$ as erroneously predicted from LRT \cite{Kryszewski92}.
Physically, the weak influence of the level broadening on the
off-resonant part of the CP force may be regarded as being a
consequence of the fact that this part corresponds to energy
nonconserving processes (the energy denominators being nonzero),
which implies that they happen on (extremely short) time scales where 
real photon emission does not play a role.

Comparing the magnitudes of the resonant and off-resonant components
of the CP force, we see that the off-resonant component is smaller
than the resonant one by about two orders of magnitude. However, this
observation should be considered with great care. While the two-level
atom is a good model for calculating the resonant part of an atom in
an excited state, such a simplification is not justified in general
when all higher levels can contribute to the off-resonant force
component. However, provided that the convergence of the
corresponding sum is sufficiently fast, we can still conclude that the
resonant part of the CP force is dominant.

%%%%%%%%%%%%%%%%%%%%%%%%%%%%%%%%%%%%%%%%%%%%%%%%%%%%%%%%%%%%%%%%%%%%%%

\section{Summary}
\label{sec6}

Basing on electromagnetic-field quantization that allows for the
presence of dispersing and absorbing linear media, and starting with
the Lorentz force acting on a neutral atom, we have extended the
concept of CP force beyond the well-known results derived on the basis
of normal-mode quantization or LRT in leading order of pertubation
theory to allow for (i) magnetodielectric bodies, (ii) an atom that is
initially prepared in an arbitrary internal (electronic) quantum
state, thereby being subjected to a time-dependent force, (iii) the
position dependence of the force via the atomic response, and (iv)
arbitrary strength of the atom-field coupling. The basic formulas also
apply to the calculation of the radiation forces arising from excited
fields such as the force acting on a driven atom.

For a first analysis, we have restricted our attention to a
nondriven atom in the weak-coupling regime, so that the internal
atomic dynamics can be treated in Markov approximation. It turns out
that the force is a superposition of force components weighted by the
time-dependent intra-atomic density-matrix elements that solve the
intra-atomic master equation. Each force component is expressed in
terms of the Green tensor of the electromagnetic field and the atomic
polarizability, which --- through the position-dependent energy level
shifts and broadenings --- now depends on the position of the atomic
system. In consequence even the force components resulting from the
electric part of the Lorentz force cannot be derived from potentials
in the usual way. Clearly, the position dependence of the atomic
polarizability become noticeable only for very small atom-body
separations. In order to illustrate the effect, we have considered a
two-level atom in the vicinity of a planar semi-infinite medium.

When the atomic system is initially prepared in an eigenstate of its
internal Hamiltonian, then only force components associated with
diagonal density-matrix elements appear. They solely result from the
electric part of the Lorentz force and reduce to the CP forces
obtained in lowest-order perturbation theory if the atomic
polarizability is replaced with its position-independent perturbative
expression. Force components that are associated with excited
intra-atomic energy levels are of course transient. As in the course
of time an initially excited level is depopulated and lower
lying levels are populated, the force that initially acts on the
atomic system in the excited state changes with time to the force that
acts on the atomic system in the ground state.

The results further show that when the atomic system is initially
prepared in an intra-atomic quantum state that is a coherent
superposition of energy eigenstates, then additional force components
associated with the corresponding off-diagonal density-matrix elements
are observed. Thus an atomic qubit would typically feel such
off-diagonal force components. It should be pointed out that not only
the electric but also the magnetic part of the Lorentz force can
contribute to the off-diagonal force components, with the magnetic
contributions being proportional to the transition frequencies.
Clearly, off-diagonal force components are transient.

In contrast to the transient force components that are associated with
excited energy levels, off-diagonal force components carry an
additional harmonic time dependence. Clearly, if the oscillations are
too fast, it can be difficult to detect them experimentally, since
they may effectively average to zero. In this case it may be advisable
to assign them to the fluctuating part of the force rather than to the
average force. The situation may be different in cases where strong
atom-field coupling (not considered here) gives rise to Rabi
oscillations.

%%%%%%%%%%%%%%%%%%%%%%%%%%%%%%%%%%%%%%%%%%%%%%%%%%%%%%%%%%%%%%%%%%%%%%

\begin{acknowledgments}
S.Y.B. acknowledges valuable discussions with O. P. Sushkov as well
as M.-P. Gorza. This work was supported by the Deutsche
Forschungsgemeinschaft. S.Y.B. is grateful for being granted a
Th\"{u}\-rin\-ger Landesgraduiertenstipendium.
\end{acknowledgments}

%%%%%%%%%%%%%%%%%%%%%%%%%%%%%%%%%%%%%%%%%%%%%%%%%%%%%%%%%%%%%%%%%%%%%%

\appendix

\section{Derivation of the multipolar Hamiltonian (\ref{eq37})}
\label{AppA}

To perform transformations of the type
\begin{equation}
 \label{A1}
  \hat{O}'=\hat{U}\hat{O}\hat{U}^\dagger,
\end{equation}
with $\hat{U}$ being given by Eq.~(\ref{eq35}) together with
Eq.~(\ref{eq36}), we apply the operator identity
\begin{equation}
\label{A2}
      e^{\hat{S}}\hat{O}e^{-\hat{S}}=
      \hat{O} + \big[\hat{S},\hat{O}\big]+\frac{1}{2!}
      \big[\hat{S},\big[\hat{S},\hat{O}\big]\big]+\ldots\,.
\end{equation}
Recalling the commutation relations (\ref{eq4}) and (\ref{eq5}), it is
not difficult to prove that the basic fields 
$\hat{\mathbf{f}}(\mathbf{r},\omega)$ are transformed as
\begin{equation}
\label{A3}
    \hat{\mathbf{f}}_\lambda'(\mathbf{r},\omega)
    = \hat{\mathbf{f}}_\lambda(\mathbf{r},\omega) 
    + \frac{1}{\hbar\omega}
    \int \mathrm{d}^3 r' \hat{\mathbf{P}}_\mathrm{A}^\perp
    (\mathbf{r}')
    \bm{G}_\lambda^\ast(\mathbf{r}',\mathbf{r},\omega).
\end{equation}
Using Eq.~(\ref{A2}) together with the commutation relation
$[\varepsilon_0\hat{E}_k(\mathbf{r}),\hat{A}_l(\mathbf{r})]$
$\!=$ $\!i\hbar\delta^\perp_{kl}(\mathbf{r}-\mathbf{r}')$, cf.
Ref.~\cite{Ho03}, we find that
\begin{equation}
\label{A4}
      \hat{\mathbf{E}}^{\prime} (\mathbf{r})
      = \hat{\mathbf{E}} (\mathbf{r})
      +\frac{1}{\varepsilon_0} \hat{\mathbf{P}}^{\perp}_\mathrm{A} 
      (\mathbf{r}).
\end{equation}
To transform the momenta of the charged particles, the identities
\begin{eqnarray}
\label{A5}
\lefteqn{
     \bm{\nabla}_{\!\!\alpha} \delta(\mathbf{r}
     -\hat{\mathbf{r}}_\mathrm{A}
              - \lambda\hat{\bar{\mathbf{r}}}_{\beta})
}
\nonumber\\&&
      = \left[\big(\lambda -1\big)\frac{m_\alpha}{m_\mathrm{A}}
      - \lambda \delta_{\alpha\beta}
      \right]
      \bm{\nabla} \delta(\mathbf{r}-\hat{\mathbf{r}}_\mathrm{A}
              - \lambda\hat{\bar{\mathbf{r}}}_{\beta}),
      \qquad\quad
\\
\label{A6}
\lefteqn{
       \int_0^1 \mathrm{d} \lambda\,
       \hat{\bar{\mathbf{r}}}_\alpha
      \bm{\nabla}
      \delta(\mathbf{r}\!-\!\hat{\mathbf{r}}_\mathrm{A}
              \!-\! \lambda\hat{\bar{\mathbf{r}}}_\alpha)
      = \delta(\mathbf{r}\!-\!\hat{\mathbf{r}}_\mathrm{A})
      - \delta(\mathbf{r}\!-\!\hat{\mathbf{r}}_\alpha),
}
\\
\label{A7}
\lefteqn{
      \int_0^1 \mathrm{d} \lambda\lambda
       \hat{\bar{\mathbf{r}}}_\alpha
      \bm{\nabla}
      \delta(\mathbf{r}-\hat{\mathbf{r}}_\mathrm{A}
              - \lambda\hat{\bar{\mathbf{r}}}_\alpha)
}
\nonumber\\&&
      = - \delta(\mathbf{r}-\hat{\mathbf{r}}_\alpha)
       +\int_0^1 \mathrm{d} \lambda\,
      \delta(\mathbf{r}-\hat{\mathbf{r}}_\mathrm{A}
              - \lambda\hat{\bar{\mathbf{r}}}_\alpha)
\end{eqnarray}
are helpful. They can be proved with the aid of the definitions
(\ref{eq27}) and (\ref{eq28}), and via (partial) integration with
respect to $\lambda$. Using Eqs.~(\ref{A5})--(\ref{A7}) we derive
\begin{equation}
\label{A8}
     \hat{\mathbf{p}}'_\alpha
     =
    \hat{\mathbf{p}}_\alpha 
    - q_\alpha \hat{\mathbf{A}}(\hat{\mathbf{r}}_\alpha)
      -\int\!\mathrm{d}^3 r
      \hat{\bm{\Xi}}_\alpha(\mathbf{r})
      \times \hat{\mathbf{B}}(\mathbf{r}),
\end{equation}
where $\hat{\bm{\Xi}}_\alpha(\mathbf{r})$ is defined as in
Eq.~(\ref{eq38}).
Further, the following quantities remain unchanged under the
transformation (\ref{A1}), because they commute with both 
$\hat{\mathbf{A}}(\mathbf{r})$ (cf. Ref.~\cite{Ho03}) and 
$\hat{\mathbf{r}}_\alpha$,
\begin{eqnarray}
\label{A9}
 &&\hspace{-8ex}
 \hat{\mathbf{A}}'(\mathbf{r})=\hat{\mathbf{A}}(\mathbf{r}),\ 
 \hat{\mathbf{B}}'(\mathbf{r})=\hat{\mathbf{B}}(\mathbf{r}),\ 
 \hat{\varphi}'(\mathbf{r})=\hat{\varphi}(\mathbf{r}), 
\\
\label{A10}
 &&\hspace{-8ex}
 \hat{\mathbf{r}}'_\alpha \!=\hat{\mathbf{r}}_\alpha,\ 
 \hat{\mathbf{r}}'_\mathrm{A} \!=\hat{\mathbf{r}}_\mathrm{A},\ 
 \hat{\rho}_\mathrm{A}'(\mathbf{r})\!=\!
 \hat{\rho}_\mathrm{A}(\mathbf{r}),\ 
 \hat{\varphi}_\mathrm{A}'(\mathbf{r})\!=\!
 \hat{\varphi}_\mathrm{A}(\mathbf{r}),
\\
\label{A11}
     &&\hspace{-8ex}
      \hat{\mathbf{P}}'_\mathrm{A}(\mathbf{r})
      \!=\!\hat{\mathbf{P}}_\mathrm{A}(\mathbf{r}),\
      \hat{\bm{\Theta}}'_\alpha(\mathbf{r})
      \!=\!\hat{\bm{\Theta}}_\alpha(\mathbf{r}),\ 
      \hat{\bm {\Xi}}'_\alpha(\mathbf{r})\!=\!
      \hat{\bm {\Xi}}_\alpha(\mathbf{r}). 
\end{eqnarray}

Applying the transformation rules (\ref{A3}), (\ref{A8}), and
(\ref{A9})--(\ref{A11}), we may now express the minimal-coupling
Hamiltonian (\ref{eq1}) in terms of the transformed variables.
Recalling Eq.~(\ref{eq20}) together with
Eqs.~(\ref{eq8})--(\ref{eq10}) and making use of the relations 
(\ref{eq18}) and
\begin{equation}
\label{A12}
    \int_0^\infty \mathrm{d}\omega\,
    \frac{\omega}{c^2}\, \mathrm{Im}\,\bm{G}
    (\mathbf{r},\mathbf{r}',\omega)
    = \frac{\pi}{2}\, \delta(\mathbf{r}-\mathbf{r}')
\end{equation}
(cf. Ref.~\cite{Knoll01}), we derive
\begin{eqnarray}
\label{A13}
      \hat{H}&=&
      \sum_{\lambda=e,m}\int\mathrm{d}^3r
      \int_0^{\infty}\mathrm{d}\omega\,
      \hbar\omega\,\hat{\mathbf{f}}^{\prime\dagger}_\lambda
      (\mathbf{r},\omega)
      \hat{\mathbf{f}}_\lambda'(\mathbf{r},\omega)
\nonumber\\&&
      + \frac{1}{2\varepsilon_0} \int\mathrm{d}^3r\,
      \hat{\mathbf{P}}^{\prime\perp}_\mathrm{A} (\mathbf{r})
      \hat{\mathbf{P}}^{\prime\perp}_\mathrm{A} (\mathbf{r})
      - \int\mathrm{d}^3r\, 
      \hat{\mathbf{P}}^{\prime\perp}_\mathrm{A}(\mathbf{r})
                   \hat{\mathbf{E}}^{\prime\perp} (\mathbf{r})
\nonumber\\
      &&+\sum_{\alpha}\frac{1}{2 m_{\alpha}}
      \left[\hat{\mathbf{p}}'_\alpha
      +  \int\!\mathrm{d}^3 r\, \hat{\bm{\Xi}}'_\alpha(\mathbf{r})
      \times \hat{\mathbf{B}}'(\mathbf{r}) \right]^2
\nonumber\\
      &&
      + {\textstyle\frac{1}{2}}
      \int\mathrm{d}^3r\,\hat{\rho}'_\mathrm{A}(\mathbf{r})
      \hat{\varphi}'_\mathrm{A}(\mathbf{r})
      +\int\mathrm{d}^3r\,\hat{\rho}'_\mathrm{A}(\mathbf{r})
      \hat{\varphi}'(\mathbf{r}).
      \quad
\end{eqnarray}
In order to simplify the last two terms of Eq.~(\ref{A13}), we recall
Eq.~(\ref{eq83}) as well as 
$\hat{\mathbf{P}}^{\prime\parallel}_\mathrm{A}(\mathbf{r})$
$\!=$ $\!\varepsilon_0\bm{\nabla}\hat{\varphi}_\mathrm{A}'
(\mathbf{r})$
and $\hat{\mathbf{E}}^{\prime\parallel}(\mathbf{r})$
$\!=$ $\!-\bm{\nabla}\hat{\varphi}'(\mathbf{r})$, obtaining with the
aid of partial integration
\begin{eqnarray}
\label{A14}
\lefteqn{
   {\textstyle\frac{1}{2}}\!\int\mathrm{d}^3r\,
   \hat{\rho}'_\mathrm{A}(\mathbf{r})
      \hat{\varphi}'_\mathrm{A}(\mathbf{r})
      +\int\mathrm{d}^3r\,\hat{\rho}'_\mathrm{A}(\mathbf{r})
      \hat{\varphi}'(\mathbf{r})
}
\nonumber\\&&\hspace{-1ex}
  ={\textstyle\frac{1}{2}}\!\int\mathrm{d}^3r\,
 \hat{\mathbf{P}}'_\mathrm{A}(\mathbf{r})
 \bm{\nabla}\hat{\varphi}'_\mathrm{A}(\mathbf{r})
  \!+\!\int\mathrm{d}^3r\,\hat{\mathbf{P}}'_\mathrm{A}(\mathbf{r})
  \bm{\nabla}\hat{\varphi}'(\mathbf{r})
\nonumber\\&&\hspace{-1ex}
 =\frac{1}{2\varepsilon_0}\int\!\mathrm{d}^3r\,
 \hat{\mathbf{P}}'_\mathrm{A}(\mathbf{r})
 \hat{\mathbf{P}}^{\prime\parallel}_\mathrm{A}(\mathbf{r}) \!-\!\!
\int\!\mathrm{d}^3r\,\hat{\mathbf{P}}'_\mathrm{A}(\mathbf{r})
 \hat{\mathbf{E}}^{\prime\parallel}(\mathbf{r}).
\qquad 
\end{eqnarray}
Combining Eqs.~(\ref{A13}) and (\ref{A14}), and noting that integrals
containing mixed products of transverse and longitudinal vector fields
vanish, we obtain Eq.~(\ref{eq37}), where we have made use of
Eqs.~(\ref{A10}) and (\ref{A11}) and hence dropped the primes of all
quantities containing the particle coordinates only.

In the simpler case in which the center-of-mass coordinate is treated
as a parameter, the transformation law (\ref{A8}) changes to
\begin{equation}
\label{A15}
     \hat{\mathbf{p}}'_\alpha
     =\hat{\mathbf{p}}_\alpha 
     - q_\alpha \hat{\mathbf{A}}(\hat{\mathbf{r}}_\alpha)
      -\int\!\mathrm{d}^3 r\,
      \hat{\bm{\Theta}}_\alpha(\mathbf{r})
      \times \hat{\mathbf{B}}(\mathbf{r}).
\end{equation}
Equations (\ref{A3}), (\ref{A4}), and (\ref{A9})--(\ref{A11}) remain
formally the same, provided that the replacement 
$\hat{\mathbf{r}}_\mathrm{A}$ $\!\mapsto$ $\!\mathbf{r}_\mathrm{A}$ is
made.

%%%%%%%%%%%%%%%%%%%%%%%%%%%%%%%%%%%%%%%%%%%%%%%%%%%%%%%%%%%%%%%%%%%%%%

\section{Orders of magnitude of interaction terms}
\label{AppB}

To estimate the order of magnitude of atom-field interactions, let us
introduce the typical atomic length and energy scales
\begin{eqnarray}
\label{B1}
 a_0
 \hspace{-1ex}&=&\hspace{-1ex}
 \frac{a_\mathrm{B}}{Z_\mathrm{eff}}
 =\frac{\hbar}{Z_\mathrm{eff}\alpha_0 m_e c}\,,\\
\label{B2}
 E_0
 \hspace{-1ex}&=&\hspace{-1ex}
 Z_\mathrm{eff}^2 E_\mathrm{R}=\frac{Z_\mathrm{eff}^2\hbar^2}
{2m_ea_\mathrm{B}^2}\approx Z_\mathrm{eff}^213.6\mathrm{eV}
\end{eqnarray}
($a_\mathrm{B}$, Bohr radius; $E_\mathrm{R}$, Rydberg energy), where
$m_e$ and $-e$ are the electron mass and charge, respectively,
$Z_\mathrm{eff}e$ is the typical effective nucleus charge felt 
by the electrons giving the main contributions to the interaction
terms to be calculated, and \mbox{$\alpha_0$ $\!=$ $\!e^2/
(4\pi\varepsilon_0\hbar c)$} is the fine-structure constant.
As a rough estimate we can then make the replacements
\begin{eqnarray}
\label{B3}
&q_\alpha \ \rightarrow\ e,
\quad
m_\alpha \ \rightarrow\ m_e,
\quad
\omega_{kl} \ \rightarrow\ E_0/\hbar,&
\\
\label{B4}
&{\hat{\bar{\mathbf{r}}}}_\alpha \ \rightarrow\ a_0,
\quad
\dot{\hat{\mathbf{r}}}_\mathrm{A} \ \rightarrow\ v,
\quad
{\hat{\bar{\mathbf{p}}}}{}^{(\prime)}_\alpha \ \rightarrow\ p= m_eE_0a_0/\hbar&
\quad
\end{eqnarray}
[for the last replacement, see Eq.~(\ref{C7})]. With regard to the
length scale of variation of the medium-assisted electromagnetic field
we may make the replacements
\begin{eqnarray}
\label{B5}
&\bm{\nabla} \ \rightarrow\  \lambda^{-1}
\sim \omega/c,
&\bm{\nabla}\hat{\varphi}\ \rightarrow\ \nabla\varphi \sim \omega A,
\\
\label{B6}
&\hat{\mathbf{E}}^{(\prime)} \ \rightarrow\ E \sim \omega A,
&\hat{\mathbf{B}}^{(\prime)} \ \rightarrow\ B \sim (\omega/c)A
\end{eqnarray}
($\hat{\mathbf{A}}^{(\prime)}$ $\to$ $A$). Noting that
materials typically become transparent for frequencies that are
greater than $20\,\mathrm{eV}$ (cf. Ref.~\cite{Adachi}),
\begin{equation}
\label{B7}
 \varepsilon(\mathbf{r},\omega)\approx 1\Rightarrow
 \bm{G}^{(1)}(\mathbf{r},\mathbf{r}',\omega)\approx 0
 \ \mbox{for}\
 \hbar\omega \gtrsim 20\,\mathrm{eV},
\end{equation}
we should require that
\begin{equation}
\label{B8}
\hbar\omega
\lesssim 20\,\mathrm{eV}
\quad\Rightarrow\quad
\frac{\hbar\omega}{E_0}
\lesssim 1.
\end{equation}

With these approximations at hand, the orders of magnitude of
$\Delta_1 E$ defined by Eq.~(\ref{eq54}) and $\Delta_2 E$ defined by
Eq.~(\ref{eq53}) in Sec.~\ref{sec3.1} can be estimated to be
\begin{equation}
\label{B9}
 \Delta_1 E \sim \frac{e^2A^2}{2m_e}
 =\frac{e^2a_0^2A^2}{\hbar^2}E_0=g^2E_0 = O\!\left(g^2\right)
\end{equation}
and
\begin{eqnarray}
\label{B10}
\lefteqn{
 \Delta_2 E \sim \frac{1}{E_0+\hbar\omega}
 \left(\frac{e^2p^2A^2}{m_e^2}+2\frac{ea_0\nabla\varphi epA}{m_e}
 +e^2a_0^2\nabla\varphi^2\right)
}
\nonumber\\&&
 = g^2\left[1+2\left(\frac{\hbar\omega}{E_0}\right)
 +\left(\frac{\hbar\omega}{E_0}\right)^2\right]
 \frac{E_0}{1+\hbar\omega/E_0}
 = O\!\left(g^2\right)\!,
\nonumber\\&&
\end{eqnarray}
where the dimensionless coupling constant
\begin{equation}
\label{B11}
 g\equiv ea_0A/\hbar
\end{equation}
has been introduced. Note that in Eq.~(\ref{B10}) we have approximated
$\hat{\mathbf{p}}_\alpha\to p$, because in Sec.~\ref{sec3} we treat an
atom at rest, hence relative and absolute momenta are identical.

In order to give a rough idea of the magnitude of the coupling
constant $g$, we need to estimate the magnitude of the field strength
$A$. In the context of the current work we consider interactions of
an atomic system with the vacuum electromagnetic field, so the
relevant quantity is the vacuum fluctuation of the field strength.
Recalling Eqs.~(\ref{eq8}) and (\ref{eq20}) and making use of the
commutation relations (\ref{eq4}) and (\ref{eq5}) as well as the
integral relation (\ref{eq18}), we find
\begin{eqnarray}
\label{B11-1}
&&\langle[\Delta\hat{\mathbf{E}}(\mathbf{r}_\mathrm{A})]^2\rangle
=\langle\{0\}|\hat{\mathbf{E}}^2(\mathbf{r}_\mathrm{A})|\{0\}\rangle-
\langle\{0\}|\hat{\mathbf{E}}(\mathbf{r}_\mathrm{A})
|\{0\}\rangle^2\nonumber\\
&&\quad=\frac{\hbar}{\pi\varepsilon_0}\int_0^\infty\mathrm{d}\omega'
\frac{\omega^{\prime2}}{c^2}\mathrm{Im}\mathrm{Tr} 
\bm{G}(\mathbf{r}_\mathrm{A},\mathbf{r}_\mathrm{A},\omega').
\end{eqnarray}
When the atomic system is placed sufficiently far away from all 
macroscopic bodies, a good estimate for the integral can be given by
using the vacuum Green tensor 
$\mathrm{Im}\bm{G}^{(0)}(\mathbf{r}, \mathbf{r}, 
\omega)=\omega/(6\pi c)\bm{I}$, leading to
\begin{eqnarray}
\label{B11-2}
\langle[\Delta\hat{\mathbf{E}}(\mathbf{r}_\mathrm{A})]^2\rangle
\sim\frac{\hbar\omega^4}{6\pi^2\varepsilon_0c^3},
\end{eqnarray}
where $\omega$ is a characteristic frequency contributing to the
interaction, cf. Eq.~(\ref{B8}). Hence making the replacement
\begin{eqnarray}
\label{B11-3}
A\sim
\sqrt{\frac{\hbar\omega^2}{6\pi^2\varepsilon_0c^3}}
\end{eqnarray}
[cf. Eq.~(\ref{B6})], we find
\begin{equation}
\label{B11-4}
g\sim Z_\mathrm{eff}\sqrt{\frac{\alpha_0}{6\pi}}
\left(\frac{\hbar\omega}{E_0}\right)\alpha_0\sim 10^{-2},
\end{equation}
depending on the specific atomic system considered and 
the characteristic frequencies of the medium. When the atom is
situated close to some macroscopic body, the scattering Green tensor
becomes much larger than the vacuum Green tensor, and the
approximation leading to Eq.~(\ref{B11-2}) is not valid anymore. The
increased value of the coupling constant $g$ is reflected by the
failure of the perturbative result for small atom-surface separations.

The orders of magnitude of the contributions of the three terms in
Eq.~(\ref{eq51}) to the eigenvalue shift in Sec.~\ref{sec3.2} can be
estimated according to
\begin{eqnarray}
\label{B12}
\lefteqn{
\frac{\bigl|\hat{\mathbf{d}}\hat{\mathbf{E}}'
(\mathbf{r}_\mathrm{A})\bigr|^2}
{\hbar(\omega_{kl}+\omega)}
\sim (e a_0 E)^2 \,\frac{E_0}{1+\hbar\omega/E_0}
}
\nonumber\\&&
= g^2 \left(\frac{\hbar\omega}{E_0}\right)^2
\frac{E_0}{1+\hbar\omega/E_0} = O\!\left(g^2\right)\!,
\end{eqnarray}
\begin{eqnarray}
\label{B13}
\lefteqn{
\frac{
\biggl|\displaystyle
\sum_\alpha \frac{q_\alpha}{2m_\alpha}\,
\hat{\mathbf{p}}'_\alpha\!\times\!\hat{\bar{\mathbf{r}}}_\alpha
\hat{\mathbf{B}}'(\mathbf{r}_\mathrm{A})\biggr|^2}
{\hbar(\omega_{kl}+\omega)}
}
\nonumber\\&&
\sim \left(\frac{e a_0 p B}{2 m_e}\right)^2\frac{E_0}{1+\hbar\omega/E_0}
= (Z_\mathrm{eff}\alpha_0g)^2 {\textstyle\frac{1}{4}}
\!\left(\frac{\hbar\omega}{E_0}\right)^2
\nonumber\\&&
\quad\times\frac{E_0}{1\!+\!\hbar\omega/E_0} = O\!\left[
(Z_\mathrm{eff}\alpha_0g)^2\right]\!,
\quad\qquad
\end{eqnarray}
\begin{eqnarray}
\label{B14}
\lefteqn{
\sum_\alpha \frac{q_\alpha^2}{8m_\alpha}
\left|\hat{\bar{\mathbf{r}}}_\alpha
\times \hat{\mathbf{B}}'(\mathbf{r}_\mathrm{A})\right|^2
\sim \frac{(e a_0 B)^2}{8m_e}}
\nonumber\\&&
= (Z_\mathrm{eff}\alpha_0g)^2 {\textstyle\frac{1}{8}}
\!\left(\frac{\hbar\omega}{E_0}\right)^2\!\! E_0
= O\!\left[(Z_\mathrm{eff}\alpha_0g)^2\right]\!.
\quad
\end{eqnarray}

Next, let us estimate the orders of magnitude of the various
contributions to the Lorentz force given in Sec.~\ref{sec4.1}. The
magnitudes of the first and third terms in curly brackets in
Eq.~(\ref{eq96}) can be approximated according to
\begin{equation}
\label{B15}
\bigl|\hat{\mathbf{d}} \hat{\mathbf{E}}
(\hat{\mathbf{r}}_\mathrm{A})\bigr|
\sim ea_0E = g \,(\hbar\omega) = O(g),
\end{equation}
\begin{eqnarray}
\label{B16}
\lefteqn{
{\textstyle \frac{1}{2}}
\Bigl|\dot{\hat{\mathbf{r}}}_\mathrm{A}
\hat{\mathbf{B}}^{(\prime)}(\hat{\mathbf{r}}_\mathrm{A})
\times\hat{\mathbf{d}}\Bigr|
}
\nonumber\\&&
\sim {\textstyle \frac{1}{2}}ea_0Bv
={\textstyle \frac{1}{2}}\left(\frac{v}{c}g\right)
(\hbar\omega) = O(gv/c).
\quad
\end{eqnarray}
In order to estimate the magnitude of the second term, we make use of
the relation
\begin{equation}
\label{newt1}
      m_\alpha\dot{\hat{\mathbf{r}}}_\alpha =
      \hat{\mathbf{p}}_{\alpha}
      -q_{\alpha}\hat{\mathbf{A}}(\hat{\mathbf{r}}_{\alpha})
\end{equation}
in order to introduce relative momenta [recall Eq.~(\ref{eq48})],
leading to
\begin{alignat}{1}
\label{B18}
&\sum_\alpha\frac{q_\alpha}{2}\,
\dot{\hat{\mathbf{r}}}_\alpha\hat{\mathbf{B}}
(\mathbf{r}_\mathrm{A})
\times\hat{\bar{\mathbf{r}}}_\alpha=
\sum_\alpha\frac{q_\alpha}{2m_\alpha}\,
\hat{\bar{\mathbf{p}}}_\alpha
      \hat{\mathbf{B}}(\hat{\mathbf{r}}_\mathrm{A})
      \times\hat{\bar{\mathbf{r}}}_\alpha
      \nonumber\\
      &\quad+\sum_\alpha
      \frac{q_\alpha^2}{2m_\alpha}\hat{\bar{\mathbf{r}}}_\alpha
      \hat{\mathbf{B}}(\hat{\mathbf{r}}_\mathrm{A})
      \times\hat{\mathbf{A}}(\hat{\mathbf{r}}_\mathrm{A})
      +{\textstyle \frac{1}{2}}
      \dot{\hat{\mathbf{r}}}_\mathrm{A}
      \hat{\mathbf{B}}(\hat{\mathbf{r}}_\mathrm{A})
      \times\hat{\mathbf{d}}.
\end{alignat}
Combining this with
\begin{eqnarray}
\label{B19}
\lefteqn{
\biggl|\sum_\alpha\frac{q_\alpha}{2m_\alpha}\,
\hat{\bar{\mathbf{p}}}_\alpha
\hat{\mathbf{B}}(\hat{\mathbf{r}}_\mathrm{A})
\times\hat{\bar{\mathbf{r}}}_\alpha
\biggr|
}
\nonumber\\&&\hspace{2ex}
\sim \frac{eBa_0p}{2m_e}
= (Z_\mathrm{eff}\alpha_0 g) \,{\textstyle\frac{1}{4}}
(\hbar\omega) = O(Z_\mathrm{eff}\alpha_0 g),\quad
\quad
\end{eqnarray}
\begin{eqnarray}
\label{B20}
\lefteqn{
\biggl|\sum_\alpha
\frac{q_\alpha^2}{2m_\alpha}\hat{\bar{\mathbf{r}}}_\alpha
\hat{\mathbf{B}}(\hat{\mathbf{r}}_\mathrm{A})
\times\hat{\mathbf{A}}(\hat{\mathbf{r}}_\mathrm{A})\biggr|
}
\nonumber\\&&
\sim \frac{e^2a_0AB}{2m_e}
= (Z_\mathrm{eff}\alpha_0 g^2)\,{\textstyle\frac{1}{2}}(\hbar\omega) 
= O(Z_\mathrm{eff}\alpha_0 g^2),\quad
\quad
\end{eqnarray}
and Eq.~(\ref{B16}), we see that the magnitude of the second term in
curly brackets in Eq.~(\ref{eq96}) is 
$O(Z_\mathrm{eff}\alpha_0 g+Z_\mathrm{eff}\alpha_0 g^2+gv/c)
=O[(Z_\mathrm{eff}\alpha_0+v/c)g]$. The magnitudes of the different
contributions to Eq.~(\ref{eq99}) are
\begin{eqnarray}
\label{B21}
\lefteqn{
\Bigl| \dot{\hat{\mathbf{d}}}\times \hat{\mathbf{B}}
(\hat{\mathbf{r}}_\mathrm{A})\Bigr|
=\biggl| \sum_\alpha\frac{q_\alpha}{m_\alpha}
\big[\hat{\bar{\mathbf{p}}}_\alpha-q_\alpha
\hat{\mathbf{A}}(\hat{\mathbf{r}}_\mathrm{A})\big]
\times \hat{\mathbf{B}}(\hat{\mathbf{r}}_\mathrm{A})\biggr|
}
\nonumber\\&&
\sim\bigg(\frac{epB}{m_e}+\frac{e^2AB}{m_e}\bigg)
=g(1+2g)\left(\frac{\omega E_0}{c}\right)= O(g),
\nonumber\\
\end{eqnarray}
\begin{eqnarray}
\label{B22}
\lefteqn{
\Bigl|\hat{\mathbf{d}}\times 
\dot{\hat{\mathbf{B}}}(\hat{\mathbf{r}}_\mathrm{A})\Bigr|
}
\nonumber\\&& 
\sim ea_0\omega B
=g\left(\frac{\hbar\omega}{E_0}\right)\left(\frac{\omega
E_0}{c}\right)
 = O(g),
\quad
\end{eqnarray}
\begin{eqnarray}
\label{B23}
\lefteqn{
{\textstyle\frac{1}{2}}\biggl|\hat{\mathbf{d}}\times \Big[
        \dot{\hat{\mathbf{r}}}_\mathrm{A}
        \bm{\nabla}_{\!\!\mathrm{A}}
        \otimes\hat{\mathbf{B}}(\hat{\mathbf{r}}_\mathrm{A})
       +\hat{\mathbf{B}}(\hat{\mathbf{r}}_\mathrm{A})
       \otimes\overleftarrow{\bm{\nabla}}_{\!\!\mathrm{A}}
       \dot{\hat{\mathbf{r}}}_\mathrm{A}\Big]
\biggr|
}
\nonumber\\&&
\sim\frac{ea_0v\omega B}{c}
=\left(\frac{v}{c}g\right)
\left(\frac{\hbar\omega}{E_0}\right)\left(\frac{\omega E_0}{c}\right)
= O(gv/c).\nonumber\\
\end{eqnarray}

Finally, let us compare the contributions of the R\"{o}nt\-gen
interaction to the temporal evolution 
$\hat{\mathbf{f}}_\lambda(\mathbf{r},\omega,t)$ with that from the
electric dipole interaction,
\begin{eqnarray}
\label{B24}
 &&\frac{\biggl|\frac{1}{2\hbar\omega}\,
       \dot{\hat{\mathbf{r}}}_\mathrm{A}(t)\hat{\mathbf{d}}(t)
       \times\big(\bm{\nabla}_{\!\!\mathrm{A}}
       \times\bm{G}_\lambda^\ast
       [\mathbf{r}_\mathrm{A}(t),\mathbf{r},\omega]\big)
       \biggr|}
       {\Bigl|\frac{i}{\hbar}
       \hat{\mathbf{d}}(t)
       \bm{G}_\lambda^\ast[\hat{\mathbf{r}}_\mathrm{A}(t),
       \mathbf{r},\omega]\Bigr|}
\nonumber\\&&\quad
\sim\left(\frac{vea_0}{\hbar c}\right){\Big /}
\left(\frac{ea_0}{\hbar}\right)=O(v/c),
\end{eqnarray}
\begin{alignat}{1}
\label{B25}
 &\frac{\Bigl|\frac{1}{\hbar\omega m_\mathrm{A}}
 \hat{\mathbf{d}}(t)\times\hat{\mathbf{B}}[\mathbf{r}_\mathrm{A}(t),t]
       \hat{\mathbf{d}}(t)\times
       \big(\bm{\nabla}_{\!\!\mathrm{A}}\times
       \bm{G}_\lambda^\ast[\mathbf{r}_\mathrm{A}(t),
       \mathbf{r},\omega]\big)\Bigr|}
       {\Bigl|\frac{i}{\hbar}
       \hat{\mathbf{d}}(t)
       \bm{G}_\lambda^\ast[\hat{\mathbf{r}}_\mathrm{A}(t),
       \mathbf{r},\omega]\Bigr|}
\nonumber\\&\quad
\sim\left(\frac{e^2a_0^2B}{\hbar m_\mathrm{A}c}\right)
{\Big /}\left(\frac{ea_0}{\hbar}\right)
=(Z_\mathrm{eff}\alpha_0)^2g{\textstyle\frac{1}{2}}
\left(\frac{\hbar\omega}{E_0}\right)
\left(\frac{m_e}{m_\mathrm{A}}\right)\nonumber\\&\quad
=O\big[(Z_\mathrm{eff}\alpha_0)^2g\big].
\end{alignat}

%%%%%%%%%%%%%%%%%%%%%%%%%%%%%%%%%%%%%%%%%%%%%%%%%%%%%%%%%%%%%%%%%%%%%%

\section{Calculation of the perturbative corrections 
(\ref{eq57}) and (\ref{eq76})}
\label{AppC}

Recalling Eq.~(\ref{eq6}) together with Eqs.~(\ref{eq8})--(\ref{eq10}),
making use of the commutation relations (\ref{eq4}) and (\ref{eq5}),
and applying Eq.~(\ref{eq18}), Eq.~(\ref{eq54}) leads to
\begin{eqnarray}
\label{C1}
\lefteqn{
\Delta_1E_l = \sum_\alpha\frac{q_\alpha^2}{2m_\alpha}
\sum_{\lambda=e,m}\int_0^\infty\mathrm{d}\omega
}
\nonumber\\&&\quad\times\,
\int\mathrm{d}^3r\frac{1}{\omega^2}
\big({}^\perp G_{\lambda}\big)_{ij}(\mathbf{r}_\mathrm{A},
\mathbf{r},\omega)
\big({}^\perp G_{\lambda}^\ast\big)_{ij}
(\mathbf{r}_\mathrm{A},\mathbf{r},\omega)
\nonumber\\&& 
= \frac{\hbar\mu_0}{\pi}\sum_\alpha
\frac{q_\alpha^2}{2m_\alpha}\int_0^\infty\mathrm{d}\omega
\mathrm{Im}\big({}^\perp G^\perp\big)_{ii}
(\mathbf{r}_\mathrm{A},\mathbf{r}_\mathrm{A},\omega),\quad
\end{eqnarray}
where we have introduced the notation
\begin{alignat}{1}
\label{C2}
     &^{\perp(\parallel)} 
     \bm{G}^{\perp(\parallel)}(\mathbf{r},\mathbf{r}',\omega)
     \nonumber\\
     &\quad\equiv \int\mathrm{d}^3 s\int\mathrm{d}^3s'
     \bm{\delta}^{\perp(\parallel)}(\mathbf{r}-\mathbf{s})
     \bm{G}(\mathbf{s},\mathbf{s}',\omega)
     \bm{\delta}^{\perp(\parallel)}(\mathbf{s}'-\mathbf{r}').
\end{alignat}
Applying the sum rule
\begin{equation}
\label{C3}
       \sum_{\alpha}\frac{q_{\alpha}^2}{2m_{\alpha}}\bm{I}
       = \frac{1}{2\hbar}\sum_k\omega_{kl}
       (\mathbf{d}_{lk}\otimes \mathbf{d}_{kl} 
       + \mathbf{d}_{kl}\otimes \mathbf{d}_{lk}),
\end{equation}
we can rewrite Eq.~(\ref{C1}) as
\begin{equation}
\label{C4}
\Delta_1E_l
= \frac{\mu_0}{\pi}\sum_k\int_0^\infty\mathrm{d}\omega\omega_{kl}
\mathbf{d}_{lk}\mathrm{Im}{}^\perp
\bm{G}^\perp(\mathbf{r}_\mathrm{A},
\mathbf{r}_\mathrm{A},\omega)\mathbf{d}_{kl}.
\end{equation}
To calculate $\Delta_2 E$, as given by Eq.~(\ref{eq53}), we first
calculate the matrix elements therein. Recalling
Eqs.~(\ref{eq4})--(\ref{eq10}), we obtain
\begin{eqnarray}
\label{C5}
&&\langle l|\langle\{0\}|
\hat{\mathbf{d}}\bm{\nabla}\hat{\varphi}(\mathbf{r})
_{\mathbf{r}=\mathbf{r}_\mathrm{A}}
     |\{\mathbf{1}_\lambda(\mathbf{r},\omega)\}\rangle|k\rangle
\nonumber\\
&&\qquad=-\mathbf{d}_{lk}{}^\parallel\bm{G}_\lambda
(\mathbf{r}_\mathrm{A},\mathbf{r},\omega),
\\
\label{C6}
&&-\langle l|\langle\{0\}|
      \sum_{\alpha}\frac{q_{\alpha}}{m_{\alpha}}
      \,\hat{\mathbf{p}}_{\alpha}\hat{\mathbf{A}}
      (\mathbf{r}_\mathrm{A})
      |\{\mathbf{1}_\lambda(\mathbf{r},\omega)\}\rangle|k\rangle
\nonumber\\
&&\qquad=\frac{\omega_{kl}}{\omega}\,\mathbf{d}_{lk}
{}^\perp\bm{G}_\lambda(\mathbf{r}_\mathrm{A},\mathbf{r},\omega),
\end{eqnarray}
where the second matrix element has been obtained by means of the
identity
\begin{equation}
\label{C7}
       \sum_{\alpha}\frac{q_{\alpha}}{m_{\alpha}}
       \langle l|\hat{\mathbf{p}}_{\alpha}|k\rangle
       =-i\omega_{kl}\mathbf{d}_{lk}.
\end{equation}
Substituting Eqs.~(\ref{C5}) and (\ref{C6}) into Eq.~(\ref{eq53}), we
then derive
\begin{eqnarray}
\label{C8}
      &&\Delta_2 E_l
      =
      -\frac{1}{\hbar}\sum_k\sum_{\lambda=e,m}\mathcal{P}\int_0^{\infty}\!\!
      \frac{\mathrm{d}\omega}{\omega_{kl}+\omega}
      \int\!\mathrm{d}^3{r}(d_{lk})_i
      \nonumber\\
      &&\quad\times\,(d_{kl})_j\Big[
      \big({}^\parallel G_\lambda\big)_{in}
      (\mathbf{r}_\mathrm{A},\mathbf{r},\omega)
      \big({}^\parallel G_\lambda^\ast\big)_{jn}
      (\mathbf{r}_\mathrm{A},\mathbf{r},\omega)
\nonumber\\
      &&\quad-\,\frac{\omega_{kl}}{\omega}
      \big({}^\parallel G_\lambda\big)_{in}(\mathbf{r}_\mathrm{A},
      \mathbf{r},\omega)
      \big({}^\perp G_\lambda^\ast\big)_{jn}
      (\mathbf{r}_\mathrm{A},\mathbf{r},\omega)
\nonumber\\
      &&\quad-\,\frac{\omega_{kl}}{\omega}
      \big({}^\perp G_\lambda\big)_{in}
      (\mathbf{r}_\mathrm{A},\mathbf{r},\omega)
      \big({}^\parallel G_\lambda^\ast\big)_{jn}
      (\mathbf{r}_\mathrm{A},\mathbf{r},\omega)
\nonumber\\
      &&\quad+\,\frac{\omega_{kl}^2}{\omega^2}
      \big({}^\perp G_\lambda\big)_{in}
      (\mathbf{r}_\mathrm{A},\mathbf{r},\omega)
      \big({}^\perp G_\lambda^\ast\big)_{jn}
      (\mathbf{r}_\mathrm{A},\mathbf{r},\omega)\Big]
\nonumber\\
      &&=\frac{\mu_0}{\pi}\sum_k\mathcal{P}\int_0^{\infty}
      \!\!\frac{\mathrm{d}\omega}{\omega_{kl}\!+\!\omega}
      \,\mathbf{d}_{lk}\bigg\{
      \!-\omega^2\mathrm{Im}\,{}^\parallel
      \bm{G}^\parallel
      (\mathbf{r}_\mathrm{A},\mathbf{r}_\mathrm{A},\omega)
\nonumber\\
      &&\quad+\,\omega_{kl}\omega\Big[\mathrm{Im}\,
      {}^\parallel\bm{G}^\perp
      (\mathbf{r}_\mathrm{A},\mathbf{r}_\mathrm{A},\omega)
      +\mathrm{Im}\,
      {}^\perp\bm{G}^\parallel
      (\mathbf{r}_\mathrm{A},\mathbf{r}_\mathrm{A},\omega)\Big]
\nonumber\\
      &&\quad-\,\omega_{kl}^2\mathrm{Im}\,
      {}^\perp\bm{G}^\perp
      (\mathbf{r}_\mathrm{A},\mathbf{r}_\mathrm{A},\omega)
      \bigg\}\mathbf{d}_{kl},
\end{eqnarray}
where we have again made use of the identity (\ref{eq18}). Adding
Eqs.~(\ref{C4}) and (\ref{C8}) according to Eq.~(\ref{eq56}), on using
the identity ${\bm G}$ $\!=$ $\!{}^{\perp}{\bm G}^{\perp}$
$\!+$ $\!{}^{\perp}{\bm G}^{\parallel}$ $\!+$
$\!{}^{\parallel}{\bm G}^{\perp}$ 
$\!+$ $\!{}^{\parallel}{\bm G}^{\parallel}$
[which directly follows from the definition (\ref{C2}) together with
$\bm{\delta}(\mathbf{r})$ $\!=$ $\!\bm{\delta}^\parallel(\mathbf{r})$
$\!+$ $\!\bm{\delta}^\perp(\mathbf{r})$], we eventually arrive at
Eq.~(\ref{eq57}).

The derivation of Eq.~(\ref{eq76}) is completely analogous. The
relevant matrix elements can be calculated with the aid of
Eq.~(\ref{eq20}) together with Eqs.~(\ref{eq8})--(\ref{eq10})
and the commutation relations (\ref{eq4}) and (\ref{eq5}), cf. the
remarks below Eq.~(\ref{eq39}). The result is
\begin{alignat}{1}
\label{C9}
-\langle l|\langle\{0'\}|
\hat{\mathbf{d}}\hat{\mathbf{E}}'(\mathbf{r}_\mathrm{A})
 |\{\mathbf{1}'_\lambda(\mathbf{r},\omega)\}\rangle|k\rangle
=-\mathbf{d}_{lk}\bm{G}_\lambda(\mathbf{r}_\mathrm{A},
\mathbf{r},\omega).
\nonumber\\
\end{alignat}
Substituting Eq.~(\ref{C9}) into Eq.~(\ref{eq75}) yields
\begin{eqnarray}
\label{C10}
\lefteqn{
      \Delta_2 E_l =
      -\frac{1}{\hbar}\sum_k\sum_{\lambda=e,m}\mathcal{P}\int_0^{\infty}\!\!
      \frac{\mathrm{d}\omega}{\omega_{kl}+\omega}
      \int \mathrm{d}^3{r}\, }
\nonumber\\&&\times\,
      (d_{lk})_i(d_{kl})_j
      \big( G_\lambda\big)_{in}
      (\mathbf{r}_\mathrm{A},\mathbf{r},\omega)
      \big( G_\lambda^\ast\big)_{jn}
      (\mathbf{r}_\mathrm{A},\mathbf{r},\omega),
\nonumber\\
\end{eqnarray}
from which Eq.~(\ref{eq76}) follows by means of Eq.~(\ref{eq18}).

%%%%%%%%%%%%%%%%%%%%%%%%%%%%%%%%%%%%%%%%%%%%%%%%%%%%%%%%%%%%%%%%%%%%%%

\section{Equivalence of Lorentz forces (\ref{eq93}) and (\ref{eq107})}
\label{AppD}

To transform the first term in Eq.~(\ref{eq107}), we apply the the
rule (\ref{A4}), recall that integrals over mixed products of
transverse and longitudinal vector fields vanish, and use the identity
for the first term in Eq.~(\ref{A14}) as well as Eqs.~(\ref{eq79}) and
(\ref{eq104}). We thus derive 
\begin{eqnarray}
\label{D1}
\lefteqn{
 \bm{\nabla}_{\!\!\mathrm{A}}
 \!\!\int\mathrm{d}^3r\, 
 \bigl[\hat{\mathbf{P}}_\mathrm{A} (\mathbf{r})
                       \hat{\mathbf{E}}' (\mathbf{r})\bigr]
}
\nonumber\\&&
 = \bm{\nabla}_{\!\!\mathrm{A}}
 \!\!\int\!\!\mathrm{d}^3r\, 
 \bigl[\hat{\mathbf{P}}_\mathrm{A}(\mathbf{r})
                       \hat{\mathbf{E}}(\mathbf{r})\bigr]
 +\frac{1}{\varepsilon_0}\bm{\nabla}_{\!\!\mathrm{A}}
 \!\!\int\!\!\mathrm{d}^3r\, 
 \bigl[\hat{\mathbf{P}}_\mathrm{A}(\mathbf{r})
 \hat{\mathbf{P}}_\mathrm{A}^\perp(\mathbf{r})\bigr]
\nonumber\\&&
 = \bm{\nabla}_{\!\!\mathrm{A}}
 \int\mathrm{d}^3r\,\bigl[ \hat{\mathbf{P}}_\mathrm{A}(\mathbf{r})
                       \hat{\mathbf{E}}(\mathbf{r})\bigr]
 +\frac{1}{\varepsilon_0}\bm{\nabla}_{\!\!\mathrm{A}}
 \int\mathrm{d}^3r\, 
 \hat{\mathbf{P}}^2_\mathrm{A}(\mathbf{r})
\nonumber\\&&\hspace{6ex}
 -\, \bm{\nabla}_{\!\!\mathrm{A}}
 \int\mathrm{d}^3r\,
  \hat{\rho}_\mathrm{A}(\mathbf{r})
  \hat{\varphi}_\mathrm{A}(\mathbf{r})
\nonumber\\&&
 = \bm{\nabla}_{\!\!\mathrm{A}}
 \int\mathrm{d}^3r \,\hat{\mathbf{P}}_\mathrm{A}(\mathbf{r})
                       \hat{\mathbf{E}}(\mathbf{r}).
\end{eqnarray}
In order to simplify the second term in Eq.~(\ref{eq107}), we use the
definitions (\ref{eq28}), (\ref{eq38}), (\ref{eq85}), and (\ref{eq86})
to calculate
\begin{eqnarray}
\lefteqn{
\label{D2}
 {\textstyle \frac{1}{2}}\sum_\alpha
  \Big[\hat{\bm{\Xi}}_\alpha (\mathbf{r})
  \times\dot{\hat{\mathbf{r}}}_\alpha
 -\dot{\hat{\mathbf{r}}}_\alpha\times\hat{\bm{\Xi}}_\alpha(\mathbf{r})
 \Big]
}
\nonumber\\&&
={\textstyle \frac{1}{2}}\sum_\alpha q_\alpha\Big[
 \hat{\bm{\Theta}}_\alpha (\mathbf{r})
 \times\dot{\hat{\mathbf{r}}}_\alpha
 -\dot{\hat{\mathbf{r}}}_\alpha
 \times\hat{\bm{\Theta}}_\alpha (\mathbf{r})\Big]
\nonumber\\&&\hspace{4ex}
-\,{\textstyle \frac{1}{2}}\sum_\beta q_\alpha\Big[
 \hat{\bm{\Theta}}_\alpha (\mathbf{r})
 \times\dot{\hat{\mathbf{r}}}_\mathrm{A}
 -\dot{\hat{\mathbf{r}}}_\mathrm{A}
 \times\hat{\bm{\Theta}}_\alpha (\mathbf{r})
 \Big]
\nonumber\\&&\hspace{4ex}
+\,{\textstyle \frac{1}{2}}\Big[
 \hat{\mathbf{P}}_\mathrm{A} (\mathbf{r})
 \times\dot{\hat{\mathbf{r}}}_\mathrm{A}
 -\dot{\hat{\mathbf{r}}}_\mathrm{A}
 \times\hat{\mathbf{P}}_\mathrm{A}(\mathbf{r})\Big]
\nonumber\\&&
=\hat{\mathbf{M}}_\mathrm{A} (\mathbf{r})
+\hat{\mathbf{M}}_\mathrm{R} (\mathbf{r}).
\end{eqnarray}
Consequently, recalling that \mbox{$\hat{\mathbf{B}}'(\mathbf{r})$ $\!=$
$\!\hat{\mathbf{B}}(\mathbf{r})$}, we may write
\begin{eqnarray}
\label{D3}
\lefteqn{
 \bm{\nabla}_{\!\!\mathrm{A}}\int\mathrm{d}^3r\,
 {\textstyle \frac{1}{2}}\sum_\alpha\Bigl[
 \hat{\bm{\Xi}}_\alpha (\mathbf{r})\times\dot{\hat{\mathbf{r}}}_\alpha
 -\dot{\hat{\mathbf{r}}}_\alpha
 \times\hat{\bm{\Xi}}_\alpha (\mathbf{r})\Bigr]
 \hat{\mathbf{B}}'(\mathbf{r})
}
\nonumber\\&&
 = \bm{\nabla}_{\!\!\mathrm{A}}
\int\mathrm{d}^3r\, \bigl[
 \hat{\mathbf{M}}_\mathrm{A} (\mathbf{r})
 +\hat{\mathbf{M}}_\mathrm{R} (\mathbf{r})\bigr]
 \hat{\mathbf{B}}(\mathbf{r})
\qquad\qquad
\end{eqnarray}
as well as
\begin{equation}
\label{D4}
\frac{\mathrm{d}}{\mathrm{d}t} \biggl[ \int\! \mathrm{d}^3 r\,
       \hat{\mathbf{P}}_\mathrm{A} (\mathbf{r}) \times
       \hat{\mathbf{B}}'(\mathbf{r})\biggr]
=\frac{\mathrm{d}}{\mathrm{d}t} \biggl[ \int\! \mathrm{d}^3 r\,
       \hat{\mathbf{P}}_\mathrm{A} (\mathbf{r}) \times
       \hat{\mathbf{B}}(\mathbf{r})\biggr].
\end{equation}
Substituting Eqs.~(\ref{D1}), (\ref{D3}), and (\ref{D4}) into
Eq.~(\ref{eq107}), we see that Eq.~(\ref{eq107}) is equivalent to
Eq.~(\ref{eq93}).

%%%%%%%%%%%%%%%%%%%%%%%%%%%%%%%%%%%%%%%%%%%%%%%%%%%%%%%%%%%%%%%%%%%%%%

\section{Equations of motion for 
$\hat{\mathbf{f}}'_\lambda(\mathbf{r},\omega,t)$}
\label{AppE}

In electric dipole approximation, the temporal evolution of the basic
fields $\hat{\mathbf{f}}'_\lambda(\mathbf{r},\omega,t)$ is governed by
the Hamiltonian given in Eq.~(\ref{eq43}) together with
Eqs.~(\ref{eq44}), (\ref{eq45}), and (\ref{eq49}). Using
Eqs.~(\ref{eq8}) and (\ref{eq20})--(\ref{eq22}) (with the unprimed
fields being replaced with the primed ones) and applying the
commutation relations (\ref{eq4}) and (\ref{eq5}), we obtain
\begin{eqnarray}
\label{E1}
\lefteqn{
       \dot{\hat{\mathbf{f}}}'_\lambda(\mathbf{r},\omega,t)
       =\frac{i}{\hbar}\bigl[\hat{H},
       \hat{\mathbf{f}}'_\lambda(\mathbf{r},\omega,t)\bigr]
}
\nonumber\\&&
       =-i\omega \hat{\mathbf{f}}'_\lambda(\mathbf{r},\omega,t)
       + \frac{i}{\hbar}\,\hat{\mathbf{d}}(t)
       \bm{G}_\lambda^\ast[\hat{\mathbf{r}}_\mathrm{A}(t),
       \mathbf{r},\omega]
\nonumber\\&&\quad
       -\,\frac{1}{2\hbar\omega}\,
       \dot{\hat{\mathbf{r}}}_\mathrm{A}(t)\hat{\mathbf{d}}(t)
       \times\big(\bm{\nabla}_{\!\!\mathrm{A}}\times
       \bm{G}_\lambda^\ast[\hat{\mathbf{r}}_\mathrm{A}(t),
       \mathbf{r},\omega]\big)
\nonumber\\&&\quad
       -\,\frac{1}{\hbar\omega m_\mathrm{A}}
       \Bigl\{
       \hat{\mathbf{d}}(t)\times
       \hat{\mathbf{B}}'[\hat{\mathbf{r}}_\mathrm{A}(t),t]
       \hat{\mathbf{d}}(t)
\nonumber\\&&\hspace{15ex}\times
       \bigl(\bm{\nabla}_{\!\!\mathrm{A}}\times
       \bm{G}_\lambda^\ast[\hat{\mathbf{r}}_\mathrm{A}(t),
       \mathbf{r},\omega]\bigr)
       \Bigr\}.
\end{eqnarray}
The third and fourth terms in Eq.~(\ref{E1}), which are due to the
R\"{o}ntgen interaction, are smaller than the second one by factors of
$v/c$ and $g(Z_\mathrm{eff}\alpha_0)^2$, respectively
[Eqs.~(\ref{B24}) and (\ref{B25}) in Appendix~\ref{AppB}], so
according to the nonrelativistic approximation, Eq.~(\ref{E1})
reduces to
\begin{equation}
\label{E2}
       \dot{\hat{\mathbf{f}}}'_\lambda(\mathbf{r},\omega,t)
       =-i\omega \hat{\mathbf{f}}'_\lambda(\mathbf{r},\omega,t)
       + \frac{i}{\hbar}\,\hat{\mathbf{d}}(t)
       \bm{G}_\lambda^\ast[\hat{\mathbf{r}}_\mathrm{A}(t),
       \mathbf{r},\omega],
\end{equation}
which can be integrated to yield 
\mbox{[$\hat{\mathbf{f}}'_\lambda(\mathbf{r},\omega,0)$ $\!\equiv$
$\!\hat{\mathbf{f}}'_\lambda(\mathbf{r},\omega)$]}
\begin{equation}
\label{E3}
       \hat{\mathbf{f}}'_\lambda(\mathbf{r},\omega,t)
       = \hat{\mathbf{f}}'_{\lambda\,\mathrm{free}}
       (\mathbf{r},\omega,t)
        +\hat{\mathbf{f}}'_{\lambda\,\mathrm{source}}
       (\mathbf{r},\omega,t),
\end{equation}
where
\begin{equation}
\label{E4}
 \hat{\mathbf{f}}'_{\lambda\,\mathrm{free}}(\mathbf{r},\omega,t)
 =e^{-i\omega t}\hat{\mathbf{f}}'_\lambda(\mathbf{r},\omega)
\end{equation}
and
\begin{equation}
\label{E5}
    \hat{\mathbf{f}}'_{\lambda\,\mathrm{source}}(\mathbf{r},\omega,t)
    =\frac{i}{\hbar}
        \int_0^t\! \mathrm{d} t'\,e^{-i\omega(t-t')}
        \hat{\mathbf{d}}(t')\,\bm{G}_\lambda^\ast
        [\hat{\mathbf{r}}_\mathrm{A}(t'),\mathbf{r},\omega].
\end{equation}
Substituting Eqs.~(\ref{E3})--(\ref{E5}) into Eq.~(\ref{eq8})
[$\hat{\underline{\mathbf{E}}}(\mathbf{r},\omega,t)$ $\!\mapsto$
$\!\hat{\underline{\mathbf{E}}}{'}(\mathbf{r},\omega,t)$]
and using the identity (\ref{eq18}), we arrive at
Eqs.~(\ref{eq111})--(\ref{eq113}).

%%%%%%%%%%%%%%%%%%%%%%%%%%%%%%%%%%%%%%%%%%%%%%%%%%%%%%%%%%%%%%%%%%%%%%

\section{Intra-atomic equations of motion}
\label{AppF}

An estimation similar to that given for the fields
$\hat{\mathbf{f}}'(\mathbf{r},\omega,t)$ shows that in the
nonrelativistic limit the second term in the interaction Hamiltonian
in electric dipole approximation (\ref{eq49}) can be disregarded in
the calculation of the temporal evolution of the intra-atomic
operators $\hat{A}_{mn}(t)$. By representing the (unperturbed)
intra-atomic Hamiltonian in the form of Eq.~(\ref{eq52}), recalling
Eqs.~(\ref{eq8}) and (\ref{eq123}), and applying standard
commutation relations, it is not difficult to prove that
the $\hat{A}_{mn}(t)$ obey the equations of motion
\begin{eqnarray}
\label{F1}
\lefteqn{
       \dot{\hat{A}}_{mn}
       = \frac{i}{\hbar}\big[\hat{H},\hat{A}_{mn}\big]
       =i\omega_{mn} \hat{A}_{mn}
}
\nonumber\\&&\hspace{-2.5ex}
       +\,\frac{i}{\hbar} \!\sum_k
       \left[
       \big(\mathbf{d}_{nk}\hat{A}_{mk}\!-
       \!\mathbf{d}_{km}\hat{A}_{kn}\big)
       \int_0^\infty\!\!  \mathrm{d}\omega\,
       \underline{\hat{\mathbf{E}}}{}'
       (\hat{\mathbf{r}}_\mathrm{A},\omega)
       \right.
\nonumber\\&&\hspace{1ex}
       \left.
       +\!\int_0^\infty\!\! \mathrm{d}\omega\,
       \underline{\hat{\mathbf{E}}}{}^{\prime\dagger}
       (\hat{\mathbf{r}}_\mathrm{A},\omega)
        \big(\mathbf{d}_{nk}\hat{A}_{mk}
       \!-\!\mathbf{d}_{km}\hat{A}_{kn}\big)
       \right].
\quad
\end{eqnarray}
We now substitute the source-quantity representation for
$\underline{\hat{\mathbf{E}}}{}'(\hat{\mathbf{r}}_\mathrm{A},\omega)
=\underline{\hat{\mathbf{E}}}{}'[\hat{\mathbf{r}}_\mathrm{A}(t),
\omega,t]$ (and its Hermitian conjugate) according to
Eqs.~(\ref{eq111})--(\ref{eq113}) into Eq.~(\ref{F1}). Carrying out
the time integral in the source-field part in Eq.~(\ref{F1}) in the
Markov approximation, we may set, on regarding
$\hat{\mathbf{r}}_\mathrm{A}=\hat{\mathbf{r}}_\mathrm{A}(t)$ as
being slowly varying,
\begin{equation}
\label{F2}
\int_0^\infty \!\! \mathrm{d}\omega\,
\underline{\hat{\mathbf{E}}}{}'_\mathrm{source}
(\hat{\mathbf{r}}_\mathrm{A},\omega)
= \sum_{m,n} \mathbf{g}_{mn}(\hat{\mathbf{r}}_\mathrm{A})
 \hat{A}_{mn},
\end{equation}
where
\begin{eqnarray}
\label{F3}
       \mathbf{g}_{mn}(\hat{\mathbf{r}}_\mathrm{A})
       &=&\frac{i\mu_0}{\pi}\int_0^\infty \!\!\mathrm{d}\omega\,
        \omega^2
        \mathrm{Im}\, \bm{G}(\hat{\mathbf{r}}_\mathrm{A},
                 \hat{\mathbf{r}}_\mathrm{A},\omega)\mathbf{d}_{mn}
\nonumber\\&&\hspace{6ex}\times\,
       \zeta[\tilde{\omega}_{nm}
       (\hat{\mathbf{r}}_\mathrm{A})\!-\!\omega]
\qquad
\end{eqnarray}
[$\zeta(x)$ $\!=$ $\!\pi\delta(x)$ $\!+$ $\!i{\cal P}/x$], with
$\tilde{\omega}_{nm}(\hat{\mathbf{r}}_\mathrm{A})$ being the shif\-ted
transition frequencies. Substituting Eq.~(\ref{F2}) into
Eq.~(\ref{F1}), we obtain
\begin{eqnarray}
\label{F3-1}
       \dot{\hat{A}}_{mn}
       &=&\Bigl\{i\omega_{mn}
        + \frac{i}{\hbar} \sum_k \big[\mathbf{d}_{nk}\mathbf{g}_{kn}
        - \mathbf{d}_{km}\mathbf{g}^\ast_{km}\big] \Bigr\}\hat{A}_{mn}
\nonumber\\&&
       +\hat{B}_{mn}+\hat{F}_{mn},
\end{eqnarray}
with
\begin{eqnarray}
\label{F3-2}
\hat{B}_{mn}&=&
          \frac{i}{\hbar} \sum_{k,l\neq n} \mathbf{d}_{nk} 
          \mathbf{g}_{kl}\hat{A}_{ml}
           - \frac{i}{\hbar} \sum_{k,l} \mathbf{d}_{km} 
          \mathbf{g}_{nl}\hat{A}_{kl}
\nonumber\\&&
         + \frac{i}{\hbar} \sum_{k,l} \mathbf{d}_{nk} 
          \mathbf{g}^\ast_{ml}\hat{A}_{lk}
           - \frac{i}{\hbar} \sum_{k,l\neq m} \mathbf{d}_{km} 
           \mathbf{g}^\ast_{kl}             \hat{A}_{ln}\qquad
\end{eqnarray}
($m$ $\!\neq$ $\!n$), and
\begin{eqnarray}
\label{F3-3}
       \dot{\hat{A}}_{mm}&=&
\frac{i}{\hbar} \sum_{k} \bigl[ \mathbf{d}_{mk} 
 \mathbf{g}_{km}
        -\mathbf{d}_{km} \mathbf{g}^\ast_{km}\bigr]\hat{A}_{mm}
       \nonumber\\&&
  - \frac{i}{\hbar} \sum_{k}\bigl[ \mathbf{d}_{km} 
  \mathbf{g}_{mk}
  - \mathbf{d}_{mk} \mathbf{g}^\ast_{mk}\bigr]\hat{A}_{kk}
  \nonumber\\&&
  +\hat{C}_{mm}+\hat{F}_{mm},
\end{eqnarray}
with
\begin{eqnarray}
\label{F3-4}
\hat{C}_{mm}
&=&\frac{i}{\hbar} \sum_{k,l\neq m}
        \Bigl\{ \mathbf{d}_{mk} \mathbf{g}_{kl}\hat{A}_{ml}
        -\mathbf{d}_{km} \mathbf{g}^\ast_{kl}\hat{A}_{lm}\Bigr\}
\nonumber\\&&
           - \frac{i}{\hbar} \sum_{k,l\neq k}
           \Bigl\{ \mathbf{d}_{km} \mathbf{g}_{ml}
             \hat{A}_{kl}
           - \mathbf{d}_{nk} \mathbf{g}^\ast_{ml}
             \hat{A}_{lk}\Bigr\},
\end{eqnarray}
where $\hat{F}_{mn}$ denotes contributions from the free-field part in
Eq.~(\ref{eq111}). Taking expectation values with respect to the
internal atomic motion and the medium-assisted electromagnetic field,
with the density-matrix given by Eq.~(\ref{eq121}), we can use the
property (\ref{eq119}), finding that the terms $\hat{F}_{mn}$ do not
contribute. In the absence of (qua\-s\mbox{i)d}e\-ge\-ner\-acies such
that 
\begin{equation}
\label{F3-5}
|\tilde{\omega}_{mn}-\tilde{\omega}_{m'n'}|\gg {\textstyle\frac{1}{2}}
|\Gamma_m+\Gamma_n-\Gamma_{m'}-\Gamma_{n'}|,
\end{equation}
we may disregard couplings between different off-dia\-gonal
transitions and between off-diagonal and diagonal transitions and thus omit the
terms $\hat{B}_{mn}$ and $\hat{C}_{mm}$, hence upon using the
decomposition
\begin{equation}
\label{F4}
   \frac{i}{\hbar}\mathbf{d}_{nk}\mathbf{g}_{kn}
   (\hat{\mathbf{r}}_\mathrm{A})
        = - i \delta\omega_n^k(\hat{\mathbf{r}}_\mathrm{A})
        - {\textstyle\frac{1}{2}} 
	\Gamma_n^k(\hat{\mathbf{r}}_\mathrm{A}),
\end{equation}
where $\delta\omega_n^k(\hat{\mathbf{r}}_\mathrm{A})$ and
$\Gamma_n^k(\hat{\mathbf{r}}_\mathrm{A})$, respectively, are defined
according to Eqs.~(\ref{eq128}) and (\ref{eq130}) [with
$\bm{G}^{(1)}(\hat{\mathbf{r}}_\mathrm{A},\hat{\mathbf{r}}_\mathrm{A},
\omega)$ instead of $\bm{G}(\hat{\mathbf{r}}_\mathrm{A},
\hat{\mathbf{r}}_\mathrm{A},\omega)$ in Eq.~(\ref{eq128})],
Eqs.~(\ref{F3-1}) and (\ref{F3-3}) lead to Eqs.~(\ref{eq125}), 
(\ref{eq133}), and (\ref{eq134}).

%%%%%%%%%%%%%%%%%%%%%%%%%%%%%%%%%%%%%%%%%%%%%%%%%%%%%%%%%%%%%%%%%%%%%%

\section{Half space medium}
\label{AppG}

The equal-position scattering Green tensor for a semi-infinite half
space which contains for $z$ $\!<$ $\!0$ a homogeneous, dispersing,
and absorbing magnetodielectric medium reads for $z$ $\!>$ $\!0$
\cite{Chew}
\begin{eqnarray}
\label{G1}
\lefteqn{
\bm{G}^{(1)}(\mathbf{r}, \mathbf{r}, \omega)
=\frac{i}{8\pi}\int_0^\infty \mathrm{d}q\,
\frac{q}{\beta_0}\,e^{2i\beta_0z}
}
\nonumber\\&&\hspace{-.5ex}\times
\left\{\!r_s\!
\left(
\begin{array}{ccc}1&0&0\\ 0&1&0\\ 0&0&0\end{array}\right)
\!+r_p\,\frac{c^2}{\omega^2}\!
\left(
\begin{array}{rrr}
-\beta_0^2&0&0\\ 0&-\beta_0^2&0\\ 0&0&2q^2
\end{array}
\right)
\!\right\}\!,
\qquad
\end{eqnarray}
where
\begin{equation}
\label{G2}
r_s=\frac{\mu\beta_0-\beta}{\mu\beta_0+\beta},\qquad
r_p=\frac{\varepsilon\beta_0-\beta}{\varepsilon\beta_0+\beta}
\end{equation}
are the reflection coefficients for $s$- and $p$-polarized waves,
respectively ($\beta_0^2$ $\!=$ $\!\omega^2/c^2$ $\!-$ $\!q^2$
with $\mathrm{Im}\,\beta_0$ $\!>$ $\!0$,
$\beta^2$ $\!=$
$\!\varepsilon\mu\omega^2/c^2$ $\!-$
$\!q^2$ with $\mathrm{Im}\,\beta$ $\!>$ $\!0$).
For $q$ $\!\gg$ $\!|\omega|/c$ and $q$ $\!\gg$
$\!\sqrt{|\varepsilon\mu|}|\omega|/c$,
respectively, the approximations
\begin{equation}
\label{G3}
\beta_0\simeq iq, \qquad \beta\simeq iq
\end{equation}
can be made. Due to the exponential factor the integration interval is
effectively limited to values $q$ $\!\lesssim$ $\!1/z$.
In the short-distance limit $z \sqrt{|\varepsilon\mu|}|\omega|/c$
$\!\ll$ $\!1$, we therefore introduce a small error, if we extrapolate
the approximations (\ref{G3}) to the whole integral, resulting in
\begin{equation}
\label{G4}
 \bm{G}^{(1)}(\mathbf{r}, \mathbf{r}, \omega)
 =\frac{c^2}{32\pi\omega^2z^3}
 \frac{\varepsilon(\omega)-1}{\varepsilon(\omega)+1}
\left(
\begin{array}{ccc}1&0&0\\ 0&1&0\\ 0&0&2\end{array}\right).
\end{equation}
Note that the magnetic properties of the medium represented by the
permeability $\mu$ begin to contribute via terms proportional to
$1/z$.
Substitution of Eq.~(\ref{G4}) into the first term of
Eq.~(\ref{eq131}) for $\delta\omega_{nk}$ $\!=$ $\!\delta\omega_{10}$
yields Eq.~(\ref{eq152}).

In order to obtain Eq.~(\ref{eq152-1}), we recall Eq.~(\ref{G1}) to
write
\begin{eqnarray}
\label{G5}
\lefteqn{
\int_0^\infty\!\!\mathrm{d}u\,f(u)\bm{G}^{(1)}(\mathbf{r}, 
\mathbf{r},iu)
}
\nonumber\\&&\hspace{-1ex}
=\frac{1}{8\pi}\int_0^\infty\mathrm{d}uf(u)
\int_{u/c}^\infty \mathrm{d}b_0\,e^{-2b_0z}
\nonumber\\&&\hspace{-1ex}\times
\left\{\!r_s\!
\left(
\begin{array}{ccc}
1&0&0\\ 0&1&0\\ 0&0&0
\end{array}
\right)
\!-r_p\frac{c^2}{u^2}\!
\left(
\begin{array}{ccc}
b_0^2&0&0\\ 0&b_0^2&0\\ 0&0&2b_0^2-(
{\textstyle\frac{u}{c}})^2
\end{array}
\right)
\!\!\right\}\!\!,
\qquad
\end{eqnarray}
having changed the integration variable to the imaginary part of
$\beta_0$ ($\beta_0$ $\!=$ $\!ib_0$). Let $\omega_\mathrm{M}$ be a
characteristic frequency of the medium such that
\begin{equation}
\label{eq154}
\varepsilon(iu)-1\ll 1 \quad\mbox{for }u>\omega_\mathrm{M}.
\end{equation}
For $u$ $\!>$ $\!\omega_\mathrm{M}$, the approximation
$\beta\sim\beta_0$ holds, and consequently the reflection
coefficients $r_s$, $r_p$ are independent of $b_0$. The frequency
integral effectively extends up to frequencies of the order $c/z$,
hence in the short-range limit $z\omega_\mathrm{M}/c$ $\!\ll$ $\!1$ 
($\Rightarrow c/z$ $\!\gg$ $\!\omega_\mathrm{M}$) we introduce only a
small error by extrapolating this approximation to the whole frequency
integral. Performing the $b_0$ integral, retaining only leading-order
terms in $uz/c$ (in consistency with $z\omega_\mathrm{M}/c\ll 1$) we
derive 
\begin{eqnarray}
\label{G6}
&&\int_0^\infty\mathrm{d}uf(u)\bm{G}^{(1)}
(\mathbf{r}, \mathbf{r},iu)\nonumber\\
&&\quad=-\frac{c^2}{32\pi z^3}\int_0^\infty\mathrm{d}u\frac{f(u)}{u^2}
\frac{\varepsilon(iu)-1}{\varepsilon(iu)+1}
\left(
\begin{array}{ccc}1&0&0\\ 0&1&0\\ 0&0&2\end{array}\right)
\!.\qquad
\end{eqnarray}
Using Eq.~(\ref{G6}) [with 
$f(u)=u^2/(\tilde{\omega}_\mathrm{A}^2+u^2)$]
together with Eq.~(\ref{eq131}), we obtain Eq.~(\ref{eq152-1}).

%%%%%%%%%%%%%%%%%%%%%%%%%%%%%%%%%%%%%%%%%%%%%%%%%%%%%%%%%%%%%%%%%%%%%%

\end{document}